\documentclass[letterpaper]{JHEP3}

\usepackage{array}
\usepackage{amsmath}
\usepackage{amssymb}
\usepackage{graphics}
\usepackage[vcentermath]{youngtab}
\usepackage{longtable}

\newcommand{\ket}[1]{\left\lvert #1 \right\rangle}

\newcommand{\stln}{\setlength{\unitlength}{10pt}}
\newcommand{\fr}{\framebox(1,1){}}
\newcommand{\sfr}{\framebox(1,1)[bl]{\begin{picture}(1,1)(0,0)
                                      \put(0,0){\line(1,1){1}}
                                     \end{picture}
                                     }
                  }

\newcommand{\onebox}{\stln \lower2.0pt \hbox{\begin{picture}(1,1)(0,0)
                                              \put(0,0){\fr}
                                             \end{picture}
                                             }
                     }
\newcommand{\oneonebox}{\stln \lower7.0pt \hbox{\begin{picture}(1,2)(0,0)
                                                 \multiput(0,0)(0,1){2}{\fr}
                                                \end{picture}
                                                }
                        }
\newcommand{\twobox}{\stln \lower2.0pt \hbox{\begin{picture}(2,1)(0,0)
                                              \multiput(0,0)(1,0){2}{\fr}
                                             \end{picture}
                                             }
                     }
\newcommand{\onedotbox}{\stln \lower2.0pt \hbox{\begin{picture}(1,1)(0,0)
                                                 \put(0,0){\fr}
                                                 \put(0.5,0.5){\circle*{.3}}
                                                \end{picture}
                                                }
                        }
\newcommand{\oneonedotbox}{\stln \lower7.0pt \hbox{\begin{picture}(1,2)(0,0)
                                                    \multiput(0,0)(0,1){2}{\fr}
                                                    \multiput(0.5,0.5)(0,1){2}{\circle*{.3}}
                                                   \end{picture}
                                                   }
                           }
\newcommand{\twodotbox}{\stln \lower2.0pt \hbox{\begin{picture}(2,1)(0,0)
                                                 \multiput(0,0)(1,0){2}{\fr}
                                                 \multiput(0.5,0.5)(1,0){2}{\circle*{.3}}
                                                \end{picture}
                                                }
                        }
\newcommand{\sonebox}{\stln \lower2.0pt \hbox{\begin{picture}(1,1)(0,0)
                                               \put(0,0){\sfr}
                                              \end{picture}
                                              }
                      }
\newcommand{\soneonebox}{\stln \lower7.0pt \hbox{\begin{picture}(1,2)(0,0)
                                                  \multiput(0,0)(0,1){2}{\sfr}
                                                 \end{picture}
                                                 }
                         }
\newcommand{\stwobox}{\stln \lower2.0pt \hbox{\begin{picture}(2,1)(0,0)
                                               \multiput(0,0)(1,0){2}{\sfr}
                                              \end{picture}
                                              }
                      }
\newcommand{\sonedotbox}{\stln \lower2.0pt \hbox{\begin{picture}(1,1)(0,0)
                                                  \put(0,0){\sfr}
                                                  \put(0.5,0.5){\circle*{.3}}
                                                 \end{picture}
                                                 }
                         }
\newcommand{\soneonedotbox}{\stln \lower7.0pt \hbox{\begin{picture}(1,2)(0,0)
                                                     \multiput(0,0)(0,1){2}{\sfr}
                                                     \multiput(0.5,0.5)(0,1){2}{\circle*{.3}}
                                                    \end{picture}
                                                    }
                            }
\newcommand{\stwodotbox}{\stln \lower2.0pt \hbox{\begin{picture}(2,1)(0,0)
                                                  \multiput(0,0)(1,0){2}{\sfr}
                                                  \multiput(0.5,0.5)(1,0){2}{\circle*{.3}}
                                                 \end{picture}
                                                 }
                         }

\title{Oscillator Construction of Spectra of PP-Wave Superalgebras in Eleven Dimensions
\footnote{Work supported in part by the National Science Foundation
  under grant number PHY-0245337
and in part by the Basic Research Program of the Korea Science and
Engineering Foundation under grant number R01-2004-000-10651-0.}}

\author{Sudarshan Fernando \\
Physics Department, Gonzaga University, Spokane, WA 99258 \\
E-mail: \email{fernando@gonzaga.edu}}
\author{Murat G\"{u}naydin \\
Physics Department, Penn State University, 104 Davey Lab, University Park, PA 16802 \\
E-mail: \email{murat@phys.psu.edu}}
\author{Seungjoon Hyun \\
Institute of Physics and Applied Physics, Yonsei University, Seoul 120-749, Korea \\
E-mail: \email{hyun@phya.yonsei.ac.kr}}

\abstract{After reviewing  the oscillator realization of the symmetry superalgebra of the BMN matrix model on its
maximally supersymmetric plane-wave background and the construction of its zero-mode spectrum, we study a large
number of non-maximally supersymmetric pp-wave algebras in eleven dimensions  which are  obtained by various
restrictions from the maximally supersymmetric case (BMN model). We also show how to obtain their zero-mode
spectra, which we explicitly construct in some chosen examples. Except for some `exotic' or degenerate special
cases, we believe our study covers all possible interesting pp-wave superalgebras of this kind in eleven
dimensions.}

\begin{document}


\section{Introduction}
\label{Sec:Introduction}

In \cite{Pen76}, Penrose showed that any spacetime (i.e., any solution of the Einstein field equations) has a limit
where it looks like a plane-wave. This limit can be thought of as a first-order approximation to the spacetime along a
null-geodesic. Although Penrose's original work focused on four dimensional spacetimes, he pointed out that his
argument could be extended to any higher dimension without any difficulties. G\"{u}ven later showed that Penrose's
idea could be applied to supergravity backgrounds in ten and eleven dimensions as well \cite{Guev00}, by extending the
limiting procedure to the other fields present in the supergravity theory.

There are precisely four types of maximally supersymmetric solutions of eleven dimensional supergravity that have
been known for some time. Three of them are the familiar cases of eleven dimensional flat space (Minkowski space
and its toroidal compactifications), $AdS_7 \times S^4$ and $AdS_4 \times S^7$, and the fourth is the
Kowalski-Glikman (KG) solution \cite{Glik84}. Similarly type IIB supergravity was originally known to have two
maximally supersymmetric solutions, one in ten dimensional flat space and the other in $AdS_5 \times S^5$, and
recently it was shown in \cite{BFHP01} that there is another maximally supersymmetric type IIB solution (BFHP
solution) which is analogous to the KG space. Even though these KG and BFHP solutions were originally constructed
by solving the equations of motion, it was later shown that they could be obtained as ``Penrose limits'' of
$AdS_{7(4)} \times S^{4(7)}$ and $AdS_5 \times S^5$ respectively \cite{BFHP02}.

In addition, it turned out that once one goes to the light-cone gauge, type IIB superstring theory $\sigma$-model
reduces to a free, massive two dimensional model in the type IIB plane-wave (BFHP) background \cite{Met01,MT02}, and
therefore becomes exactly solvable (unlike its counterpart in $AdS_5 \times S^5$).

Then in \cite{BMN02} it was argued that in the AdS/CFT context, taking the Penrose limit on the string theory side
corresponds to restricting the gauge theory to operators with a large charge under a certain $U(1)$ subgroup of the
$R$-symmetry group $SU(4) \approx SO(6)$ and simultaneously taking the large $N$ limit. This sector of the four
dimensional $\mathcal{N} = 4$ $SU(N)$ super Yang-Mills theory, named the BMN sector, consists of operators that have
large conformal dimension $\Delta$ and large $R$-charge $J$, such that $\Delta - J$ remains finite.

Interestingly the plane-wave string / gauge theory duality is perturbatively accessible from either side of the
correspondence \cite{SS03,Plef03}. This provides a novel opportunity to set up a ``dictionary'' between string
states and operators in the super Yang-Mills theory, and compare their spectra in a perturbative expansion on
both sides of the duality. The plane-wave string / gauge theory duality was the first successful attempt to test
the AdS/CFT correspondence outside the low energy supergravity regime.

For a detailed account of the development of the subject and various aspects of the duality, we refer the reader to
\cite{SS03,Plef03} and the references therein.

In this paper, we take an algebraic approach and study a broad class of pp-wave superalgebras originating from eleven
dimensional supergravity and construct their spectra. Our study includes not only the well-known maximally
supersymmetric pp-wave algebra in eleven dimensions, but an extensive list of non-maximally supersymmetric pp-wave
algebras as well. Furthermore, we explicitly identify the symmetry superalgebras of some pp-wave solutions that have
been constructed in the literature.

The organization of the paper is as follows. In section \ref{Sec:Oscillators} we give a general review of the
oscillator method, which we use to realize the pp-wave algebras and construct their spectra, and also the oscillator
realization of $\mathfrak{osp}(8^*|4)$ and $\mathfrak{osp}(8|4,\mathbb{R})$, the symmetry superalgebras of $AdS_7
\times S^4$ and $AdS_4 \times S^7$ compactifications of the eleven dimensional supergravity. Readers who are familiar
with the oscillator method may proceed directly to section \ref{Sec:MaxSusy}.

We present our main results in sections \ref{Sec:MaxSusy} and \ref{Sec:NonMaxSusy}. Section \ref{Sec:MaxSusy} is a
review of our earlier work on maximally supersymmetric pp-wave algebra in eleven dimensions, $\mathfrak{su}(4|2) ~
\circledS ~ \mathfrak{h}^{9,8}$ (where $\mathfrak{h}^{9,8}$ denotes a super-Heisenberg algebra) \cite{FGP02}. We also
review how to obtain the pp-wave limit of \emph{any} given superalgebra with a 3-grading (with respect to a maximal
compact subsuperalgebra as defined in equations \eqref{3grading1}-\eqref{3grading3}) in oscillator method, as an
In\"{o}n\"{u}-Wigner contraction \cite{IW53}.

Then by restricting the maximal compact part, $\mathfrak{su}(4|2)$, to its various semi-simple subsuperalgebras,
in section \ref{Sec:NonMaxSusy} we obtain a large number of non-maximally supersymmetric pp-wave algebras in
eleven dimensions. It should be noted that some of them can be compactified to ten dimensions to their respective
type IIA theories. In all the pp-wave algebras, elements of $\mathfrak{g}^{(\pm 1)}$ subspace generate a
super-Heisenberg algebra with a central charge. It is this generic feature of all pp-wave algebras that allows us
to take various decompositions of $\mathfrak{g}^{(0)}$ and obtain a variety of different pp-wave superalgebras
with different numbers of supersymmetries.

Section \ref{Sec:Discussion} finally concludes this paper with a short summary and a discussion. Appendix
\ref{ApSec:SupergravitySpectra} contains the spectra of the eleven dimensional supergravity with maximal
supersymmetry, as reference, obtained by using the oscillator method (section \ref{Sec:Oscillators}) as first
presented in \cite{GvNW85,GW86}.


\section{Oscillator Realization of Positive Energy UIRs of $OSp(8^*|4)$ and $OSp(8|4,\mathbb{R})$}
\label{Sec:Oscillators}

In this section, we give a short account of the oscillator construction of positive energy unitary irreducible
representations (UIRs) of the supergroups $OSp(8^*|4)$ and $OSp(8|4,\mathbb{R})$. These two supergroups are,
respectively, the symmetry groups of $AdS_7 \times S^4$ and $AdS_4 \times S^7$ compactifications of the eleven
dimensional supergravity. Readers who are familiar with the oscillator method \cite{GS82a,BG83,mg88} may skip
this section and go directly to section \ref{Sec:MaxSusy}.

A simple noncompact (super)group $G$ that has a maximal compact sub(super)group $G^{(0)}$, such that $G/G^{(0)}$
is a Hermitian (super)symmetric space,  admits UIRs of the lowest weight type \footnote{ For noncompact groups
the condition that $G/G^{(0)}$ be Hermitian symmetric is both necessary and sufficient. For noncompact
supergroups this is sufficient , but not necessary. See \cite{mg88} for a generalization of the oscillator method
to those noncompact supergroups for which this condition is not satisfied.}. This maximal compact sub(super)group
has an abelian factor, i.e. $G^{(0)} = H \times U(1)$, and the Lie (super)algebra $\mathfrak{g}$ of $G$ has a
3-grading with respect to the Lie (super)algebra $\mathfrak{g}^{(0)}$ of $G^{(0)}$:

\begin{equation}
\mathfrak{g} = \mathfrak{g}^{(-1)} \oplus \mathfrak{g}^{(0)} \oplus \mathfrak{g}^{(+1)} \,.
\label{3grading1}
\end{equation}
This simply means that the (super-)commutators of elements of grade $k$ and $l$ ($= 0, \pm 1$) spaces satisfy
\begin{equation}
\left[ \mathfrak{g}^{(k)} , \mathfrak{g}^{(l)} \right\} \subseteq \mathfrak{g}^{(k+l)}
\end{equation}
with $\mathfrak{g}^{(k+l)} = 0$ for $\left| k+l \right| > 1$.

The 3-grading is determined by the generator $E$ of the $U(1)$ factor of the maximal compact sub(super)group
i.e.:
\begin{equation}
\begin{split}
\left[ E , \mathfrak{g}^{(+1)} \right] &=   \mathfrak{g}^{(+1)} \\
\left[ E , \mathfrak{g}^{(-1)} \right] &= - \mathfrak{g}^{(-1)} \\
\left[ E , \mathfrak{g}^{(0)} \right]  &=   0 \,.
\end{split}
\label{3grading3}
\end{equation}

Typically, in physical applications,  $E$ is the energy operator and the lowest weight UIRs correspond to
positive energy representations. To construct these representations (of the lowest weight type) in the Fock space
$\mathcal{F}$ of all the oscillators, one chooses a set of states $\ket{\Omega}$, called a ``lowest weight
vector'' or a ``ground state'', which transforms irreducibly under $G^{(0)} = H \times U(1)$ and is annihilated
by all the generators in $\mathfrak{g}^{(-1)}$ subspace.\footnote{We should  note that  $\ket{\Omega}$ consists ,
in general, of a set of states transforming irreducibly under $G^{(0)}$ which can be uniquely labelled by a
single vector that is referred to as the lowest weight vector of that
irrep in the mathematics literature. By an
abuse of terminology the entire irrep $\ket{\Omega}$ of $G^{(0)}$ is sometimes referred to as the "lowest weight
vector" of the unitary representation of $G$.} Then by successively acting on $\ket{\Omega}$ with the generators
in $\mathfrak{g}^{(+1)}$, one obtains an infinite set of states
\begin{equation}
\ket{\Omega}
 \quad , \quad
\mathfrak{g}^{(+1)} \ket{\Omega}
 \quad , \quad
\mathfrak{g}^{(+1)} \mathfrak{g}^{(+1)} \ket{\Omega}
 \quad , \quad \dots \quad
\label{UIRconstruction}
\end{equation}
which forms a UIR of the lowest weight (positive energy) type of $G$. Any two $\ket{\Omega}$ that transform in
the same irreducible representation of $G^{(0)} = H \times U(1)$ will lead to  equivalent UIRs of $G$. The
irreducibility of the resulting representation of the noncompact (super)group $G$ follows from the irreducibility
of the ``lowest weight vector'' $\ket{\Omega}$ with respect to the maximal compact sub(super)group $G^{(0)}$.


\subsection{Symmetry supergroup $OSp(8^*|4)$ of $AdS_7 \times S^4$ compactification}
\label{SubSec:AdS7S4}

The compactification of eleven dimensional simple supergravity on four-sphere leads to maximal $\mathcal{N} = 4$
supergravity in seven dimensional AdS space. The symmetry supergroup of this $d = 7$ supergravity is $OSp(8^*|4)
\approx OSp(6,2|4)$, whose even subgroup is $SO^*(8) \times USp(4) \approx SO(6,2) \times SO(5)$. It was shown in
\cite{GvNW85}, how to construct  the positive energy unitary representations of this supergroup using the
oscillator method, and here we outline a summary of their results that are relevant to the current discussion.


\subsubsection{Representations of $SO^*(8) \approx SO(6,2)$ via the oscillator method}
\label{SubSubSec:SO*8}

The noncompact group $SO^*(8)$, which is isomorphic to $SO(6,2)$ acts as  the conformal group in $d=6$ and as
the anti-de Sitter group in $d=7$. The 3-grading (as in equations \eqref{3grading1}-\eqref{3grading3}) of the Lie
algebra $\mathfrak{so}^*(8) \approx \mathfrak{so}(6,2)$ is defined with respect to its maximal compact subalgebra
$\mathfrak{u}(4) = \mathfrak{su}(4) \oplus \mathfrak{u}(1)$.

To construct the positive energy UIRs of $SO^*(8)$, one introduces an arbitrary number $P$ pairs (``generations''
or ``colors'') of \emph{bosonic} annihilation and creation operators $a_i(K)$, $b_i(K)$ and $a^i(K) :=
a_i(K)^\dag$, $b^i(K) := b_i(K)^\dag$ ($i=1,2,3,4$ \,; $K=1,\dots,P$), which transform as $\overline{\mathbf{4}}
= \onedotbox$ and $\mathbf{4} = \onebox$ representations\footnote{~The Young tableau $\onebox$ corresponds to the
\emph{contravariant} fundamental representation of $SU(n)$, and $\onedotbox$ corresponds to the complex conjugate
\emph{covariant} fundamental representation of $SU(n)$.} of the maximal compact subgroup $U(4) = SU(4) \times
U(1)$ and satisfy the usual canonical commutation relations:
\begin{equation}
\begin{split}
\left[ a_i(K) , a^j(L) \right] = \delta_i^j \delta_{KL}
 \qquad &
\left[ b_i(K) , b^j(L) \right] = \delta_i^j \delta_{KL} \\
\left[ a_i(K) , a_j(L) \right] = 0 & = \left[ a^i(K) , a^j(L) \right] \\
\left[ b_i(K) , b_j(L) \right] = 0 & = \left[ b^i(K) , b^j(L) \right] \,.
\end{split}
\label{bosonicSU4oscillators}
\end{equation}

The Lie algebra $\mathfrak{so}^*(8)$ is now realized as bilinears of these bosonic oscillators in the following manner:
\begin{equation}
\begin{split}
A_{ij}   & = \vec{a}_i \cdot \vec{b}_j - \vec{a}_j \cdot \vec{b}_i
           = \vec{a}_{[i} \cdot \vec{b}_{j]}
         ~ = ~ \oneonedotbox \\
M^i_{~j} & = \vec{a}^i \cdot \vec{a}_j + \vec{b}_j \cdot \vec{b}^i \\
A^{ij}   & = \vec{a}^i \cdot \vec{b}^j - \vec{a}^j \cdot \vec{b}^i
           = \vec{a}^{[i} \cdot \vec{b}^{j]}
         ~ = ~ \oneonebox \,.
\label{SO*8generators}
\end{split}
\end{equation}
Note that we denote the sum over all ``colors'' by $\vec{a}_i \cdot \vec{b}_j := \sum_{K=1}^P a_i(K) b_j(K)$.

The generators of $\mathfrak{g}^{(-1)}$ and $\mathfrak{g}^{(+1)}$ subspaces commute to give
\begin{equation}
\left[ A_{ij} , A^{kl} \right] = \delta_i^k M^l_{~j} - \delta_i^l M^k_{~j}
                                 - \delta_j^k M^l_{~i} + \delta_j^l M^k_{~i} \,.
\end{equation}
The generators $M^i_{~j}$ form the Lie algebra $\mathfrak{u}(4)$, while $A_{ij}$ and $A^{ij}$, both transforming as
$\mathbf{6}$ of $\mathfrak{su}(4)$ with opposite charges under the $\mathfrak{u}(1)$ generator $M^i_{~i}$, extend this
$\mathfrak{u}(4)$ to the Lie algebra $\mathfrak{so}^*(8)$. The $\mathfrak{u}(1)$ charge $M^i_{~i}$ gives the AdS
energy
\begin{equation}
E = \frac{1}{2} M^i_{~i} = \frac{1}{2} N_B + 2 P \,,
\label{E74}
\end{equation}
where $N_B = \vec{a}^i \cdot \vec{a}_i + \vec{b}^i \cdot \vec{b}_i$ is the bosonic number operator.

The vacuum state $\ket{0}$ is defined by:
\begin{equation}
a_i(K) \ket{0} = 0 = b_i(K) \ket{0}
\end{equation}
where $i=1,2,3,4$ \,; $K=1,\dots,P$.

Now the lowest weight UIRs of $SO^*(8) \approx SO(6,2)$ can be constructed by choosing sets of states $\ket{\Omega}$,
that transform irreducibly under the maximal compact subgroup $SU(4) \times U(1)$ and are annihilated by all the
generators in $\mathfrak{g}^{(-1)}$ subspace, $A_{ij}$. These UIRs, constructed by acting on $\ket{\Omega}$ repeatedly
with the elements of $\mathfrak{g}^{(+1)}$, $A^{ij}$ (as in equation \eqref{UIRconstruction}), are uniquely determined
by these lowest weight vectors $\ket{\Omega}$ and can be identified with AdS fields in $d=7$ or conformal fields in
$d=6$ \cite{GvNW85,GT99,FGT01}.

If one chooses only one pair of these oscillators (i.e. $P=1$) we obtain all the doubleton representations of
$SO^*(8)$  which, as $AdS_7$ representations,   do not have a Poincar\'{e} limit in $d=7$. The Poincar\'{e} mass
operator in $d=6$ vanishes identically for these representations, and therefore they correspond to massless
conformal fields in $d=6$. The tensoring of two copies of these doubletons, or in other words taking $P=2$,
produces massless representations of $AdS_7$, but in the $CFT_6$ sense they correspond to massive conformal
fields. Tensoring more than two copies of doubletons ($P>2$) leads to representations that are massive both in
the $AdS_7$ and $CFT_6$ sense \cite{GvNW85,GT99,FGT01}.


\subsubsection{Representations of $USp(4) \approx SO(5)$ via the oscillator method}
\label{SubSubSec:SO5}

Unlike $SO^*(8) \approx SO(6,2)$, the group $USp(4) \approx SO(5)$ is compact. Therefore, the UIRs of $SO(5)$ are
finite dimensional. In constructing the UIRs of $USp(4) \approx SO(5)$, one introduces $P$ pairs of
\emph{fermionic} annihilation and creation operators $\alpha_\mu(K)$, $\beta_\mu(K)$ and $\alpha^\mu(K) :=
\alpha_\mu(K)^\dag$, $\beta^\mu(K) := \beta_\mu(K)^\dag$ ($\mu =1,2$ \,; $K=1,\dots,P$), which transform as
$\overline{\mathbf{2}} = \onedotbox$ and $\mathbf{2} = \onebox$, respectively, with respect to the subgroup $U(2)
= SU(2) \times U(1)$. They satisfy the canonical anti-commutation relations:
\begin{equation}
\begin{split}
\left\{ \alpha_\mu(K) , \alpha^\nu(L) \right\} = \delta_\mu^\nu \delta_{KL}
 \qquad &
\left\{ \beta_\mu(K) , \beta^\nu(L) \right\}   = \delta_\mu^\nu \delta_{KL} \\
\left\{ \alpha_\mu(K) , \alpha_\nu(L) \right\} = 0 & = \left\{ \alpha^\mu(K) , \alpha^\nu(L) \right\} \\
\left\{ \beta_\mu(K) , \beta_\nu(L) \right\}   = 0 & = \left\{ \beta^\mu(K) , \beta^\nu(L) \right\} \,.
\end{split}
\label{fermionicSU2oscillators}
\end{equation}

The Lie algebra $\mathfrak{so}(5)$ is now realized as the following bilinears of these fermionic oscillators:
\begin{equation}
\begin{split}
A_{\mu\nu}   & = \vec{\alpha}_\mu \cdot \vec{\beta}_\nu + \vec{\alpha}_\nu \cdot \vec{\beta}_\mu
               = \vec{\alpha}_{(\mu} \cdot \vec{\beta}_{\nu)}
             ~ = ~ \twodotbox \\
M^\mu_{~\nu} & = \vec{\alpha}^\mu \cdot \vec{\alpha}_\nu - \vec{\beta}_\nu \cdot \vec{\beta}^\mu \\
A^{\mu\nu}   & = \vec{\alpha}^\mu \cdot \vec{\beta}^\nu + \vec{\alpha}^\nu \cdot \vec{\beta}^\mu
               = \vec{\alpha}^{(\mu} \cdot \vec{\beta}^{\nu)}
             ~ = ~ \twobox \,.
\label{SO5generators}
\end{split}
\end{equation}
Therefore, the generators of $\mathfrak{g}^{(-1)}$ and $\mathfrak{g}^{(+1)}$ subspaces satisfy
\begin{equation}
\left[ A_{\mu\nu} , A^{\sigma\rho} \right] = \delta_\mu^\sigma M^\rho_{~\nu} + \delta_\mu^\rho M^\sigma_{~\nu}
                                             + \delta_\nu^\sigma M^\rho_{~\mu} + \delta_\nu^\rho M^\sigma_{~\mu} \,.
\end{equation}
The above $M^\mu_{~\nu}$ generate the Lie algebra $\mathfrak{u}(2)$, and $A_{\mu\nu}$ and $A^{\mu\nu}$, both
transforming as $\mathbf{3}$ of $\mathfrak{su}(2)$ with opposite charges with respect to the $\mathfrak{u}(1)$
generator, extend it to that of $\mathfrak{so}(5)$. This $\mathfrak{u}(1)$ charge,  with respect to which the
3-grading is defined, is
\begin{equation}
J = \frac{1}{2} M^\mu_{~\mu} = \frac{1}{2} N_F - P \,,
\label{J74}
\end{equation}
where $N_F = \vec{\alpha}^\mu \cdot \vec{\alpha}_\mu + \vec{\beta}^\mu \cdot \vec{\beta}_\mu$ is the fermionic number
operator.

Once again, the vacuum state is annihilated by all the annihilation operators:
\begin{equation}
\alpha_\mu(K) \ket{0} = 0 = \beta_\mu(K) \ket{0}
\end{equation}
for all $\mu=1,2$ \,; $K=1,\dots,P$.

The choice of the lowest weight vectors $\ket{\Omega}$ (that transform irreducibly under $U(2)$ and are annihilated
by the generators of $\mathfrak{g}^{(-1)}$ subspace) and the construction of the representations of $USp(4) \approx
SO(5)$ are now done analogous to the previous section. However, as mentioned before, due to the fermionic nature of
the oscillators in this case, equation \eqref{UIRconstruction} produces only finite dimensional representations.


\subsubsection{Representations of $OSp(8^*|4)$ via the oscillator method}
\label{SubSubSec:OSp8*4}

The superalgebra $\mathfrak{osp}(8^*|4)$ has a 3-grading with respect to its maximal compact subsuperalgebra
$\mathfrak{u}(4|2)$, which has an even subalgebra $\mathfrak{u}(4) \oplus \mathfrak{u}(2)$.

Therefore, to construct the UIRs of $OSp(8^*|4)$, one defines the $U(4|2)$ covariant super-oscillators as
follows: \footnote{ We shall use the Young supertableaux for labelling the representations of the compact
supergroups $SU(n|m)$ developed by Bars and collaborators \cite{bars}.}
\begin{equation}
\begin{split}
\xi_A(K)  & = \left( \begin{matrix} a_i(K) \cr \alpha_\mu(K) \cr \end{matrix} \right)
          ~ = ~ \sonedotbox
 \qquad
\xi^A(K)   := \xi_A(K)^\dag
            = \left( \begin{matrix} a^i(K) \cr \alpha^\mu(K) \cr \end{matrix} \right)
          ~ = ~ \sonebox \\
\eta_A(K) & = \left( \begin{matrix} b_i(K) \cr \beta_\mu(K) \cr \end{matrix} \right)
          ~ = ~ \sonedotbox
 \qquad
\eta^A(K)  := \eta_A(K)^\dag
            = \left( \begin{matrix} b^i(K) \cr \beta^\mu(K) \cr \end{matrix} \right)
          ~ = ~ \sonebox
\end{split}
\end{equation}
where $A=1,2,3,4|1,2$ \,; $K=1,\dots,P$. They satisfy the super-commutation relations:
\begin{equation}
\left[ \xi_A(K) , \xi^B(L) \right\}   = \delta_A^B \delta_{KL}
 \qquad
\left[ \eta_A(K) , \eta^B(L) \right\} = \delta_A^B \delta_{KL}
\end{equation}
where the super-commutators are defined as
\begin{equation*}
\left[ \xi_A(K) , \xi^B(L) \right\} := \xi_A(K) \xi^B(L) - (-1)^{(\mathrm{deg}A)(\mathrm{deg}B)}\xi^B(L) \xi_A(K) \,,
\end{equation*}
etc., with $\mathrm{deg}A = 0$ ($\mathrm{deg}A = 1$) if $A$ is a bosonic (fermionic) index.

Now, the Lie superalgebra $\mathfrak{osp}(8^*|4)$ can be realized as the following bilinears:
\begin{equation}
\begin{split}
\mathcal{A}_{AB}   & = \vec{\xi}_A \cdot \vec{\eta}_B - \vec{\eta}_A \cdot \vec{\xi}_B
                     = \vec{\xi}_{[A} \cdot \vec{\eta}_{B]}
              \qquad = ~ \soneonedotbox \\
\mathcal{M}^A_{~B} & = \vec{\xi}^A \cdot \vec{\xi}_B +
                       (-1)^{(\mathrm{deg}A)(\mathrm{deg}B)} \vec{\eta}_B \cdot \vec{\eta}^A \\
\mathcal{A}^{AB}   & = \vec{\xi}^A \cdot \vec{\eta}^B - \vec{\eta}^A \cdot \vec{\xi}^B
                     = \vec{\xi}^{[A} \cdot \vec{\eta}^{B]}
              \qquad = ~ \soneonebox \,.
\label{OSp8*4generators}
\end{split}
\end{equation}
Clearly, $\mathcal{M}^A_{~B}$ generate the $\mathfrak{g}^{(0)}$ subsuperalgebra $\mathfrak{u}(4|2)$, while
$\mathcal{A}_{AB}$ and $\mathcal{A}^{AB}$, which correspond to $\mathfrak{g}^{(-1)}$ and $\mathfrak{g}^{(+1)}$
subspaces respectively, extend this to the full $\mathfrak{osp}(8^*|4)$ superalgebra. It is worth noting that the
above 3-grading is defined with respect to the abelian $\mathfrak{u}(1)$ charge in $\mathfrak{u}(4|2) =
\mathfrak{su}(4|2) \oplus \mathfrak{u}(1)$:
\begin{equation}
C = \frac{1}{2} \mathcal{M}^A_{~A}
  = \frac{1}{2} \left( N_B + N_F \right) + P
  = E + J \,.
\label{OSp8*4C}
\end{equation}

Furthermore we should stress that sixteen of the supersymmetry generators ($Q^i_{~\mu} := \mathcal{M}^i_{~\mu}$
and $Q^\nu_{~j} := \mathcal{M}^\nu_{~j}$) belong to the subspace $\mathfrak{g}^{(0)}$, and the other sixteen
($Q_{i\mu} := \mathcal{A}_{i\mu}$ and $Q^{j\nu} := \mathcal{A}^{j\nu}$) belong to the subspace
$\mathfrak{g}^{(-1)} \oplus \mathfrak{g}^{(+1)}$.

The super-Fock space vacuum $\ket{0}$ is defined by
\begin{equation}
\xi_A(K) \ket{0} = 0 = \eta_A(K) \ket{0}
\end{equation}
where $A=1,2,3,4|1,2$ \,; $K=1,\dots,P$. Given this super-oscillator realization, one can easily construct the
positive energy UIRs of $OSp(8^*|4)$ by first choosing sets of states $\ket{\Omega}$ in the super-Fock space
$\mathcal{F}$ that transform irreducibly under $U(4|2)$ and are annihilated by all the generators of
$\mathfrak{g}^{(-1)}$ subspace, i.e. $\mathcal{A}_{AB}$, and then repeatedly acting on them with the generators
of $\mathfrak{g}^{(+1)}$, $\mathcal{A}^{AB}$.

By choosing only one pair ($P=1$) of super-oscillators, one can build all the doubleton representations of
$OSp(8^*|4)$, and as $AdS_7$ supermultiplets they do not have a Poincar\'{e} limit in $d=7$. On the other hand by
choosing two pairs ($P=2$) and more  ($P>2$), one can obtain the massless and massive $AdS_7$ supermultiplets ,
respectively.

In \cite{GvNW85}, the spectrum of the eleven dimensional supergravity compactified to $AdS_7$ over $S^4$
\cite{PTvN84} was  fitted into an infinite tower of UIRs of $OSp(8^*|4)$. The ``$CPT$ self-conjugate'' doubleton
supermultiplet, obtained by starting from $\ket{\Omega} = \ket{0}$ for $P=1$, decouples from the spectrum as
local gauge degrees of freedom, but the entire physical spectrum can be obtained by tensoring an arbitrary number
of its copies and restricting ourselves to the ``$CPT$ self-conjugate'' vacuum supermultiplets. The spectrum of
the eleven dimensional supergravity on $AdS_7 \times S^4$ \cite{GvNW85} is reproduced in Table \ref{Table:AdS7S4}
of appendix \ref{ApSec:SupergravitySpectra}.


\subsection{Symmetry supergroup $OSp(8|4,\mathbb{R})$ of $AdS_4 \times S^7$ compactification}
\label{SubSec:AdS4S7}

The compactification of eleven dimensional supergravity to $AdS_4$ space on the seven-sphere, $S^7$ has the
symmetry supergroup $OSp(8|4,\mathbb{R})$, which has an even subgroup $SO(8) \times Sp(4,\mathbb{R})$. The
compact group $SO(8)$ is the isometry group of the seven sphere $S^7$ and  while $Sp(4,\mathbb{R})$ acts as the
isometry group of the AdS space in $d=4$. In \cite{GW86}, a detailed account of how to construct all the positive
energy UIRs of this supergroup was given, and here we briefly outline their results that are pertinent to our
work.


\subsubsection{Representations of $Sp(4,\mathbb{R}) \approx SO(3,2)$ via the oscillator method}
\label{SubSubSec:Sp4R}

The noncompact group $Sp(4,\mathbb{R}) $, which is the covering group of $SO(3,2)$,  acts as AdS group in four
dimensions, as well as the conformal group in three dimensions. The 3-grading of the Lie algebra
$\mathfrak{sp}(4,\mathbb{R}) \approx \mathfrak{so}(3,2)$, as usual, is defined with respect to its maximal
compact subalgebra $\mathfrak{u}(2) = \mathfrak{su}(2) \oplus \mathfrak{u}(1)$.

Therefore, to construct the positive energy UIRs of $Sp(4,\mathbb{R})$, one introduces an arbitrary number $n$
``colors'' of bosonic annihilation and creation operators. However, unlike in the previous case of $SO^*(8)$,
where one had to choose an even number of oscillators $a_i(1),\dots,a_i(P)$ \,; $b_i(1),\dots,b_i(P)$, here one
also has the freedom of choosing an odd as well as an even number of colors.

Choosing an even number of oscillators constitutes taking $n=2P$ annihilation operators $a_i(K)$, $b_i(K)$ and
their Hermitian conjugate creation operators $a^i(K)$, $b^i(K)$ ($i=1,2$ \,; $K=1,\dots,P$), which transform
covariantly and contravariantly in $\overline{\mathbf{2}} = \onedotbox$ and $\mathbf{2} = \onebox$
representations, respectively, with respect to $SU(2)$. On the other hand, an odd number $n=2P+1$ of oscillators
can be chosen by taking an extra oscillator $c_i$ and its Hermitian conjugate $c^i$, in addition to the above
$2P$ oscillators. They satisfy  the following canonical commutation relations :
\begin{equation}
\left[ a_i(K) , a^j(L) \right] = \delta_i^j \delta_{KL}
 \qquad
\left[ b_i(K) , b^j(L) \right] = \delta_i^j \delta_{KL}
 \qquad
\left[ c_i , c^j \right]       = \delta_i^j
 \quad \mbox{(if present)} \,.
\label{bosonicSU2oscillators}
\end{equation}

Thus, the Lie algebra $\mathfrak{sp}(4,\mathbb{R})$ is realized as bilinears of these bosonic oscillators in the
following manner:
\begin{equation}
\begin{split}
A_{ij}   & = \vec{a}_i \cdot \vec{b}_j + \vec{a}_j \cdot \vec{b}_i + \epsilon \; c_i c_j
           = \vec{a}_{(i} \cdot \vec{b}_{j)} + \frac{\epsilon}{2} \, c_{(i} c_{j)}
         ~ = ~ \twodotbox \\
M^i_{~j} & = \vec{a}^i \cdot \vec{a}_j + \vec{b}_j \cdot \vec{b}^i
             + \frac{\epsilon}{2} \left( c^i c_j + c_j c^i \right) \\
A^{ij}   & = \vec{a}^i \cdot \vec{b}^j + \vec{a}^j \cdot \vec{b}^i + \epsilon \; c^i c^j
           = \vec{a}^{(i} \cdot \vec{b}^{j)} + \frac{\epsilon}{2} \, c^{(i} c^{j)}
         ~ = ~ \twobox
\label{Sp4Rgenerators}
\end{split}
\end{equation}
where $\epsilon = 0$ ($\epsilon = 1$) if the number of oscillators $n$ is even (odd). The generators of
$\mathfrak{g}^{(-1)}$ and $\mathfrak{g}^{(+1)}$ subspaces commute as follows:
\begin{equation}
\left[ A_{ij} , A^{kl} \right] = \delta_i^k M^l_{~j} + \delta_i^l M^k_{~j} + \delta_j^k M^l_{~i} + \delta_j^l M^k_{~i}
 \,.
\end{equation}
The generators $M^i_{~j}$ form the maximal compact subalgebra $\mathfrak{u}(2)$ of $\mathfrak{sp}(4,\mathbb{R})$,
while $A_{ij}$ and $A^{ij}$, both transforming as $\mathbf{3}$ of $\mathfrak{su}(2)$ with opposite charges under the
$\mathfrak{u}(1)$ generator $M^i_{~i}$, extend it to the Lie algebra $\mathfrak{sp}(4,\mathbb{R})$. This
$\mathfrak{u}(1)$ charge $M^i_{~i}$ is given by
\begin{equation}
E = \frac{1}{2} M^i_{~i} = \frac{1}{2} N_B + P + \frac{\epsilon}{2}
\label{E47}
\end{equation}
where $N_B = \vec{a}^i \cdot \vec{a}_i + \vec{b}^i \cdot \vec{b}_i + \epsilon \; c^i c_i$ is the bosonic number
operator.

The vacuum state $\ket{0}$ is annihilated by all $a_i(K)$ and $b_i(K)$ ($i=1,2$ \,; $K=1,\dots,P$), as well as, if
present, by all $c_i$ ($i=1,2$). The lowest weight UIRs of $Sp(4,\mathbb{R}) \approx SO(3,2)$ can now be constructed,
as explained before, by choosing sets of states $\ket{\Omega}$ that transform irreducibly under $U(2) = SU(2) \times
U(1)$ and are annihilated by all the generators in $\mathfrak{g}^{(-1)}$ subspace, and acting repeatedly on them with
the generators of $\mathfrak{g}^{(+1)}$.

By choosing a single oscillator (which corresponds to $P=0$, and hence $n=1$), one can construct both singleton
representations of $Sp(4,\mathbb{R})$, and they do not have a Poincar\'{e} limit in $d=4$ \cite{FF78,Frons75}.
These are the same representations of $AdS_4$ group $SO(3,2)$ that were discovered by Dirac \cite{Dirac63} who
referred to them as the ``remarkable representations''. The singleton  representations correspond to massless
conformal fields in three dimensions. Tensoring two copies of these singletons (taking $n=2$) produces massless
representations of $AdS_4$, which are, however, massive in the $CFT_3$ sense. When more than two copies are
tensored  ($n>2$) they lead to representations that correspond to  massive  fields in $AdS_4$ as well as being
massive as conformal fields in $d=3$ \cite{GW86}.


\subsubsection{Representations of $SO(8)$ via the oscillator method}
\label{SubSubSec:SO8}

The compact group $SO(8)$ has a 3-grading structure with respect to its subgroup $U(4) = SU(4) \times U(1)$.
Therefore, to construct the  UIRs of $SO(8)$, one introduces fermionic annihilation and creation operators that
transform as $\overline{\mathbf{4}} = \onedotbox$ and $\mathbf{4} = \onebox$ representations of $SU(4)$.

To work with an even number of fermionic oscillators ($n=2P$), one may take $\alpha_\mu(K)$, $\beta_\mu(K)$
($\mu=1,2,3,4$ \,; $K=1,\dots,P$) and their Hermitian conjugates $\alpha^\mu(K)$, $\beta^\mu(K)$ , but on the
other hand, to work with an odd number of colors of oscillators ($n=2P+1$), one needs,  in addition to the above,
another oscillator $\gamma_\mu$ and its conjugate $\gamma^\mu$. They satisfy the canonical anti-commutation
relations:
\begin{equation}
\left\{ \alpha_\mu(K) , \alpha^\nu(L) \right\} = \delta_\mu^\nu \delta_{KL}
 \qquad
\left\{ \beta_\mu(K) , \beta^\nu(L) \right\} = \delta_\mu^\nu \delta_{KL}
 \qquad
\left\{ \gamma_\mu , \gamma^\nu \right\} = \delta_\mu^\nu \quad \mbox{(if present)}
\label{fermionicSU4oscillators}
\end{equation}
while all the other anti-commutators vanish.

The Lie algebra $\mathfrak{so}(8)$ is realized as bilinears of these fermionic oscillators as follows:
\begin{equation}
\begin{split}
A_{\mu\nu}   & = \vec{\alpha}_\mu \cdot \vec{\beta}_\nu - \vec{\alpha}_\nu \cdot \vec{\beta}_\mu
                 + \epsilon \; \gamma_\mu \gamma_\nu
               = \vec{\alpha}_{[\mu} \cdot \vec{\beta}_{\nu]} + \frac{\epsilon}{2} \, \gamma_{[\mu} \gamma_{\nu]}
             ~ = ~ \oneonedotbox \\
M^\mu_{~\nu} & = \vec{\alpha}^\mu \cdot \vec{\alpha}_\nu - \vec{\beta}_\nu \cdot \vec{\beta}^\mu
                 + \frac{\epsilon}{2} \left( \gamma^\mu \gamma_\nu - \gamma_\nu \gamma^\mu \right) \\
A^{\mu\nu}   & = \vec{\alpha}^\mu \cdot \vec{\beta}^\nu - \vec{\alpha}^\nu \cdot \vec{\beta}^\mu
                 + \epsilon \; \gamma^\mu \gamma^\nu
               = \vec{\alpha}^{[\mu} \cdot \vec{\beta}^{\nu]} + \frac{\epsilon}{2} \, \gamma^{[\mu} \gamma^{\nu]}
             ~ = ~ \oneonebox \,.
\label{SO8generators}
\end{split}
\end{equation}
The generators of $\mathfrak{g}^{(\pm 1)}$ subspaces satisfy
\begin{equation}
\left[ A_{\mu\nu} , A^{\sigma\rho} \right] = \delta_\mu^\sigma M^\rho_{~\nu} - \delta_\mu^\rho M^\sigma_{~\nu}
                                             - \delta_\nu^\sigma M^\rho_{~\mu} + \delta_\nu^\rho M^\sigma_{~\mu} \,.
\end{equation}
The generators $M^\mu_{~\nu}$ form the subalgebra $\mathfrak{u}(4)$ of $\mathfrak{so}(8)$ and $A_{\mu\nu}$ and
$A^{\mu\nu}$, both of which transform as $\mathbf{6}$ with respect to $\mathfrak{su}(4)$ with opposite charges under
the $\mathfrak{u}(1)$ generator $M^\mu_{~\mu}$, extend it to the Lie algebra $\mathfrak{so}(8)$. The $\mathfrak{u}(1)$
charge, with respect to which the 3-grading is defined, is given by
\begin{equation}
J = \frac{1}{2} M^\mu_{~\mu} = \frac{1}{2} N_F - 2P - \epsilon
\label{J47}
\end{equation}
where $N_F = \vec{\alpha}^\mu \cdot \vec{\alpha}_\mu + \vec{\beta}^\mu \cdot \vec{\beta}_\mu + \epsilon \; \gamma^\mu
\gamma_\mu$.

As usual, the vacuum state $\ket{0}$ is annihilated by all the annihilation operators $\alpha_\mu(K)$, $\beta_\mu(K)$
and $\gamma_\mu$ (if present) for all values of $\mu$ and $K$. The choice of the lowest weight vectors $\ket{\Omega}$
(that transform irreducibly under $SU(4) \times U(1)$ and are annihilated by the generators in $\mathfrak{g}^{(-1)}$)
and the construction of the representations of $SO(8)$ now proceed analogous to the previous cases. Once again,
because of the fermionic nature of the oscillators, equation \eqref{UIRconstruction} produces only finite dimensional
representations, as one would expect for the compact group $SO(8)$.


\subsubsection{Representations of $OSp(8|4,\mathbb{R})$ via the oscillator method}
\label{SubSubSec:OSp84R}

The superalgebra $\mathfrak{osp}(8|4,\mathbb{R})$ has a 3-grading with respect to its maximal compact
subsuperalgebra $\mathfrak{u}(2|4)$, which has an even subalgebra  $\mathfrak{u}(2) \oplus \mathfrak{u}(4)$.

Therefore, to construct the UIRs of $OSp(8|4,\mathbb{R})$, one defines the $U(2|4)$ covariant super-oscillators as
follows:
\begin{equation}
\begin{split}
\xi_A(K)  & = \left( \begin{matrix} a_i(K) \cr \alpha_\mu(K) \cr \end{matrix} \right)
          ~ = ~ \sonedotbox
 \qquad
\xi^A(K)   := \xi_A(K)^\dag
            = \left( \begin{matrix} a^i(K) \cr \alpha^\mu(K) \cr \end{matrix} \right)
          ~ = ~ \sonebox \\
\eta_A(K) & = \left( \begin{matrix} b_i(K) \cr \beta_\mu(K) \cr \end{matrix} \right)
          ~ = ~ \sonedotbox
 \qquad
\eta^A(K)  := \eta_A(K)^\dag
            = \left( \begin{matrix} b^i(K) \cr \beta^\mu(K) \cr \end{matrix} \right)
          ~ = ~ \sonebox \\
\zeta_A   & = \left( \begin{matrix} c_i \cr \gamma_\mu \cr \end{matrix} \right)
  \qquad \; = ~ \sonedotbox
 \qquad \quad \;\;
\zeta^A    := {\zeta_A}^\dag
     \qquad = \left( \begin{matrix} c^i \cr \gamma^\mu \cr \end{matrix} \right)
  \, \qquad = ~ \sonebox \\
\end{split}
\end{equation}
where $A=1,2|1,2,3,4$ \,; $K=1,\dots,P$. They satisfy the canonical super-commutation relations:
\begin{equation}
\left[ \xi_A(K) , \xi^B(L) \right\}   = \delta_A^B \delta_{KL}
 \qquad
\left[ \eta_A(K) , \eta^B(L) \right\} = \delta_A^B \delta_{KL}
 \qquad
\left[ \zeta_A , \zeta^B \right\}     = \delta_A^B \,.
\end{equation}

Now, in terms of these super-oscillators, the Lie superalgebra $\mathfrak{osp}(8|4,\mathbb{R})$ has the following
realization:
\begin{equation}
\begin{split}
\mathcal{A}_{AB}   & = \vec{\xi}_A \cdot \vec{\eta}_B + \vec{\eta}_A \cdot \vec{\xi}_B + \epsilon \; \zeta_A \zeta_B \\
                   & = \vec{\xi}_{(A} \cdot \vec{\eta}_{B)} + \frac{\epsilon}{2} \, \zeta_{(A} \zeta_{B)}
              \qquad = ~ \stwodotbox \\
\mathcal{M}^A_{~B} & = \vec{\xi}^A \cdot \vec{\xi}_B
                       + (-1)^{(\mathrm{deg}A)(\mathrm{deg}B)} \vec{\eta}_B \cdot \vec{\eta}^A \\
                   &   \quad + \frac{\epsilon}{2} \left( \zeta^A \zeta_B + (-1)^{(\mathrm{deg}A)(\mathrm{deg}B)}
                                                                           \zeta_B \zeta^A \right) \\
\mathcal{A}^{AB}   & = \vec{\xi}^A \cdot \vec{\eta}^B + \vec{\eta}^A \cdot \vec{\xi}^B + \epsilon \; \zeta^A \zeta^B \\
                   & = \vec{\xi}^{(A} \cdot \vec{\eta}^{B)} + \frac{\epsilon}{2} \, \zeta^{(A} \zeta^{B)}
              \qquad = ~ \stwobox \,.
\label{OSp84Rgenerators}
\end{split}
\end{equation}
It is easy to see that, $\mathcal{M}^A_{~B}$ generate the subsuperalgebra $\mathfrak{g}^{(0)} = \mathfrak{u}(2|4)$,
and $\mathcal{A}_{AB}$ and $\mathcal{A}^{AB}$ extend it to the full superalgebra $\mathfrak{osp}(8|4,\mathbb{R})$. The
abelian $\mathfrak{u}(1)$ charge which defines the above 3-grading is given by
\begin{equation}
C = \frac{1}{2} \mathcal{M}^A_{~A}
  = \frac{1}{2} \left( N_B + N_F \right) - P - \frac{\epsilon}{2}
  = E + J \,.
\label{OSp84RC}
\end{equation}

Once again in this case, sixteen of the supersymmetry generators ($Q^i_{~\mu} := \mathcal{M}^i_{~\mu}$ and $Q^\nu_{~j}
:= \mathcal{M}^\nu_{~j}$) reside in the subspace $\mathfrak{g}^{(0)}$, and the other sixteen ($Q_{i\mu} :=
\mathcal{A}_{i\mu}$ and $Q^{j\nu} := \mathcal{A}^{j\nu}$) reside in $\mathfrak{g}^{(-1)} \oplus \mathfrak{g}^{(+1)}$.

The super-Fock space vacuum $\ket{0}$ is defined, as usual, by
\begin{equation}
\xi_A(K) \ket{0} = 0 = \eta_A(K) \ket{0}
\end{equation}
where $A=1,2|1,2,3,4$ \,; $K=1,\dots,P$. Therefore, one can construct the positive energy UIRs of
$OSp(8|4,\mathbb{R})$ by first choosing sets of states $\ket{\Omega}$ in the super-Fock space $\mathcal{F}$ that
transform irreducibly under $U(2|4)$ and are annihilated by $\mathfrak{g}^{(-1)}$, and then by repeatedly acting with
the generators of $\mathfrak{g}^{(+1)}$.

By choosing a single set of super-oscillators (i.e. by choosing $\epsilon = 1$, $P=0$, and therefore making
$n=1$), one can build both singleton supermultiplets of $OSp(8|4,\mathbb{R})$. They do not have a Poincar\'{e}
limit in $d=4$ and decouple as local gauge modes from the Kaluza-Klein of 11 dimensional supergravity over $AdS_4
\times S^7$. On the other hand, if one chooses two pairs ($n=2$) of super-oscillators (i.e. tensoring of two
copies of singletons), one obtains  the massless supermultiplets of $OSp(8|4,\mathbb{R})$. By considering more
than two copies ($n = 2P + \epsilon > 2$), one obtains the massive supermultiplets \cite{GW86}.

The spectrum of the $S^7$ compactification of the eleven dimensional supergravity as an infinite tower of
supermultiplets with  the lowest weight vector $\ket{\Omega} = \ket{0}$, as obtained in \cite{GW86},  is
presented in Table \ref{Table:AdS4S7} of appendix \ref{ApSec:SupergravitySpectra} for reference.


\section{Maximally supersymmetric pp-wave algebra in eleven dimensions}
\label{Sec:MaxSusy}

Taking a Penrose limit of any supergravity solution leads , in general,  to a plane-wave background. In
particular, Penrose limits of $AdS \times S$ type backgrounds result in plane-wave geometries with supersymmetry.
As is well-known (see for example \cite{HKS02b,FGP02,dago}), the symmetry superalgebra of a supersymmetric
pp-wave background can be obtained, from an algebraic point of view, by an In\"{o}n\"{u}-Wigner contraction
\cite{IW53} from the corresponding symmetry superalgebra of the $AdS \times S$ space.

Now, to obtain the corresponding maximally supersymmetric pp-wave superalgebra in eleven dimensions, one performs the
following contraction of $\mathfrak{osp}(8^*|4)$ or $\mathfrak{osp}(8|4,\mathbb{R})$ \cite{FGP02}. First of all, it is
clear that, since the general oscillator realization of superalgebras corresponds to taking the direct sum of an
arbitrary number of `colors' $P$ of oscillators, the only free parameter available for an In\"{o}n\"{u}-Wigner
contraction is $P$. When one normal orders all the generators, this parameter $P$ appears explicitly in the
super-commutators of the form $\left[ \mathfrak{g}^{(-1)} , \mathfrak{g}^{(+1)} \right\}$. More specifically, the
generators that explicitly depend on $P$ after normal ordering are precisely the $\mathfrak{u}(1)$ generators $E$ and
$J$, that determine the 3-grading of the AdS and internal symmetry subalgebras of $\mathfrak{osp}(8^*|4)$ and
$\mathfrak{osp}(8|4,\mathbb{R})$ (as given in equations \eqref{E74}, \eqref{J74} and \eqref{E47}, \eqref{J47}
respectively). At this point, it is important to recall that the maximal compact subsuperalgebra of both
$\mathfrak{osp}(8^*|4)$ and $\mathfrak{osp}(8|4,\mathbb{R})$ is $\mathfrak{u}(4|2)$. These $\mathfrak{u}(4|2)$
subsuperalgebras have abelian factors $\mathfrak{u}(1)_C$:
\begin{equation}
\mathfrak{u}(4|2) = \mathfrak{su}(4|2) \oplus \mathfrak{u}(1)_C
\end{equation}
given by $C = E + J$ in both cases (see equations \eqref{OSp8*4C} and \eqref{OSp84RC}) that have explicit
$P$-dependence. It is also clear that the following unique linear combinations of $E$ and $J$, which reside
\emph{inside} the $\mathfrak{su}(4|2)$ part:
\begin{equation}
\mathfrak{su}(4|2) \supset \mathfrak{su}(4) \oplus \mathfrak{su}(2) \oplus \mathfrak{u}(1)_G
\label{SU42decompSU4SU2U1}
\end{equation}
are $P$-independent:
\begin{equation}
G = \begin{cases}
     \frac{1}{2} E + J = \frac{1}{4} \left( N_B + 2 N_F \right)
      & \qquad \mbox{for } \mathfrak{osp}(8^*|4) \\
     E + \frac{1}{2} J = \frac{1}{4} \left( 2 N_B + N_F \right)
      & \qquad \mbox{for } \mathfrak{osp}(8|4,\mathbb{R})
    \end{cases}
\label{SU42SU24G}
\end{equation}

We shall then define re-normalized generators
\begin{equation}
\mathcal{A}_{AB} \longmapsto \hat{\mathcal{A}}_{AB} = \sqrt{\frac{\lambda}{2P}} \mathcal{A}_{AB}
 \qquad \qquad
\mathcal{A}^{AB} \longmapsto \hat{\mathcal{A}}^{AB} = \sqrt{\frac{\lambda}{2P}} \mathcal{A}^{AB}
\end{equation}
belonging to $\mathfrak{g}^{(\pm 1)}$ subspaces and the $P$-dependent generator
\begin{equation*}
C \longmapsto \frac{\lambda}{P} C
\end{equation*}
belonging to $\mathfrak{g}^{(0)} = \mathfrak{u}(4|2)$ subspace, and take the limit $P \to \infty$ to obtain the
pp-wave superalgebra ($\lambda$ being a freely adjustable parameter) corresponding to each case \cite{FGP02}. Note
that $\mathfrak{su}(4|2)$ part of $\mathfrak{g}^{(0)}$ is unchanged.

It is evident that in this limit, the generators belonging to the re-normalized subspace $\hat{\mathfrak{g}}^{(-1)}
\oplus \hat{\mathfrak{g}}^{(+1)}$ form a Heisenberg superalgebra:
\begin{equation}
\left[ \hat{\mathcal{A}}_{AB} , \hat{\mathcal{A}}^{CD} \right\} = \lambda (-1)^{(\mathrm{deg}C)(\mathrm{deg}D)}
                                                                   \delta_A^{[C} \delta_B^{D\}}
\end{equation}
along with $C \stackrel{P \rightarrow \infty}{\longrightarrow} \lambda$, which becomes the central charge. One
may denote this Heisenberg superalgebra by $\mathfrak{h}^{9,8}$, since it contains 9 pairs of bosonic generators
and 8 pairs of fermionic generators. On the other hand, the generators in $\mathfrak{g}^{(0)}$ subspace (modulo
$P$-dependent $C$), that do not depend explicitly on $P$ (assuming all the generators are in normal ordered
form), will survive this limit intact. The $P$-independent $\mathfrak{u}(1)$ generator $G$ is  the Hamiltonian
(up to an overall scale factor), and $SU(4) \times SU(2)$  becomes the covering group of  the rotation group $
SO(6) \times SO(3)$ in $9=6+3$ transverse dimensions.

Both superalgebras $\mathfrak{osp}(8^*|4)$ and $\mathfrak{osp}(8|4,\mathbb{R})$ lead to the same pp-wave
superalgebra under the above contraction. This is not  surprising,  since both these superalgebras have
isomorphic maximal  compact subsuperalgebras $\mathfrak{g}^{(0)}$ , and re-normalizing and taking the limit $P
\to \infty$ are done in both cases the same way.

Thus the maximally supersymmetric pp-wave algebra in eleven dimensions is unique, which could be obtained by starting
from either of the eleven dimensional maximally supersymmetric $AdS \times S$ algebras ($\mathfrak{osp}(8^*|4)$ or
$\mathfrak{osp}(8|4,\mathbb{R})$), and it is the semi-direct sum of a compact subsuperalgebra $\mathfrak{u}(4|2) / C$
and a Heisenberg superalgebra $\mathfrak{h}^{9,8}$:
\begin{equation*}
\mathfrak{su}(4|2) ~ \circledS ~ \mathfrak{h}^{9,8} \,.
\end{equation*}

Therefore, from now on we only work with the $AdS_7 \times S^4$ symmetry superalgebra $\mathfrak{osp}(8^*|4)$ with the
understanding that the same results can be equally established by starting from the $AdS_4 \times S^7$ symmetry
superalgebra $\mathfrak{osp}(8|4,\mathbb{R})$.

To construct a UIR of the resulting pp-wave superalgebra in our case in discussion, namely $\mathfrak{su}(4|2) ~
\circledS ~ \mathfrak{h}^{9,8}$, we choose a set of states $| \hat{\Omega} \rangle$ that transforms irreducibly under
$\mathfrak{su}(4|2)$ and is annihilated by $\hat{\mathfrak{g}}^{(-1)}$ generators. Then by acting on $| \hat{\Omega}
\rangle$ with $\hat{\mathfrak{g}}^{(+1)}$ generators repeatedly, we obtain a UIR of the pp-wave superalgebra.

There are infinitely many such lowest weight vectors $| \hat{\Omega} \rangle$, but $| \hat{\Omega} \rangle =
\ket{0}$ is the only $\hat{\mathfrak{g}}^{(0)}$-invariant state with zero $U(1)_G$ charge (i.e. with a zero
eigenvalue of the Hamiltonian). Since the entire Kaluza-Klein spectrum of the eleven dimensional supergravity
over $AdS_7 \times S^4$ fits into short unitary supermultiplets of $OSp(8^*|4)$ with the ground state
$\ket{\Omega} = \ket{0}$ , the zero-mode spectrum of the pp-wave superalgebra relevant to supergravity is the
unitary supermultiplet obtained by starting from $| \hat{\Omega} \rangle = \ket{0}$.

Note that, since $SU(4|2) \supset SU(4) \times SU(2)$, $\hat{\mathfrak{g}}^{(+1)}$ generators can be decomposed
as follows (in $SU(4) \times SU(2)$ Young tableau notation) :
\begin{equation}
\soneonebox \Big|_{SU(4|2)} = \left( \yng(1,1) \,,\, 1 \right) \oplus
                                     \left( \yng(1) \,,\, \yng(1) \right) \oplus
                                     \left( 1 \,,\, \yng (2) \right)
                              \Big|_{SU(4) \times SU(2)}
\end{equation}
Since $\hat{\mathfrak{g}}^{(-1)}$ generators are the Hermitian conjugates of $\hat{\mathfrak{g}}^{(+1)}$ generators,
they have a similar decomposition in the $SU(4) \times SU(2)$ basis. In this decomposition, it is easy to identify
$\mathcal{A}^{ij} = \left( \yng(1,1) \,,\, 1 \right)$ and $\mathcal{A}^{\mu\nu} = \left( 1 \,,\, \yng (2) \right)$ as
the (6 + 3 =) 9 bosonic generators in $\hat{\mathfrak{g}}^{(+1)}$, which together with their Hermitian conjugate
counterparts in $\hat{\mathfrak{g}}^{(-1)}$, produce translations ($\hat{\mathfrak{g}}^{(+1)} + i \,
\hat{\mathfrak{g}}^{(-1)}$) and boosts ($\hat{\mathfrak{g}}^{(+1)} - i \, \hat{\mathfrak{g}}^{(-1)}$) in the 9
transverse directions of eleven dimensional pp-wave spacetime.

The term $\left( \yng(1) \,,\, \yng(1) \right)$ represent  the 8 supersymmetries $Q^{i\mu}$ in
$\hat{\mathfrak{g}}^{(+1)}$ with which we act on the lowest weight vector $| \hat{\Omega} \rangle = \ket{0}$ to
obtain the entire unitary supermultiplet \cite{FGP02}. We present our results in Table \ref{Table:SU42} below.
The first column gives the $SU(4) \times SU(2)$ Young tableau of the state. Then we list the eigenvalues of the
Hamiltonian $\mathcal{H}$ (i.e. rescaled $G$, in order to obtain energy increments in integers - see equation
\eqref{SU42SU24G}):
\begin{equation}
\mathcal{H} = \frac{1}{3} \left( N_B + 2 N_F \right) \,,
\end{equation}
the number of bosonic/fermionic degrees of freedom $N_{\mathrm{dof}}$, $SU(4)$ Dynkin labels and the $SU(2)$ spin
of these states. Our definition of Dynkin labels is such that, the fundamental representation of $SU(4)$
corresponds to (1,0,0).\footnote{  $SU(n)$ Young tableau have only $(n-1)$ rows. However, throughout this
paper,we shall use $U(n)$ tableaux with $n$ rows in order  to show explicitly the full oscillator content of each
state as well as their $U(1)$ charges.}

\Yboxdim5.0pt
\begin{longtable}[c]{|l|c|c|c|c||l|c|c|c|c|}
\kill
\caption[The zero-mode spectrum of the maximally supersymmetric pp-wave algebra in eleven dimensions,
$\mathfrak{su}(4|2) ~ \circledS ~ \mathfrak{h}^{9,8}$]{\normalsize The zero-mode spectrum of the maximally
supersymmetric pp-wave algebra in eleven dimensions, $\mathfrak{su}(4|2) ~ \circledS ~ \mathfrak{h}^{9,8}$.
\label{Table:SU42}} \\
\hline
\hline
& & & & & & & & & \\
{\scriptsize $SU(4) \times SU(2)$} &
 &
 &
{\scriptsize $SU(4)$} &
 &
{\scriptsize $SU(4) \times SU(2)$} &
 &
 &
{\scriptsize $SU(4)$} &
 \\
{\scriptsize Young tableau} &
 &
{\scriptsize $N_{\mathrm{dof}}$} &
{\scriptsize Dynkin} &
{\scriptsize $SU(2)$} &
{\scriptsize Young tableau} &
 &
{\scriptsize $N_{\mathrm{dof}}$} &
{\scriptsize Dynkin} &
{\scriptsize $SU(2)$} \\
{\scriptsize (bosonic states)} &
{\scriptsize $\mathcal{H}$} &
{\scriptsize (B)} &
{\scriptsize labels} &
{\scriptsize spin} &
{\scriptsize (fermionic states)} &
{\scriptsize $\mathcal{H}$} &
{\scriptsize (F)} &
{\scriptsize labels} &
{\scriptsize spin} \\
& & & & & & & & & \\
\hline
& & & & & & & & & \\
\endfirsthead
\caption[]{(continued)} \\
\hline
& & & & & & & & & \\
{\scriptsize $SU(4) \times SU(2)$} &
 &
 &
{\scriptsize $SU(4)$} &
 &
{\scriptsize $SU(4) \times SU(2)$} &
 &
 &
{\scriptsize $SU(4)$} &
 \\
{\scriptsize Young tableau} &
 &
{\scriptsize $N_{\mathrm{dof}}$} &
{\scriptsize Dynkin} &
{\scriptsize $SU(2)$} &
{\scriptsize Young tableau} &
 &
{\scriptsize $N_{\mathrm{dof}}$} &
{\scriptsize Dynkin} &
{\scriptsize $SU(2)$} \\
{\scriptsize (bosonic states)} &
{\scriptsize $\mathcal{H}$} &
{\scriptsize (B)} &
{\scriptsize labels} &
{\scriptsize spin} &
{\scriptsize (fermionic states)} &
{\scriptsize $\mathcal{H}$} &
{\scriptsize (F)} &
{\scriptsize labels} &
{\scriptsize spin} \\
& & & & & & & & & \\
\hline
& & & & & & & & & \\
\endhead
& & & & & & & & & \\
\hline
\endfoot
& & & & & & & & & \\
\hline
& & & & & & & & & \\
& & 128 & & & & & 128 & & \\
& & & & & & & & & \\
\hline
\hline
\endlastfoot
$\ket{1,1}$                     & 0 &  1 & $(0,0,0)$ & 0             &
$\ket{\yng(1),\yng(1)}$         & 1 &  8 & $(1,0,0)$ & $\frac{1}{2}$ \\[8pt]
$\ket{\yng(2),\yng(1,1)}$       & 2 & 10 & $(2,0,0)$ & 0             &
$\ket{\yng(2,1),\yng(2,1)}$     & 3 & 40 & $(1,1,0)$ & $\frac{1}{2}$ \\[8pt]
$\ket{\yng(1,1),\yng(2)}$       & 2 & 18 & $(0,1,0)$ & 1             &
$\ket{\yng(1,1,1),\yng(3)}$     & 3 & 16 & $(0,0,1)$ & $\frac{3}{2}$ \\[8pt]
$\ket{\yng(2,2),\yng(2,2)}$     & 4 & 20 & $(0,2,0)$ & 0             &
$\ket{\yng(2,2,1),\yng(3,2)}$   & 5 & 40 & $(0,1,1)$ & $\frac{1}{2}$ \\[8pt]
$\ket{\yng(2,1,1),\yng(3,1)}$   & 4 & 45 & $(1,0,1)$ & 1             &
$\ket{\yng(2,1,1,1),\yng(4,1)}$ & 5 & 16 & $(1,0,0)$ & $\frac{3}{2}$ \\[8pt]
$\ket{\yng(1,1,1,1),\yng(4)}$   & 4 &  5 & $(0,0,0)$ & 2             &
$\ket{\yng(2,2,2,1),\yng(4,3)}$ & 7 &  8 & $(0,0,1)$ & $\frac{1}{2}$ \\[8pt]
$\ket{\yng(2,2,2),\yng(3,3)}$   & 6 & 10 & $(0,0,2)$ & 0             &
                                &   &    &           &               \\[8pt]
$\ket{\yng(2,2,1,1),\yng(4,2)}$ & 6 & 18 & $(0,1,0)$ & 1             &
                                &   &    &           &               \\[8pt]
$\ket{\yng(2,2,2,2),\yng(4,4)}$ & 8 &  1 & $(0,0,0)$ & 0             &
                                &   &    &           &               \\[8pt]
\end{longtable}
\Yboxdim7.0pt


\section{Non-maximally supersymmetric pp-wave algebras in eleven dimensions}
\label{Sec:NonMaxSusy}

From the discussion in section \ref{Sec:MaxSusy}, it is clear that a generic pp-wave superalgebra is the
semi-direct sum $\mathfrak{k}\circledS \mathfrak{h}$  of a compact subsuperalgebra $\mathfrak{k}$ and a
Heisenberg superalgebra $\mathfrak{h}$. In the above maximally supersymmetric case in eleven dimensions, 16
(kinematical) supersymmetries belong to the Heisenberg superalgebra $\mathfrak{h}=\mathfrak{h}^{9,8}$, and the
other 16 (dynamical) supersymmetries belong to the compact subsuperalgebra $\mathfrak{k} = \mathfrak{su}(4|2)$.
We shall refer to the subsuperalgebra $\mathfrak{k}$ as the dynamical subsuperalgebra of the corresponding
pp-wave algebra. The term " dynamical subsuperalgebra" is also justified by the fact that the symmetry
superalgebra governing the interacting part of the matrix model for M-theory on the maximally supersymmetric
pp-wave is the basic classical Lie superalgebra $SU(4|2)$ \cite{dsvr}.

 Now starting from this maximally supersymmetric eleven dimensional pp-wave algebra
$\mathfrak{su}(4|2) ~ \circledS ~ \mathfrak{h}^{9,8}$, one can obtain a number of non-maximally supersymmetric
pp-wave algebras, by taking $\mathfrak{k}$ to be a subsuperalgebra of $\mathfrak{su}(4|2)$. In this section, we
present an extensive list of such cases and give the corresponding zero-mode pp-wave spectra for some of them. In
all these cases,  all 16 kinematical supersymmetries belonging to the Heisenberg superalgebra $\mathfrak{h}$ are
 preserved.

Once again, we note that in the remainder of this section we adhere to the oscillator realization of
$\mathfrak{su}(4|2) ~ \circledS ~ \mathfrak{h}^{9,8}$ that came from the AdS superalgebra
$\mathfrak{osp}(8^*|4)$, where the oscillators that transform in the fundamental representation of $SU(4)$ are
bosonic , while those that transform in the fundamental representation of $SU(2)$ are fermionic. The conclusions,
including the structure of the non-maximally supersymmetric pp-wave algebras we obtain and their zero-mode
spectra, will not change if one followed the other possibility (i.e. work with $SU(4)$ fermionic oscillators and
$SU(2)$ bosonic oscillators).


\subsection{$\left\{ \mathfrak{su}(2|2) \oplus \mathfrak{su}(2) \right\} ~ \circledS ~ \mathfrak{h}^{8,8}$}
\label{SubSec:SU22SU2}

We first consider the decomposition of $\mathfrak{su}(4|2)$ into its even subalgebra $\mathfrak{su}(4) \oplus
\mathfrak{su}(2) \oplus \mathfrak{u}(1)$ as in equation \eqref{SU42decompSU4SU2U1}. Since this $\mathfrak{su}(2)$
comes from the internal symmetry algebra $\mathfrak{usp}(4) \approx \mathfrak{so}(5)$ of $\mathfrak{osp}(8^*|4)$,
and is realized in terms of fermionic oscillators, we denote it by $\mathfrak{su}(2)_F$. The $\mathfrak{u}(1)$
charge is given by the equation \eqref{SU42SU24G}:
\begin{equation}
G = \frac{1}{4} \left( N_B + 2 N_F \right) \,.
\label{SU42G}
\end{equation}

Then we decompose $\mathfrak{su}(4)$ into
\begin{equation}
\mathfrak{su}(4) \supset \mathfrak{su}(2)_{B_1} \oplus \mathfrak{su}(2)_{B_2} \oplus \mathfrak{u}(1)_D
\label{SU4decompSU2SU2U1}
\end{equation}
and relabel $a_i(K)$, $a^i(K)$, $b_i(K)$, $b^i(K)$, for $i=1,2$ as $\grave{a}_m(K)$, $\grave{a}^m(K)$,
$\grave{b}_m(K)$, $\grave{b}^m(K)$ ($m=1,2$) and for $i=3,4$ as $\tilde{a}_r(K)$, $\tilde{a}^r(K)$, $\tilde{b}_r(K)$,
$\tilde{b}^r(K)$ ($r=3,4$). Thus, the generators of $\mathfrak{su}(2)_{B_1}$ are realized in terms of the $\grave{a}$
and $\grave{b}$ type oscillators, while the generators of $\mathfrak{su}(2)_{B_2}$ are realized in terms of the
$\tilde{a}$ and $\tilde{b}$ type oscillators. The $\mathfrak{u}(1)$ charge that appears in this decomposition is given
by
\begin{equation}
D = \frac{1}{2} \left( N_{B_1} - N_{B_2} \right)
\label{SU2SU2D}
\end{equation}
where $N_{B_1} = \vec{\grave{a}}^m \cdot \vec{\grave{a}}_m + \vec{\grave{b}}^m \cdot \vec{\grave{b}}_m$ and $N_{B_2} =
\vec{\tilde{a}}^r \cdot \vec{\tilde{a}}_r + \vec{\tilde{b}}^r \cdot \vec{\tilde{b}}_r$. Therefore, $N_B = N_{B_1} +
N_{B_2}$.

Now  $\mathfrak{su}(2)_{B_1}$ and $\mathfrak{su}(2)_F$ along with the following linear combination of $G$ and
$D$:
\begin{equation}
G + \frac{1}{2} D = \frac{1}{2} \left( N_{B_1} + N_F \right)
\end{equation}
can be embedded  into an $\mathfrak{su}(2|2)$ subalgebra  \footnote{Any $\mathfrak{su}(n|n)$ type superalgebra
has a $\mathfrak{u}(1)$ charge that commutes with all the other generators, and hence acting as a central charge.
The superalgebra obtained by modding out this $\mathfrak{u}(1)$ is denoted by $\mathfrak{psu}(n|n)$.}
\begin{equation}
\mathfrak{su}(2)_{B_1} \oplus \mathfrak{su}(2)_F \oplus \mathfrak{u}(1)_{G+\frac{1}{2}D}
 \subset \mathfrak{su}(2|2)
       = \mathfrak{psu}(2|2) \oplus \mathfrak{u}(1)_{G+\frac{1}{2}D} \,.
\end{equation}

Now the dynamical compact part of the pp-wave superalgebra has the decomposition
\begin{equation}
\mathfrak{g}^{(0)} = \mathfrak{su}(4|2)
             \supset \mathfrak{su}(2|2) \oplus \mathfrak{su}(2)_{B_2} \oplus \mathfrak{u}(1)
\label{SU22SU2}
\end{equation}
and therefore, the $SU(4|2)$ covariant super-oscillators must be decomposed in the $SU(2|2) \times SU(2)$ basis as
\begin{equation}
\begin{split}
\xi^A(K)  = \left( \begin{matrix} a^i(K) \cr \alpha^\mu(K) \cr \end{matrix} \right)
 \qquad & \longrightarrow \qquad
  \grave{\xi}^M(K) \oplus \tilde{a}^r(K) \\
\eta^A(K) = \left( \begin{matrix} b^i(K) \cr \beta^\mu(K) \cr \end{matrix} \right)
 \qquad & \longrightarrow \qquad
  \grave{\eta}^M(K) \oplus \tilde{b}^r(K) \,, \qquad \mbox{etc.}
\end{split}
\end{equation}
where
\begin{equation}
\grave{\xi}^M(K)  = \left( \begin{matrix} \grave{a}^m(K) \cr \alpha^\mu(K) \cr \end{matrix} \right)
 \qquad \qquad
\grave{\eta}^M(K) = \left( \begin{matrix} \grave{b}^m(K) \cr \beta^\mu(K) \cr \end{matrix} \right) \,.
\end{equation}

It is now easy to see how $\hat{\mathfrak{g}}^{(\pm 1)}$ subspaces decompose with respect to  $SU(2|2) \times
SU(2)$. For example,
\begin{equation}
\begin{split}
\hat{\mathfrak{g}}^{(+1)}
   = \soneonebox \Big|_{SU(4|2)}
 & = \left( \: \soneonebox \,,\, 1 \right) \oplus
     \left( \: \sonebox \,,\, \onebox \right) \oplus
     \left( 1 \,,\, \oneonebox \right)
     \Big|_{SU(2_{B_1}|2_F) \times SU(2)_{B_2}} \\
\vec{\xi}^{[A} \cdot \vec{\eta}^{B\}}
 & = \vec{\grave{\xi}}^{[M} \cdot \vec{\grave{\eta}}^{N\}} \oplus
     \left( \left( \vec{\grave{\xi}}^M \cdot \vec{\tilde{b}}^s -
                   \vec{\tilde{a}}^s \cdot \vec{\grave{\eta}}^M
            \right)
     \right. \\
 &   \qquad \qquad \qquad \quad
     \left. + \left( \vec{\tilde{a}}^r \cdot \vec{\grave{\eta}}^N -
                     \vec{\grave{\xi}}^N \cdot \vec{\tilde{b}}^r
              \right)
     \right) \oplus
     \vec{\tilde{a}}^{[r} \cdot \vec{\tilde{b}}^{s]} \,.
\end{split}
\end{equation}

Thus there is a singlet in $\hat{\mathfrak{g}}^{(+1)}$ with respect to the new dynamical compact subsuperalgebra
$\mathfrak{su}(2|2) \oplus \mathfrak{su}(2)$. In the literature, a compactification of this eleven dimensional
pp-wave solution to ten dimensions has been considered \cite{HS02a,HS02b,KS03}. It is this single noncompact
generator $\vec{\tilde{a}}^{[r} \cdot \vec{\tilde{b}}^{s]} = \left( 1 \,,\, {\yng(1,1)} \right)$ in
$\hat{\mathfrak{g}}^{(+1)}$ (and the corresponding Hermitian conjugate generator $\vec{\tilde{a}}_{[r} \cdot
\vec{\tilde{b}}_{s]}$ in $\hat{\mathfrak{g}}^{(-1)}$) that corresponds to the transverse direction in eleven
dimensions, along which the compactification was performed. Hence in ten dimensional superalgebra, these two
singlets in $\hat{\mathfrak{g}}^{(\pm 1)}$ drop out.

We also discard the compact generators of the type $\mathcal{M}^M_{~~r}$ and $\mathcal{M}^{s}_{~N}$
($8_{\mathrm{bosonic}} + 8_{\mathrm{fermionic}}$) - see equation \eqref{OSp8*4generators} - and the
$\mathfrak{u}(1)$ generator in $\mathfrak{g}^{(0)}$ that commutes with $\mathfrak{su}(2|2) \oplus
\mathfrak{su}(2)$ (equation \eqref{SU22SU2}). We denote this new dynamical compact subsuperalgebra
$\mathfrak{su}(2|2) \oplus \mathfrak{su}(2)$ as $\hat{\mathfrak{g}}^{(0)}$. Thus, there are only 8 supersymmetry
generators and 10 bosonic generators in $\hat{\mathfrak{g}}^{(0)}$. Nine of these bosonic generators are rotation
generators, that belong to the rotation group $SU(2) \times SU(2) \times SU(2) \approx SO(3) \times SO(3) \times
SO(3)$ and $\mathfrak{u}(1)_{G + \frac{1}{2} D}$ in $\mathfrak{su}(2|2)$ plays the role of the Hamiltonian (up to
a rescaling factor):
\begin{equation}
\mathcal{H} = N_{B_1} + N_F \,.
\label{SU22SU2H}
\end{equation}

It is useful now to write the above decomposition of the $\hat{\mathfrak{g}}^{(+1)}$ space after compactification
( i.e. dropping the extra singlets )  in the $SU(2)_{B_1} \times SU(2)_F \times SU(2)_{B_2}$ basis:
\begin{equation}
\begin{split}
\hat{\mathfrak{g}}^{(+1)}
   = \soneonebox \Big|_{SU(4|2)}
 & \Rightarrow \left( \: \soneonebox \,,\, 1 \right) \oplus
     \left( \: \sonebox \,,\, \onebox \right)
     \Big|_{SU(2_{B_1}|2_F) \times SU(2)_{B_2}} \\
 & = \left( \yng(1,1) \,,\, 1 \,,\, 1 \right) \oplus
     \left( \yng(1) \,,\, \yng(1) \,,\, 1 \right) \oplus
     \left( 1 \,,\, \yng(2) \,,\, 1 \right) \\
 &   \quad \oplus
     \left( \yng(1) \,,\, 1 \,,\, \yng(1) \right) \oplus
     \left( 1 \,,\, \yng(1) \,,\, \yng(1) \right)
     \Big|_{SU(2)_{B_1} \times SU(2)_F \times SU(2)_{B_2}}
\end{split}
\end{equation}
Clearly, $\left( \yng(1) \,,\, \yng(1) \,,\, 1 \right) \oplus \left( 1 \,,\, \yng(1) \,,\, \yng(1) \right)$ are the
supersymmetries in $\hat{\mathfrak{g}}^{(+1)}$ of the pp-wave superalgebra (i.e. kinematical supersymmetries), with
which we act on $| \hat{\Omega} \rangle$ to construct the entire unitary supermultiplet. Now they transform as
$\mathbf{4} + \mathbf{4}$ in the $SU(2)_{B_1} \times SU(2)_F \times SU(2)_{B_2}$ basis, where
$Q^{m\mu} = \left( \yng(1) \,,\, \yng(1) \,,\, 1 \right)$ increase energy by 2 units, while
$Q^{r\mu} = \left( 1 \,,\, \yng(1) \,,\, \yng(1) \right)$ increase energy by 1 unit (see equation \eqref{SU22SU2H}).

The 8 bosonic generators remaining in $\hat{\mathfrak{g}}^{(+1)}$ and their hermitian conjugate counterparts in
$\hat{\mathfrak{g}}^{(-1)}$ produce translations ($\hat{\mathfrak{g}}^{(+1)} + i \, \hat{\mathfrak{g}}^{(-1)}$) and
boosts ($\hat{\mathfrak{g}}^{(+1)} - i \, \hat{\mathfrak{g}}^{(-1)}$) in the 8 transverse directions in this ten
dimensional type IIA background.

This pp-wave superalgebra has a total of 24 supersymmetries (8 in $\hat{\mathfrak{g}}^{(+1)}$, 8 in
$\hat{\mathfrak{g}}^{(-1)}$ and 8 in $\hat{\mathfrak{g}}^{(0)}$), and the symmetry superalgebra of this type IIA
pp-wave solution is,
\begin{equation*}
\left[ \mathfrak{su}(2|2) \oplus \mathfrak{su}(2) \right] ~ \circledS ~ \mathfrak{h}^{8,8} \,.
\end{equation*}

Now we construct the zero-mode spectrum of this pp-wave superalgebra in the basis $SU(2)_{B_1} \times SU(2)_F \times
SU(2)_{B_2}$, by starting from the ground state $| \hat{\Omega} \rangle = \ket{0}$ (Table \ref{Table:SU22SU2}). We
note that this spectrum is in agreement with the zero-mode spectrum that was first given in \cite{KS03}. The first
column on each side gives the $SU(2)_{B_1} \times SU(2)_F \times SU(2)_{B_2}$ Young tableau of the state. Then we list
the eigenvalues of the Hamiltonian $\mathcal{H}$ (according to equation \eqref{SU22SU2H}), the number of degrees of
freedom and the $SU(2)$ spin of these states.

\Yboxdim5.0pt
\begin{longtable}[c]{|l|c|c|c||l|c|c|c|}
\kill
\caption[The zero-mode spectrum of the type IIA pp-wave superalgebra with 24 supersymmetries, $\left\{
\mathfrak{su}(2|2) \oplus \mathfrak{su}(2) \right\} ~ \circledS ~ \mathfrak{h}^{8,8}$]{\normalsize The zero-mode
spectrum of the type IIA pp-wave superalgebra with 24 supersymmetries, $\left[ \mathfrak{su}(2|2) \oplus
\mathfrak{su}(2) \right] ~ \circledS ~ \mathfrak{h}^{8,8}$.
\label{Table:SU22SU2}} \\
\hline
\hline
& & & & & & & \\
{\scriptsize $SU(2)_{B_1} \times SU(2)_F$} &
 &
 &
 &
{\scriptsize $SU(2)_{B_1} \times SU(2)_F$} &
 &
 &
 \\
{\scriptsize $\times ~ SU(2)_{B_2}$} &
 &
 &
 &
{\scriptsize $\times ~ SU(2)_{B_2}$} &
 &
 &
 \\
{\scriptsize Young tableau} &
 &
{\scriptsize $N_{\mathrm{dof}}$} &
{\scriptsize $SU(2)$} &
{\scriptsize Young tableau} &
 &
{\scriptsize $N_{\mathrm{dof}}$} &
{\scriptsize $SU(2)$} \\
{\scriptsize (bosonic states)} &
{\scriptsize $\mathcal{H}$} &
{\scriptsize (B)} &
{\scriptsize spin} &
{\scriptsize (fermionic states)} &
{\scriptsize $\mathcal{H}$} &
{\scriptsize (F)} &
{\scriptsize spin} \\
& & & & & & & \\
\hline
& & & & & & & \\
\endfirsthead
\caption[]{(continued)} \\
\hline
& & & & & & & \\
{\scriptsize $SU(2)_{B_1} \times SU(2)_F$} &
 &
 &
 &
{\scriptsize $SU(2)_{B_1} \times SU(2)_F$} &
 &
 &
 \\
{\scriptsize $\times ~ SU(2)_{B_2}$} &
 &
 &
 &
{\scriptsize $\times ~ SU(2)_{B_2}$} &
 &
 &
 \\
{\scriptsize Young tableau} &
 &
{\scriptsize $N_{\mathrm{dof}}$} &
{\scriptsize $SU(2)$} &
{\scriptsize Young tableau} &
 &
{\scriptsize $N_{\mathrm{dof}}$} &
{\scriptsize $SU(2)$} \\
{\scriptsize (bosonic states)} &
{\scriptsize $\mathcal{H}$} &
{\scriptsize (B)} &
{\scriptsize spin} &
{\scriptsize (fermionic states)} &
{\scriptsize $\mathcal{H}$} &
{\scriptsize (F)} &
{\scriptsize spin} \\
& & & & & & & \\
\hline
& & & & & & & \\
\endhead
& & & & & & & \\
\hline
\endfoot
& & & & & & & \\
\hline
& & & & & & & \\
& & 128 & & & & 128 & \\
& & & & & & & \\
\hline
\hline
\endlastfoot
$\ket{1,1,1}$                               &  0 &  1 & $(0,0,0)$                     &
$\ket{1,{\yng(1)},{\yng(1)}}$               &  1 &  4 & $(0,\frac{1}{2},\frac{1}{2})$ \\[8pt]
$\ket{1,{\yng(2)},{\yng(1,1)}}$             &  2 &  3 & $(0,1,0)$                     &
$\ket{{\yng(1)},{\yng(1)},1}$               &  2 &  4 & $(\frac{1}{2},\frac{1}{2},0)$ \\[8pt]
$\ket{1,{\yng(1,1)},{\yng(2)}}$             &  2 &  3 & $(0,0,1)$                     &
$\ket{1,{\yng(2,1)},{\yng(2,1)}}$           &  3 &  4 & $(0,\frac{1}{2},\frac{1}{2})$ \\[8pt]
$\ket{{\yng(1)},{\yng(2)},{\yng(1)}}$       &  3 & 12 & $(\frac{1}{2},1,\frac{1}{2})$ &
$\ket{{\yng(1)},{\yng(3)},{\yng(1,1)}}$     &  4 &  8 & $(\frac{1}{2},\frac{3}{2},0)$ \\[8pt]
$\ket{{\yng(1)},{\yng(1,1)},{\yng(1)}}$     &  3 &  4 & $(\frac{1}{2},0,\frac{1}{2})$ &
$\ket{{\yng(1)},{\yng(2,1)},{\yng(2)}}$     &  4 & 12 & $(\frac{1}{2},\frac{1}{2},1)$ \\[8pt]
$\ket{1,{\yng(2,2)},{\yng(2,2)}}$           &  4 &  1 & $(0,0,0)$                     &
$\ket{{\yng(1)},{\yng(2,1)},{\yng(1,1)}}$   &  4 &  4 & $(\frac{1}{2},\frac{1}{2},0)$ \\[8pt]
$\ket{{\yng(2)},{\yng(1,1)},1}$             &  4 &  3 & $(1,0,0)$                     &
$\ket{{\yng(2)},{\yng(2,1)},{\yng(1)}}$     &  5 & 12 & $(1,\frac{1}{2},\frac{1}{2})$ \\[8pt]
$\ket{{\yng(1,1)},{\yng(2)},1}$             &  4 &  3 & $(0,1,0)$                     &
$\ket{{\yng(1,1)},{\yng(3)},{\yng(1)}}$     &  5 &  8 & $(0,\frac{3}{2},\frac{1}{2})$ \\[8pt]
$\ket{{\yng(1)},{\yng(3,1)},{\yng(2,1)}}$   &  5 & 12 & $(\frac{1}{2},1,\frac{1}{2})$ &
$\ket{{\yng(1,1)},{\yng(2,1)},{\yng(1)}}$   &  5 &  4 & $(0,\frac{1}{2},\frac{1}{2})$ \\[8pt]
$\ket{{\yng(1)},{\yng(2,2)},{\yng(2,1)}}$   &  5 &  4 & $(\frac{1}{2},0,\frac{1}{2})$ &
$\ket{{\yng(1)},{\yng(3,2)},{\yng(2,2)}}$   &  6 &  4 & $(\frac{1}{2},\frac{1}{2},0)$ \\[8pt]
$\ket{{\yng(2)},{\yng(3,1)},{\yng(1,1)}}$   &  6 &  9 & $(1,1,0)$                     &
$\ket{{\yng(2,1)},{\yng(2,1)},1}$           &  6 &  4 & $(\frac{1}{2},\frac{1}{2},0)$ \\[8pt]
$\ket{{\yng(2)},{\yng(2,2)},{\yng(2)}}$     &  6 &  9 & $(1,0,1)$                     &
$\ket{{\yng(2)},{\yng(3,2)},{\yng(2,1)}}$   &  7 & 12 & $(1,\frac{1}{2},\frac{1}{2})$ \\[8pt]
$\ket{{\yng(1,1)},{\yng(4)},{\yng(1,1)}}$   &  6 &  5 & $(0,2,0)$                     &
$\ket{{\yng(1,1)},{\yng(4,1)},{\yng(2,1)}}$ &  7 &  8 & $(0,\frac{3}{2},\frac{1}{2})$ \\[8pt]
$\ket{{\yng(1,1)},{\yng(3,1)},{\yng(2)}}$   &  6 &  9 & $(0,1,1)$                     &
$\ket{{\yng(1,1)},{\yng(3,2)},{\yng(2,1)}}$ &  7 &  4 & $(0,\frac{1}{2},\frac{1}{2})$ \\[8pt]
$\ket{{\yng(1,1)},{\yng(3,1)},{\yng(1,1)}}$ &  6 &  3 & $(0,1,0)$                     &
$\ket{{\yng(2,1)},{\yng(4,1)},{\yng(1,1)}}$ &  8 &  8 & $(\frac{1}{2},\frac{3}{2},0)$ \\[8pt]
$\ket{{\yng(1,1)},{\yng(2,2)},{\yng(1,1)}}$ &  6 &  1 & $(0,0,0)$                     &
$\ket{{\yng(2,1)},{\yng(3,2)},{\yng(2)}}$   &  8 & 12 & $(\frac{1}{2},\frac{1}{2},1)$ \\[8pt]
$\ket{{\yng(2,1)},{\yng(3,1)},{\yng(1)}}$   &  7 & 12 & $(\frac{1}{2},1,\frac{1}{2})$ &
$\ket{{\yng(2,1)},{\yng(3,2)},{\yng(1,1)}}$ &  8 &  4 & $(\frac{1}{2},\frac{1}{2},0)$ \\[8pt]
$\ket{{\yng(2,1)},{\yng(2,2)},{\yng(1)}}$   &  7 &  4 & $(\frac{1}{2},0,\frac{1}{2})$ &
$\ket{{\yng(2,2)},{\yng(3,2)},{\yng(1)}}$   &  9 &  4 & $(0,\frac{1}{2},\frac{1}{2})$ \\[8pt]
$\ket{{\yng(2)},{\yng(3,3)},{\yng(2,2)}}$   &  8 &  3 & $(1,0,0)$                     &
$\ket{{\yng(2,1)},{\yng(4,3)},{\yng(2,2)}}$ & 10 &  4 & $(\frac{1}{2},\frac{1}{2},0)$ \\[8pt]
$\ket{{\yng(1,1)},{\yng(4,2)},{\yng(2,2)}}$ &  8 &  3 & $(0,1,0)$                     &
$\ket{{\yng(2,2)},{\yng(4,3)},{\yng(2,1)}}$ & 11 &  4 & $(0,\frac{1}{2},\frac{1}{2})$ \\[8pt]
$\ket{{\yng(2,2)},{\yng(2,2)},1}$           &  8 &  1 & $(0,0,0)$                     &
                                            &    &    &                               \\[8pt]
$\ket{{\yng(2,1)},{\yng(4,2)},{\yng(2,1)}}$ &  9 & 12 & $(\frac{1}{2},1,\frac{1}{2})$ &
                                            &    &    &                               \\[8pt]
$\ket{{\yng(2,1)},{\yng(3,3)},{\yng(2,1)}}$ &  9 &  4 & $(\frac{1}{2},0,\frac{1}{2})$ &
                                            &    &    &                               \\[8pt]
$\ket{{\yng(2,2)},{\yng(4,2)},{\yng(1,1)}}$ & 10 &  3 & $(0,1,0)$                     &
                                            &    &    &                               \\[8pt]
$\ket{{\yng(2,2)},{\yng(3,3)},{\yng(2)}}$   & 10 &  3 & $(0,0,1)$                     &
                                            &    &    &                               \\[8pt]
$\ket{{\yng(2,2)},{\yng(4,4)},{\yng(2,2)}}$ & 12 &  1 & $(0,0,0)$                     &
                                            &    &    &                               \\[8pt]
\end{longtable}
\Yboxdim7.0pt


\subsection{$\left\{ \mathfrak{su}(3|2) \oplus \mathfrak{u}(1) \right\} ~ \circledS ~ \mathfrak{h}^{9,8}$}
\label{SubSec:SU32U1}

Once again, we first decompose $\mathfrak{su}(4|2)$ into its even subgroup $\mathfrak{su}(4) \oplus \mathfrak{su}(2)
\oplus \mathfrak{u}(1)_G$ as in equations \eqref{SU42decompSU4SU2U1} and \eqref{SU42G}), and then break
$\mathfrak{su}(4)$ into
\begin{equation}
\mathfrak{su}(4) \supset \mathfrak{su}(3) \oplus \mathfrak{u}(1)_D
\label{SU4decompSU3U1}
\end{equation}
and relabel $a_i(K)$, $a^i(K)$, $b_i(K)$, $b^i(K)$ for $i=1,2,3$ as $\grave{a}_m(K)$, $\grave{a}^m(K)$,
$\grave{b}_m(K)$, $\grave{b}^m(K)$ ($m=1,2,3$) and for $i=4$ as $\tilde{a}_4(K)$, $\tilde{a}^4(K)$, $\tilde{b}_4(K)$,
$\tilde{b}^4(K)$. Therefore, it is useful to define the number operators $N_{B_1} = \vec{\grave{a}}^m \cdot
\vec{\grave{a}}_m + \vec{\grave{b}}^m \cdot \vec{\grave{b}}_m$ and $N_{B_2} = \vec{\tilde{a}}^4 \cdot
\vec{\tilde{a}}_4 + \vec{\tilde{b}}^4 \cdot \vec{\tilde{b}}_4$, such that we have $N_B = N_{B_1} + N_{B_2}$. The
$\mathfrak{u}(1)_D$ charge can be written as
\begin{equation}
D = M^4_{~4}
  = N_{B_2} - \frac{1}{4} N_B
  = \frac{1}{4} \left( 3 N_{B_2} - N_{B_1} \right) \,.
\label{SU3D}
\end{equation}

Now  $\mathfrak{su}(3)_{B_1}$ and $\mathfrak{su}(2)_F$ along with the following linear combination of $G$ and
$D$:
\begin{equation}
G - \frac{1}{3} D = \frac{1}{3} N_{B_1} + \frac{1}{2} N_F
\end{equation}
can be embedded in  $\mathfrak{su}(3|2)$ subalgebra
\begin{equation}
\mathfrak{su}(3) \oplus \mathfrak{su}(2) \oplus \mathfrak{u}(1)_{G-\frac{1}{3}D}
 \subset \mathfrak{su}(3|2) \,.
\end{equation}

Thus, the dynamical compact subsuperalgebra of the pp-wave superalgebra now has the decomposition
\begin{equation}
\mathfrak{g}^{(0)} = \mathfrak{su}(4|2)
             \supset \mathfrak{su}(3|2) \oplus \mathfrak{u}(1)
\end{equation}
where the $\mathfrak{u}(1)$ charge that commutes with $\mathfrak{su}(3|2)$ is given by
\begin{equation}
G - D = \frac{1}{2} \left( N_{B_1} + N_F \right) - \frac{1}{2} N_{B_2} \,.
\end{equation}

Therefore, the $SU(4|2)$ covariant super-oscillators need to be decomposed, this time, in the $SU(3|2) \times
U(1)$ basis as
\begin{equation}
\begin{split}
\xi^A(K)  = \left( \begin{matrix} a^i(K) \cr \alpha^\mu(K) \cr \end{matrix} \right)
 \qquad & \longrightarrow \qquad
  \grave{\xi}^M(K) \oplus \tilde{a}^4(K) \\
\eta^A(K) = \left( \begin{matrix} b^i(K) \cr \beta^\mu(K) \cr \end{matrix} \right)
 \qquad & \longrightarrow \qquad
  \grave{\eta}^M(K) \oplus \tilde{b}^4(K) \,, \qquad \mbox{etc.}
\end{split}
\end{equation}
where
\begin{equation}
\grave{\xi}^M(K)  = \left( \begin{matrix} \grave{a}^m(K) \cr \alpha^\mu(K) \cr \end{matrix} \right)
 \qquad \qquad
\grave{\eta}^M(K) = \left( \begin{matrix} \grave{b}^m(K) \cr \beta^\mu(K) \cr \end{matrix} \right) \,.
\end{equation}

The subspace $\hat{\mathfrak{g}}^{(+1)}$ (and similarly $\hat{\mathfrak{g}}^{(-1)}$) now decomposes with respect to
this new compact basis $SU(3|2) \times U(1)_{G-D}$ as
\begin{equation}
\begin{split}
\hat{\mathfrak{g}}^{(+1)}
   = \soneonebox \Big|_{SU(4|2)}
 & = \left( \: \soneonebox \,,\, 1 \right) \oplus
     \left( \: \sonebox \,,\, 0 \right)
     \Big|_{SU(3|2) \times U(1)_{G-D}} \\
\vec{\xi}^{[A} \cdot \vec{\eta}^{B\}}
 & = \vec{\grave{\xi}}^{[M} \cdot \vec{\grave{\eta}}^{N\}} \oplus
     \left( \left( \vec{\grave{\xi}}^M \cdot \vec{\tilde{b}}^4 -
                   \vec{\tilde{a}}^4 \cdot \vec{\grave{\eta}}^M
            \right)
     \right. \\
 &   \qquad \qquad \qquad \quad
     \left. + \left( \vec{\tilde{a}}^4 \cdot \vec{\grave{\eta}}^N -
                     \vec{\grave{\xi}}^N \cdot \vec{\tilde{b}}^4
              \right)
     \right) \,.
\end{split}
\end{equation}
In the $SU(3) \times SU(2) \times U(1)_{G-D}$ basis, the above decomposition takes the form:
\begin{equation}
\begin{split}
\hat{\mathfrak{g}}^{(+1)}
   = \soneonebox \Big|_{SU(4|2)}
 & = \left( \: \soneonebox \,,\, 1 \right) \oplus
     \left( \: \sonebox \,,\, 0 \right)
     \Big|_{SU(3|2) \times U(1)_{G-D}} \\
 & = \left( \yng(1,1) \,,\, 1 \,,\, 1 \right) \oplus
     \left( \yng(1) \,,\, \yng(1) \,,\, 1 \right) \oplus
     \left( 1 \,,\, \yng(2) \,,\, 1 \right) \\
 &   \quad \oplus
     \left( \yng(1) \,,\, 1 \,,\, 0 \right) \oplus
     \left( 1 \,,\, \yng(1) \,,\, 0 \right)
     \Big|_{SU(3) \times SU(2) \times U(1)_{G-D}}
\end{split}
\label{SU3SU2U1decomp}
\end{equation}

The generators $Q^{m\mu} = \left( \yng(1) \,,\, \yng(1) \,,\, 1 \right)$ and $Q^{4\mu} = \left( 1 \,,\, \yng(1)
\,,\, 0 \right)$ are the supersymmetries in $\hat{\mathfrak{g}}^{(+1)}$ of this pp-wave superalgebra (i.e.
kinematical supersymmetries), and they now transform as $\mathbf{(6,1)} \oplus \mathbf{(1,2)}$ under $SU(3)
\times SU(2) \times U(1)_{G-D}$.

From $\mathfrak{g}^{(0)}$ subspace, since we only retain the $\mathfrak{su}(3|2) \oplus \mathfrak{u}(1)_{G-D}$
part, we must eliminate generators of the form $\mathcal{M}^M_{~~4}$ and $\mathcal{M}^4_{~N}$ as usual (6 bosonic
and 4 fermionic), and therefore, this pp-wave superalgebra has 28 supersymmetries (8 in
$\hat{\mathfrak{g}}^{(+1)}$, 8 in $\hat{\mathfrak{g}}^{(-1)}$ and 12 in $\hat{\mathfrak{g}}^{(0)} =
\mathfrak{su}(3|2) \oplus \mathfrak{u}(1)$).

The number of bosonic generators that remain from the maximal compact subsuperalgebra $\hat{\mathfrak{g}}^{(0)}$
is 13. Twelve of them are rotation generators, that belong to $SU(3) \times SU(2) \times U(1)$ and the other one
is the Hamiltonian. Once again we rescale it, so that we obtain energy increases of the states in integer steps:
\begin{equation}
\mathcal{H} = 2 N_{B_1} + 3 N_F
\end{equation}

Thus, it is clear that the 6 of the kinematical supersymmetries,
$Q^{m\mu} = \left( \yng(1) \,,\, \yng(1) \,,\, 1 \right)$ increase energy by 5 units, while the other 2,
$Q^{4\mu} = \left( 1 \,,\, \yng(1) \,,\, 0 \right)$ increase energy by 3 units.

The symmetry superalgebra of this pp-wave solution is
\begin{equation*}
\left[ \mathfrak{su}(3|2) \oplus \mathfrak{u}(1) \right] ~ \circledS ~ \mathfrak{h}^{9,8} \,.
\end{equation*}

To our knowledge  this solution has not yet  appeared  in the literature as a possible pp-wave superalgebra in
eleven dimensions. We expect it to be obtainable by taking the  Penrose limit(s)  of one (or more) of the
compactifications of 11-dimensional supergravity with $SU(3)\times SU(2) \times U(1)$ symmetry ( $M(m,n)$ spaces)
and/or $SU(3)\times U(1)$ symmetry ( $N(m,n)$ spaces) \cite{sase}.

Now we construct the zero-mode spectrum of this pp-wave superalgebra, in the basis $SU(3) \times SU(2) \times
U(1)_{G-D}$, by starting from the ground state $| \hat{\Omega} \rangle = \ket{0}$ and present our results in Table
\ref{Table:SU32U1}.

\Yboxdim5.0pt
\begin{longtable}[c]{|l|c|c|c|c||l|c|c|c|c|}
\kill
\caption[The zero-mode spectrum of the eleven dimensional pp-wave superalgebra with 28 supersymmetries, $\left\{
\mathfrak{su}(3|2) \oplus \mathfrak{u}(1) \right\} ~ \circledS ~ \mathfrak{h}^{9,8}$]{\normalsize The zero-mode
spectrum of the eleven dimensional pp-wave superalgebra with 28 supersymmetries, $\left[ \mathfrak{su}(3|2) \oplus
\mathfrak{u}(1) \right] ~ \circledS ~ \mathfrak{h}^{9,8}$.
\label{Table:SU32U1}} \\
\hline
\hline
& & & & & & & & & \\
{\scriptsize $SU(3) \times SU(2)$} &
 &
 &
 &
 &
{\scriptsize $SU(3) \times SU(2)$} &
 &
 &
 &
 \\
{\scriptsize $\times ~ U(1)_{G-D}$} &
 &
 &
{\scriptsize $SU(3)$} &
 &
{\scriptsize $\times ~ U(1)_{G-D}$} &
 &
 &
{\scriptsize $SU(3)$} &
 \\
{\scriptsize Young tableau} &
 &
{\scriptsize $N_{\mathrm{dof}}$} &
{\scriptsize Dynkin} &
{\scriptsize $SU(2)$} &
{\scriptsize Young tableau} &
 &
{\scriptsize $N_{\mathrm{dof}}$} &
{\scriptsize Dynkin} &
{\scriptsize $SU(2)$} \\
{\scriptsize (bosonic states)} &
{\scriptsize $\mathcal{H}$} &
{\scriptsize (B)} &
{\scriptsize labels} &
{\scriptsize spin} &
{\scriptsize (fermionic states)} &
{\scriptsize $\mathcal{H}$} &
{\scriptsize (F)} &
{\scriptsize labels} &
{\scriptsize spin} \\
& & & & & & & & & \\
\hline
& & & & & & & & & \\
\endfirsthead
\caption[]{(continued)} \\
\hline
& & & & & & & & & \\
{\scriptsize $SU(3) \times SU(2)$} &
 &
 &
 &
 &
{\scriptsize $SU(3) \times SU(2)$} &
 &
 &
 &
 \\
{\scriptsize $\times ~ U(1)_{G-D}$} &
 &
 &
{\scriptsize $SU(3)$} &
 &
{\scriptsize $\times ~ U(1)_{G-D}$} &
 &
 &
{\scriptsize $SU(3)$} &
 \\
{\scriptsize Young tableau} &
 &
{\scriptsize $N_{\mathrm{dof}}$} &
{\scriptsize Dynkin} &
{\scriptsize $SU(2)$} &
{\scriptsize Young tableau} &
 &
{\scriptsize $N_{\mathrm{dof}}$} &
{\scriptsize Dynkin} &
{\scriptsize $SU(2)$} \\
{\scriptsize (bosonic states)} &
{\scriptsize $\mathcal{H}$} &
{\scriptsize (B)} &
{\scriptsize labels} &
{\scriptsize spin} &
{\scriptsize (fermionic states)} &
{\scriptsize $\mathcal{H}$} &
{\scriptsize (F)} &
{\scriptsize labels} &
{\scriptsize spin} \\
& & & & & & & & & \\
\hline
& & & & & & & & & \\
\endhead
& & & & & & & & & \\
\hline
\endfoot
& & & & & & & & & \\
\hline
& & & & & & & & & \\
& & 128 & & & & & 128 & & \\
& & & & & & & & & \\
\hline
\hline
\endlastfoot
$\ket{1,1,0}$                   &  0 &  1 & $(0,0)$ & 0             &
$\ket{1,\yng(1),0}$             &  3 &  2 & $(0,0)$ & $\frac{1}{2}$ \\[8pt]
$\ket{1,\yng(1,1),0}$           &  6 &  1 & $(0,0)$ & 0             &
$\ket{\yng(1),\yng(1),1}$       &  5 &  6 & $(1,0)$ & $\frac{1}{2}$ \\[8pt]
$\ket{\yng(1),\yng(2),1}$       &  8 &  9 & $(1,0)$ & 1             &
$\ket{\yng(1),\yng(2,1),1}$     & 11 &  6 & $(1,0)$ & $\frac{1}{2}$ \\[8pt]
$\ket{\yng(1),\yng(1,1),1}$     &  8 &  3 & $(1,0)$ & 0             &
$\ket{\yng(2),\yng(2,1),2}$     & 13 & 12 & $(2,0)$ & $\frac{1}{2}$ \\[8pt]
$\ket{\yng(2),\yng(1,1),2}$     & 10 &  6 & $(2,0)$ & 0             &
$\ket{\yng(1,1),\yng(3),2}$     & 13 & 12 & $(0,1)$ & $\frac{3}{2}$ \\[8pt]
$\ket{\yng(1,1),\yng(2),2}$     & 10 &  9 & $(0,1)$ & 1             &
$\ket{\yng(1,1),\yng(2,1),2}$   & 13 &  6 & $(0,1)$ & $\frac{1}{2}$ \\[8pt]
$\ket{\yng(2),\yng(2,2),2}$     & 16 &  6 & $(2,0)$ & 0             &
$\ket{\yng(2,1),\yng(2,1),3}$   & 15 & 16 & $(1,1)$ & $\frac{1}{2}$ \\[8pt]
$\ket{\yng(1,1),\yng(3,1),2}$   & 16 &  9 & $(0,1)$ & 1             &
$\ket{\yng(1,1,1),\yng(3),3}$   & 15 &  4 & $(0,0)$ & $\frac{3}{2}$ \\[8pt]
$\ket{\yng(2,1),\yng(3,1),3}$   & 18 & 24 & $(1,1)$ & 1             &
$\ket{\yng(2,1),\yng(3,2),3}$   & 21 & 16 & $(1,1)$ & $\frac{1}{2}$ \\[8pt]
$\ket{\yng(2,1),\yng(2,2),3}$   & 18 &  8 & $(1,1)$ & 0             &
$\ket{\yng(1,1,1),\yng(4,1),3}$ & 21 &  4 & $(0,0)$ & $\frac{3}{2}$ \\[8pt]
$\ket{\yng(1,1,1),\yng(4),3}$   & 18 &  5 & $(0,0)$ & 2             &
$\ket{\yng(2,2),\yng(3,2),4}$   & 23 & 12 & $(0,2)$ & $\frac{1}{2}$ \\[8pt]
$\ket{\yng(1,1,1),\yng(3,1),3}$ & 18 &  3 & $(0,0)$ & 1             &
$\ket{\yng(2,1,1),\yng(4,1),4}$ & 23 & 12 & $(1,0)$ & $\frac{3}{2}$ \\[8pt]
$\ket{\yng(2,2),\yng(2,2),4}$   & 20 &  6 & $(0,2)$ & 0             &
$\ket{\yng(2,1,1),\yng(3,2),4}$ & 23 &  6 & $(1,0)$ & $\frac{1}{2}$ \\[8pt]
$\ket{\yng(2,1,1),\yng(3,1),4}$ & 20 &  9 & $(1,0)$ & 1             &
$\ket{\yng(2,2,1),\yng(3,2),5}$ & 25 &  6 & $(0,1)$ & $\frac{1}{2}$ \\[8pt]
$\ket{\yng(2,2),\yng(3,3),4}$   & 26 &  6 & $(0,2)$ & 0             &
$\ket{\yng(2,2,1),\yng(4,3),5}$ & 31 &  6 & $(0,1)$ & $\frac{1}{2}$ \\[8pt]
$\ket{\yng(2,1,1),\yng(4,2),4}$ & 26 &  9 & $(1,0)$ & 1             &
$\ket{\yng(2,2,2),\yng(4,3),6}$ & 33 &  2 & $(0,0)$ & $\frac{1}{2}$ \\[8pt]
$\ket{\yng(2,2,1),\yng(4,2),5}$ & 28 &  9 & $(0,1)$ & 1             &
                                &    &    &         &               \\[8pt]
$\ket{\yng(2,2,1),\yng(3,3),5}$ & 28 &  3 & $(0,1)$ & 0             &
                                &    &    &         &               \\[8pt]
$\ket{\yng(2,2,2),\yng(3,3),6}$ & 30 &  1 & $(0,0)$ & 0             &
                                &    &    &         &               \\[8pt]
$\ket{\yng(2,2,2),\yng(4,4),6}$ & 36 &  1 & $(0,0)$ & 0             &
                                &    &    &         &               \\[8pt]
\end{longtable}
\Yboxdim7.0pt


\subsection{$\left\{ \mathfrak{su}(1|2) \oplus \mathfrak{su}(3) \right\} ~ \circledS ~ \mathfrak{h}^{9,8}$}
\label{SubSec:SU12SU3}

In this case, after the decomposition of $\mathfrak{su}(4)$ into $\mathfrak{su}(3) \oplus \mathfrak{u}(1)_D$
(according to equations \eqref{SU4decompSU3U1} and \eqref{SU3D}) and relabeling the bosonic oscillators as
$\grave{a}_m(K)$, $\grave{a}^m(K)$, $\grave{b}_m(K)$, $\grave{b}^m(K)$ (for $i=1,2,3$) and $\tilde{a}_4(K)$,
$\tilde{a}^4(K)$, $\tilde{b}_4(K)$, $\tilde{b}^4(K)$ (for $i=4$), we use the following linear combination of
$G = \frac{1}{4} \left( N_B + 2 N_F \right)$ and $D = \frac{1}{4} \left( 3 N_{B_2} - N_{B_1} \right)$:
\begin{equation}
G + D  = N_{B_2} + \frac{1}{2} N_F \,,
\end{equation}
and $\mathfrak{su}(2)_F$ to form $\mathfrak{su}(1|2)$.

Now, the dynamical compact part of the pp-wave superalgebra has the decomposition
\begin{equation}
\mathfrak{g}^{(0)} = \mathfrak{su}(4|2)
             \supset \mathfrak{su}(1|2) \oplus \mathfrak{su}(3) \oplus \mathfrak{u}(1)_{G + \frac{1}{3} D}
\label{SU42decompSU12SU3}
\end{equation}
where
\begin{equation}
G + \frac{1}{3} D = \frac{1}{2} ( N_{B_2} + N_F ) + \frac{1}{6} N_{B_1} \,.
\end{equation}
Therefore, we must decompose the $SU(4|2)$ covariant super-oscillators in the $SU(1|2) \times SU(3)$ basis as
\begin{equation}
\begin{split}
\xi^A(K)  = \left( \begin{matrix} a^i(K) \cr \alpha^\mu(K) \cr \end{matrix} \right)
 \qquad & \longrightarrow \qquad
  \tilde{\xi}^R(K) \oplus \grave{a}^m(K) \\
\eta^A(K) = \left( \begin{matrix} b^i(K) \cr \beta^\mu(K) \cr \end{matrix} \right)
 \qquad & \longrightarrow \qquad
  \tilde{\eta}^R(K) \oplus \grave{b}^m(K) \,, \qquad \mbox{etc.}
\end{split}
\end{equation}
where
\begin{equation}
\tilde{\xi}^R(K)  = \left( \begin{matrix} \tilde{a}^4(K) \cr \alpha^\mu(K) \cr \end{matrix} \right)
 \qquad \qquad
\tilde{\eta}^R(K) = \left( \begin{matrix} \tilde{b}^4(K) \cr \beta^\mu(K) \cr \end{matrix} \right) \,.
\end{equation}

The decomposition of $\hat{\mathfrak{g}}^{(+1)}$ space (and similarly $\hat{\mathfrak{g}}^{(-1)}$ space) with
respect to this new dynamical compact basis $SU(1|2) \times SU(3)$ takes the form
\begin{equation}
\begin{split}
\hat{\mathfrak{g}}^{(+1)}
   = \soneonebox \Big|_{SU(4|2)}
 & = \left( \: \soneonebox \,,\, 1 \right) \oplus
     \left( \: \sonebox \,,\, \onebox \right) \oplus
     \left( 1 \,,\, \oneonebox \right)
     \Big|_{SU(1|2) \times SU(3)} \\
\vec{\xi}^{[A} \cdot \vec{\eta}^{B\}}
 & = \vec{\tilde{\xi}}^{[R} \cdot \vec{\tilde{\eta}}^{S\}} \oplus
     \left( \left( \vec{\tilde{\xi}}^R \cdot \vec{\grave{b}}^n -
                   \vec{\grave{a}}^n \cdot \vec{\tilde{\eta}}^R
            \right)
     \right. \\
 &   \qquad \qquad \qquad \quad
     \left. + \left( \vec{\grave{a}}^m \cdot \vec{\tilde{\eta}}^S -
                     \vec{\tilde{\xi}}^S \cdot \vec{\grave{b}}^m
              \right)
     \right) \oplus
     \vec{\grave{a}}^{[m} \cdot \vec{\grave{b}}^{n]}
\end{split}
\end{equation}
In the $SU(2) \times SU(3)$ basis, this decomposition is obviously identical to equation
\eqref{SU3SU2U1decomp}:
\begin{equation}
\begin{split}
\hat{\mathfrak{g}}^{(+1)}
   = \soneonebox \Big|_{SU(4|2)}
 & = \left( \soneonebox \,,\, 1 \right) \oplus
     \left( \sonebox \,,\, \onebox \right) \oplus
     \left( 1 \,,\, \oneonebox \right)
     \Big|_{SU(1|2) \times SU(3)} \\
 & = \left( \yng(1) \,,\, 1 \right) \oplus
     \left( \yng(2) \,,\, 1 \right) \oplus
     \left( 1 \,,\, \yng(1) \right) \oplus
     \left( \yng(1) \,,\, \yng(1) \right) \oplus
     \left( 1 \,,\, \yng(1,1) \right)
     \Big|_{SU(2) \times SU(3)}
\end{split}
\end{equation}

Again in this case, the supersymmetries in $\hat{\mathfrak{g}}^{(+1)}$, $Q^{4\mu} = \left( \yng(1) \,,\, 1
\right)$ and $Q^{m\mu} = \left( \yng(1) \,,\, \yng(1) \right)$, transform as $\mathbf{(2,1)} \oplus
\mathbf{(1,6)}$ under $SU(2) \times SU(3)$.

From the subspace $\mathfrak{g}^{(0)}$, this time we eliminate generators of the form $\mathcal{M}^R_{~~m}$ and
$\mathcal{M}^n_{~S}$ (6 bosonic and 12 fermionic), and the $\mathfrak{u}(1)$ generator in equation
\eqref{SU42decompSU12SU3}, since we want to retain only the $\mathfrak{su}(1|2) \oplus \mathfrak{su}(3)$
subalgebra. Therefore, this pp-wave superalgebra has only 20 supersymmetries (8 in $\hat{\mathfrak{g}}^{(+1)}$, 8
in $\hat{\mathfrak{g}}^{(-1)}$ and 4 in $\hat{\mathfrak{g}}^{(0)} = \mathfrak{su}(1|2) \oplus \mathfrak{su}(3)$).

The number of bosonic generators that remain in the new dynamical compact subsuperalgebra
$\hat{\mathfrak{g}}^{(0)}$ is 12. Eleven of them are rotation generators, that belong to $SU(2) \times SU(3)$ and
the other one is the Hamiltonian ($= G + D$). Once again we rescale it, so that we obtain energy increments of
the states in integers:
\begin{equation}
\mathcal{H} = 2 N_{B_2} + N_F \,.
\label{SU12SU3H}
\end{equation}

Therefore, two kinematical supersymmetries, $Q^{4\mu} = \left( \yng(1) \,,\, 1 \right)$ increase energy by 3 units,
and the remaining six, $Q^{m\mu} = \left( \yng(1) \,,\, \yng(1) \right)$ increase energy by 1 unit.

We must mention again, that this solution is not found in literature on the study of possible pp-wave solutions in
eleven dimensions either, for the same reason that its rotation group contains an $SU(3)$ part, which does not have an
isomorphic orthogonal group $SO(m)$.

The symmetry superalgebra of this pp-wave solution is clearly
\begin{equation*}
\left[ \mathfrak{su}(1|2) \oplus \mathfrak{su}(3) \right] ~ \circledS ~ \mathfrak{h}^{9,8} \,.
\end{equation*}

Its zero-mode spectrum, in the basis $SU(2) \times SU(3)$, obtained by starting from the ground state $|
\hat{\Omega} \rangle = \ket{0}$ is identical to that in Table \ref{Table:SU32U1} as far as the transformation
properties of the states under $SU(2)\times SU(3)$ are concerned. The states, however, differ in energies from
those in Table \ref{Table:SU32U1} (see equation \eqref{SU12SU3H}).


\subsection{$\left\{ \mathfrak{su}(4|1) \oplus \mathfrak{u}(1) \right\} ~ \circledS ~ \mathfrak{h}^{9,8}$}
\label{SubSec:SU41U1}

This time, we break the $\mathfrak{su}(2)$ subalgebra of $\mathfrak{su}(4|2) \supset \mathfrak{su}(4) \oplus
\mathfrak{su}(2) \oplus \mathfrak{u}(1)_G$ and rename the fermionic oscillators as $\grave{\alpha}_1(K)$,
$\grave{\alpha}^1(K)$, $\grave{\beta}_1(K)$, $\grave{\beta}^1(K)$ (for $\mu=1$) and $\tilde{\alpha}_2(K)$,
$\tilde{\alpha}^2(K)$, $\tilde{\beta}_2(K)$, $\tilde{\beta}^2(K)$ (for $\mu=2$). The $\mathfrak{u}(1)_F$ charge that
arises in the breaking of $\mathfrak{su}(2)$ is
\begin{equation}
F = N_{F_1} - N_{F_2}
\end{equation}
where $N_{F_1} = \vec{\grave{\alpha}}^1 \cdot \vec{\grave{\alpha}}_1 + \vec{\grave{\beta}}^1 \cdot
\vec{\grave{\beta}}_1$ and $N_{F_2} = \vec{\tilde{\alpha}}^2 \cdot \vec{\tilde{\alpha}}_2 + \vec{\tilde{\beta}}^2
\cdot \vec{\tilde{\beta}}_2$. Then we form $\mathfrak{su}(4|1)$ (which is spanned by all the bosonic oscillators $a$,
$b$ and the fermionic oscillators $\grave{\alpha}$, $\grave{\beta}$) by combining $\mathfrak{su}(4)$ and the following
linear combination of $G = \frac{1}{4} \left( N_B + 2 N_F \right)$ (equation \eqref{SU42G}) and $F$:
\begin{equation}
G + \frac{1}{2} F = \frac{1}{4} N_B + N_{F_1}
\label{SU4U1}
\end{equation}
so that we can write
\begin{equation}
\mathfrak{su}(4) \oplus \mathfrak{u}(1)_{G + \frac{1}{2} F} \subset \mathfrak{su}(4|1) \,.
\label{SU41SU4U1}
\end{equation}
We must therefore decompose $SU(4|2)$ covariant super-oscillators, in the $SU(4|1) \times U(1)$ basis, into:
\begin{equation}
\begin{split}
\xi^A(K)  = \left( \begin{matrix} a^i(K) \cr \alpha^\mu(K) \cr \end{matrix} \right)
 \qquad & \longrightarrow \qquad
  \grave{\xi}^M(K) \oplus \tilde{\alpha}^2(K) \\
\eta^A(K) = \left( \begin{matrix} b^i(K) \cr \beta^\mu(K) \cr \end{matrix} \right)
 \qquad & \longrightarrow \qquad
  \grave{\eta}^M(K) \oplus \tilde{\beta}^2(K) \,, \qquad \mbox{etc.}
\end{split}
\end{equation}
where
\begin{equation}
\grave{\xi}^M(K)  = \left( \begin{matrix} a^i(K) \cr \grave{\alpha}^1(K) \cr \end{matrix} \right)
 \qquad \qquad
\grave{\eta}^M(K) = \left( \begin{matrix} b^i(K) \cr \grave{\beta}^1(K) \cr \end{matrix} \right) \,.
\end{equation}

The maximal dynamical compact subsuperalgebra $\mathfrak{g}^{(0)}$ has now decomposed into
\begin{equation}
\mathfrak{su}(4|2) \supset \mathfrak{su}(4|1) \oplus \mathfrak{u}(1)
\label{SU42decompSU41U1}
\end{equation}
where the $\mathfrak{u}(1)$ charge is given by
\begin{equation}
2 G - \frac{1}{2} F = \frac{1}{2} \left( N_B + N_{F_1} \right) + \frac{3}{2} N_{F_2} \,.
\end{equation}

With respect to this new dynamical compact basis $SU(4|1) \times U(1)_{2 G - \frac{1}{2} F}$,
$\hat{\mathfrak{g}}^{(+1)}$ space (and similarly $\hat{\mathfrak{g}}^{(-1)}$ space) has the following
decomposition:
\begin{equation}
\begin{split}
\hat{\mathfrak{g}}^{(+1)}
   = \soneonebox \Big|_{SU(4|2)}
 & = \left( \: \soneonebox \,,\, 1 \right) \oplus
     \left( \: \sonebox \,,\, 2 \right) \oplus
     \left( 1 \,,\, 3 \right)
     \Big|_{SU(4|1) \times U(1)_{2 G - \frac{1}{2} F}} \\
\vec{\xi}^{[A} \cdot \vec{\eta}^{B\}}
 & = \vec{\grave{\xi}}^{[M} \cdot \vec{\grave{\eta}}^{N\}} \oplus
     \left( \left( \vec{\grave{\xi}}^M \cdot \vec{\tilde{\beta}}^2 -
                   \vec{\tilde{\alpha}}^2 \cdot \vec{\grave{\eta}}^M
            \right)
     \right. \\
 &   \qquad \qquad \qquad \quad
     \left. + \left( \vec{\tilde{\alpha}}^2 \cdot \vec{\grave{\eta}}^N -
                     \vec{\grave{\xi}}^N \cdot \vec{\tilde{\beta}}^2
              \right)
     \right) \oplus
     \vec{\tilde{\alpha}}^{(2} \cdot \vec{\tilde{\beta}}^{2)}
\end{split}
\end{equation}
In the $SU(4) \times U(1)_{2 G - \frac{1}{2} F}$ basis, this decomposition takes the form:
\begin{equation}
\begin{split}
\hat{\mathfrak{g}}^{(+1)}
   = \soneonebox \Big|_{SU(4|2)}
 & = \left( \: \soneonebox \,,\, 1 \right) \oplus
     \left( \: \sonebox \,,\, 2 \right) \oplus
     \left( 1 \,,\, 3 \right)
     \Big|_{SU(4|1) \times U(1)_{2 G - \frac{1}{2} F}} \\
 & = \left( \yng(1,1) \,,\, 1 \right) \oplus
     \left( \yng(1) \,,\, 1 \right) \oplus
     \left( 1 \,,\, 1 \right) \\
 &   \quad \oplus
     \left( \yng(1) \,,\, 2 \right) \oplus
     \left( 1 \,,\, 2 \right) \oplus
     \left( 1 \,,\, 3 \right)
     \Big|_{SU(4) \times U(1)_{2 G - \frac{1}{2} F}}
\end{split}
\end{equation}

Naturally, all 8 supersymmetries in the $\hat{\mathfrak{g}}^{(+1)}$ space (and similarly, all 8 in the
$\hat{\mathfrak{g}}^{(-1)}$ space) are preserved, but now they transform as $\mathbf{4} + \mathbf{4}$ of $SU(4) \times
U(1)$: $Q^{i1} = \left( \yng(1) \,,\, 1 \right)$ and $Q^{i2} = \left( \yng(1) \,,\, 2 \right)$.

From $\mathfrak{g}^{(0)}$ subspace, we eliminate generators of the form $\mathcal{M}^M_{~~2}$ and
$\mathcal{M}^2_{~N}$ (2 bosonic and 8 fermionic) to retain only the $\mathfrak{su}(4|1) \oplus \mathfrak{u}(1)_{2
G - \frac{1}{2} F}$ part, and therefore, this pp-wave superalgebra also has 24 supersymmetries (8 in
$\hat{\mathfrak{g}}^{(+1)}$, 8 in $\hat{\mathfrak{g}}^{(-1)}$ and 8 in $\hat{\mathfrak{g}}^{(0)} =
\mathfrak{su}(4|1) \oplus \mathfrak{u}(1)$).

The number of bosonic generators that remain in $\hat{\mathfrak{g}}^{(0)}$ is 17. Sixteen of them are rotation
generators, that belong to $SU(4) \times U(1) \approx SO(6) \times SO(2)$ and the other one, which is the
$\mathfrak{u}(1)$ charge inside $\mathfrak{su}(4|1)$ (see equations \eqref{SU41SU4U1} and \eqref{SU4U1}) is the
Hamiltonian. Once again we rescale it, so that we obtain energy increases of the states in integer steps:
\begin{equation}
\mathcal{H} = N_B + 4 N_{F_1}
\label{SU41U1H}
\end{equation}

Therefore, it is clear that half of the kinematical supersymmetries in $\hat{\mathfrak{g}}^{(+1)}$,
$\left( \yng(1) \,,\, 1 \right)$ increase energy by 5 units, while the other half $\left( \yng(1) \,,\, 2 \right)$
increase energy by only 1 unit.

The symmetry superalgebra of this pp-wave solution in eleven dimensions is \footnote{If we compactify this
geometry along the direction which corresponds to the single noncompact generator $\vec{\tilde{\alpha}}^{(2}
\cdot \vec{\tilde{\beta}}^{2)}$ in $\hat{\mathfrak{g}}^{(+1)}$ and its Hermitian conjugate generator
$\vec{\tilde{\alpha}}_{(2} \cdot \vec{\tilde{\beta}}_{2)}$ in $\hat{\mathfrak{g}}^{(-1)}$, we obtain another ten
dimensional pp-wave geometry with 24 supercharges. The superalgebra becomes $\left[ \mathfrak{su}(4|1) \oplus
\mathfrak{u}(1) \right] ~ \circledS ~ \mathfrak{h}^{8,8} $. }
\begin{equation*}
\left[ \mathfrak{su}(4|1) \oplus \mathfrak{u}(1) \right] ~ \circledS ~ \mathfrak{h}^{9,8} \,.
\end{equation*}

Now we construct the zero-mode spectrum of this pp-wave superalgebra, in the basis $SU(4) \times U(1)_{2 G -
\frac{1}{2} F}$, by starting from the ground state $| \hat{\Omega} \rangle = \ket{0}$ (Table \ref{Table:SU41U1}).

\Yboxdim5.0pt
\begin{longtable}[c]{|l|c|c|c||l|c|c|c|}
\kill
\caption[The zero-mode spectrum of the eleven dimensional pp-wave superalgebra with 24 supersymmetries, $\left\{
\mathfrak{su}(4|1) \oplus \mathfrak{u}(1) \right\} ~ \circledS ~ \mathfrak{h}^{9,8}$]{\normalsize The zero-mode
spectrum of the eleven dimensional pp-wave superalgebra with 24 supersymmetries, $\left[ \mathfrak{su}(4|1) \oplus
\mathfrak{u}(1) \right] ~ \circledS ~ \mathfrak{h}^{9,8}$.
\label{Table:SU41U1}} \\
\hline
\hline
& & & & & & & \\
{\scriptsize $SU(4) \times U(1)_{2 G - \frac{1}{2} F}$} &
 &
 &
{\scriptsize $SU(4)$} &
{\scriptsize $SU(4) \times U(1)_{2 G - \frac{1}{2} F}$} &
 &
 &
{\scriptsize $SU(4)$} \\
{\scriptsize Young tableau} &
 &
{\scriptsize $N_{\mathrm{dof}}$} &
{\scriptsize Dynkin} &
{\scriptsize Young tableau} &
 &
{\scriptsize $N_{\mathrm{dof}}$} &
{\scriptsize Dynkin} \\
{\scriptsize (bosonic states)} &
{\scriptsize $\mathcal{H}$} &
{\scriptsize (B)} &
{\scriptsize labels} &
{\scriptsize (fermionic states)} &
{\scriptsize $\mathcal{H}$} &
{\scriptsize (F)} &
{\scriptsize labels} \\
& & & & & & & \\
\hline
& & & & & & & \\
\endfirsthead
\caption[]{(continued)} \\
\hline
& & & & & & & \\
{\scriptsize $SU(4) \times U(1)_{2 G - \frac{1}{2} F}$} &
 &
 &
{\scriptsize $SU(4)$} &
{\scriptsize $SU(4) \times U(1)_{2 G - \frac{1}{2} F}$} &
 &
 &
{\scriptsize $SU(4)$} \\
{\scriptsize Young tableau} &
 &
{\scriptsize $N_{\mathrm{dof}}$} &
{\scriptsize Dynkin} &
{\scriptsize Young tableau} &
 &
{\scriptsize $N_{\mathrm{dof}}$} &
{\scriptsize Dynkin} \\
{\scriptsize (bosonic states)} &
{\scriptsize $\mathcal{H}$} &
{\scriptsize (B)} &
{\scriptsize labels} &
{\scriptsize (fermionic states)} &
{\scriptsize $\mathcal{H}$} &
{\scriptsize (F)} &
{\scriptsize labels} \\
& & & & & & & \\
\hline
& & & & & & & \\
\endhead
& & & & & & & \\
\hline
\endfoot
& & & & & & & \\
\hline
& & & & & & & \\
& & 128 & & & & 128 & \\
& & & & & & & \\
\hline
\hline
\endlastfoot
$\ket{1,0}$              &  0 &  1 & $(0,0,0)$ &
$\ket{\yng(1),2}$        &  1 &  4 & $(1,0,0)$ \\[8pt]
$\ket{\yng(1,1),4}$      &  2 &  6 & $(0,1,0)$ &
$\ket{\yng(1,1,1),6}$    &  3 &  4 & $(0,0,1)$ \\[8pt]
$\ket{\yng(1,1,1,1),8}$  &  4 &  1 & $(0,0,0)$ &
$\ket{\yng(1),1}$        &  5 &  4 & $(1,0,0)$ \\[8pt]
$\ket{\yng(2),3}$        &  6 & 10 & $(2,0,0)$ &
$\ket{\yng(2,1),5}$      &  7 & 20 & $(1,1,0)$ \\[8pt]
$\ket{\yng(1,1),3}$      &  6 &  6 & $(0,1,0)$ &
$\ket{\yng(1,1,1),5}$    &  7 &  4 & $(0,0,1)$ \\[8pt]
$\ket{\yng(2,1,1),7}$    &  8 & 15 & $(1,0,1)$ &
$\ket{\yng(2,1,1,1),9}$  &  9 &  4 & $(1,0,0)$ \\[8pt]
$\ket{\yng(1,1,1,1),7}$  &  8 &  1 & $(0,0,0)$ &
$\ket{\yng(2,1),4}$      & 11 & 20 & $(1,1,0)$ \\[8pt]
$\ket{\yng(1,1),2}$      & 10 &  6 & $(0,1,0)$ &
$\ket{\yng(1,1,1),4}$    & 11 &  4 & $(0,0,1)$ \\[8pt]
$\ket{\yng(2,2),6}$      & 12 & 20 & $(0,2,0)$ &
$\ket{\yng(2,2,1),8}$    & 13 & 20 & $(0,1,1)$ \\[8pt]
$\ket{\yng(2,1,1),6}$    & 12 & 15 & $(1,0,1)$ &
$\ket{\yng(2,1,1,1),8}$  & 13 &  4 & $(1,0,0)$ \\[8pt]
$\ket{\yng(1,1,1,1),6}$  & 12 &  1 & $(0,0,0)$ &
$\ket{\yng(1,1,1),3}$    & 15 &  4 & $(0,0,1)$ \\[8pt]
$\ket{\yng(2,2,1,1),10}$ & 14 &  6 & $(0,1,0)$ &
$\ket{\yng(2,2,1),7}$    & 17 & 20 & $(0,1,1)$ \\[8pt]
$\ket{\yng(2,1,1),5}$    & 16 & 15 & $(1,0,1)$ &
$\ket{\yng(2,1,1,1),7}$  & 17 &  4 & $(1,0,0)$ \\[8pt]
$\ket{\yng(1,1,1,1),5}$  & 16 &  1 & $(0,0,0)$ &
$\ket{\yng(2,2,2,1),11}$ & 19 &  4 & $(0,0,1)$ \\[8pt]
$\ket{\yng(2,2,2),9}$    & 18 & 10 & $(0,0,2)$ &
$\ket{\yng(2,1,1,1),6}$  & 21 &  4 & $(1,0,0)$ \\[8pt]
$\ket{\yng(2,2,1,1),9}$  & 18 &  6 & $(0,1,0)$ &
$\ket{\yng(2,2,2,1),10}$ & 23 &  4 & $(0,0,1)$ \\[8pt]
$\ket{\yng(1,1,1,1),4}$  & 20 &  1 & $(0,0,0)$ &
                         &    &    &           \\[8pt]
$\ket{\yng(2,2,1,1),8}$  & 22 &  6 & $(0,1,0)$ &
                         &    &    &           \\[8pt]
$\ket{\yng(2,2,2,2),12}$ & 24 &  1 & $(0,0,0)$ &
                         &    &    &           \\[8pt]
\end{longtable}
\Yboxdim7.0pt


\subsection{$\left\{ \mathfrak{su}(2|1) \oplus \mathfrak{su}(2|1) \oplus \mathfrak{u}(1) \right\} ~ \circledS ~
\mathfrak{h}^{8,8}$}
\label{SubSec:SU21SU21U1}

In this case, again we take the decomposition of $\mathfrak{su}(4|2)$ into its even subalgebra
\begin{equation*}
\mathfrak{su}(4)_B \oplus \mathfrak{su}(2)_F \oplus \mathfrak{u}(1)_{G = \frac{1}{4} ( N_B + 2 N_F )}
\end{equation*}
as in equations \eqref{SU42decompSU4SU2U1} and \eqref{SU42G}, and then break $\mathfrak{su}(4)_B$ into
\begin{equation*}
\mathfrak{su}(2)_{B_1} \oplus \mathfrak{su}(2)_{B_2} \oplus \mathfrak{u}(1)_{D = \frac{1}{2} ( N_{B_1} - N_{B_2} )}
\end{equation*}
as in equations \eqref{SU4decompSU2SU2U1} and \eqref{SU2SU2D}. Next we break $\mathfrak{su}(2)_F$ as well, into
$F_1$ and $F_2$ parts, as done in the previous case in section \ref{SubSec:SU41U1}.

Now we combine the $F_1$ part (spanned by $\grave{\alpha}$, $\grave{\beta}$ oscillators) with $\mathfrak{su}(2)_{B_1}$
(spanned by $\grave{a}$, $\grave{b}$ oscillators) to form one $\mathfrak{su}(2|1)$. It should be noted
that, the $\mathfrak{u}(1)$ charge inside this $\mathfrak{su}(2_{B_1}|1_{F_1})$ is:
\begin{equation}
2 G + D + F = N_{B_1} + 2 N_{F_1} \,.
\label{firstSU21}
\end{equation}

Similarly, we combine the $F_2$ part (spanned by $\tilde{\alpha}$, $\tilde{\beta}$) with $\mathfrak{su}(2)_{B_2}$
(spanned by $\tilde{a}$, $\tilde{b}$) to form a second $\mathfrak{su}(2|1)$. The $\mathfrak{u}(1)$ charge inside this
$\mathfrak{su}(2_{B_2}|1_{F_2})$ is given by
\begin{equation}
2 G - D - F = N_{B_2} + 2 N_{F_2} \,.
\label{secondSU21}
\end{equation}

Now, the decomposition of the dynamical compact part of the pp-wave superalgebra reads as
\begin{equation}
\mathfrak{g}^{(0)} = \mathfrak{su}(4|2)
             \supset \mathfrak{su}(2|1) \oplus \mathfrak{su}(2|1) \oplus \mathfrak{u}(1)_{2 D + F}
\label{SU42decompSU21SU21U1}
\end{equation}
where the $\mathfrak{u}(1)$ charge
\begin{equation}
2 D + F = N_{B_1} + N_{F_1} - N_{B_2} - N_{F_2}
\label{outsideU1}
\end{equation}
commutes with both $\mathfrak{su}(2|1)$.

Now it clear that we must decompose the $SU(4|2)$ covariant super-oscillators in this new $SU(2|1) \times SU(2|1)$
basis as:
\begin{equation}
\begin{split}
\xi^A(K)  = \left( \begin{matrix} a^i(K) \cr \alpha^\mu(K) \cr \end{matrix} \right)
 \qquad & \longrightarrow \qquad
  \grave{\xi}^M(K) \oplus \tilde{\xi}^R(K) \\
\eta^A(K) = \left( \begin{matrix} b^i(K) \cr \beta^\mu(K) \cr \end{matrix} \right)
 \qquad & \longrightarrow \qquad
  \grave{\eta}^M(K) \oplus \tilde{\eta}^R(K) \,, \qquad \mbox{etc.}
\end{split}
\label{SU42decompSU21SU21}
\end{equation}
where
\begin{equation}
\begin{split}
\grave{\xi}^M(K)  = \left( \begin{matrix} \grave{a}^m(K) \cr \grave{\alpha}^1(K) \cr \end{matrix} \right)
 \qquad \qquad &
\grave{\eta}^M(K) = \left( \begin{matrix} \grave{b}^m(K) \cr \grave{\beta}^1(K) \cr \end{matrix} \right) \\
\tilde{\xi}^R(K)  = \left( \begin{matrix} \tilde{a}^r(K) \cr \tilde{\alpha}^2(K) \cr \end{matrix} \right)
 \qquad \qquad &
\tilde{\eta}^R(K) = \left( \begin{matrix} \tilde{b}^r(K) \cr \tilde{\beta}^2(K) \cr \end{matrix} \right) \,.
\end{split}
\label{SU21Oscillators}
\end{equation}

The subspace $\hat{\mathfrak{g}}^{(+1)}$ then decomposes with respect to the new basis $SU(2|1) \times SU(2|1) \times
U(1)$ as
\begin{equation}
\begin{split}
\hat{\mathfrak{g}}^{(+1)}
   = \soneonebox \Big|_{SU(4|2)}
 & = \left( \: \soneonebox \,,\, 1 \,,\, 2 \right) \oplus
     \left( \: \sonebox \,,\, \sonebox \,,\, 0 \right) \oplus
     \left( 1 \,,\, \soneonebox \,,\, -2 \right)
     \Big|_{SU(2|1) \times SU(2|1) \times U(1)} \\
\vec{\xi}^{[A} \cdot \vec{\eta}^{B\}}
 & = \vec{\grave{\xi}}^{[M} \cdot \vec{\grave{\eta}}^{N\}} \oplus
     \left( \left( \vec{\grave{\xi}}^M \cdot \vec{\tilde{\eta}}^S -
                   \vec{\grave{\eta}}^M \cdot \vec{\tilde{\xi}}^S
            \right)
     \right. \\
 &   \qquad \qquad \qquad \quad
     \left. + \left( \vec{\tilde{\xi}}^R \cdot \vec{\grave{\eta}}^N -
                     \vec{\tilde{\eta}}^R \cdot \vec{\grave{\xi}}^N
              \right)
     \right) \oplus
     \vec{\tilde{\xi}}^{[R} \cdot \vec{\tilde{\eta}}^{S\}} \,.
\end{split}
\end{equation}

In the $SU(2)_{B_1} \times U(1)_{N_{B_1} + 2 N_{F_1}} \times SU(2)_{B_2} \times U(1)_{N_{B_2} + 2 N_{F_2}} \times
U(1)_{N_{B_1} + N_{F_1} - N_{B_2} - N_{F_2}}$ basis, the above decomposition takes the form:
\begin{equation}
\begin{split}
\hat{\mathfrak{g}}^{(+1)}
   = \soneonebox \Big| _{SU(4|2)}
 & = \left( \: \soneonebox \,,\, 1 \,,\, 2 \right) \oplus
     \left( \: \sonebox \,,\, \sonebox \,,\, 0 \right) \oplus
     \left( 1 \,,\, \soneonebox \,,\, -2 \right)
     \Big|_{SU(2|1) \times SU(2|1) \times U(1)} \\
 & = \left( \yng(1,1) \,,\, 2 \,,\, 1 \,,\, 0 \,,\, 2 \right) \oplus
     \left( \yng(1) \,,\, 3 \,,\, 1 \,,\, 0 \,,\, 2 \right) \oplus
     \left( 1 \,,\, 4 \,,\, 1 \,,\, 0 \,,\, 2 \right) \\
 &   \quad \oplus \left( \yng(1) \,,\, 1 \,,\, \yng(1) \,,\, 1 \,,\, 0 \right)
           \oplus \left( \yng(1) \,,\, 1 \,,\, 1 \,,\, 2 \,,\, 0 \right)
           \oplus \left( 1 \,,\, 2 \,,\, \yng(1) \,,\, 1 \,,\, 0 \right)
           \oplus \left( 1 \,,\, 2 \,,\, 1 \,,\, 2 \,,\, 0 \right) \\
 &   \quad \oplus \left( 1 \,,\, 0 \,,\, \yng(1,1) \,,\, 2 \,,\, -2 \right)
           \oplus \left( 1 \,,\, 0 \,,\, \yng(1) \,,\, 3 \,,\, -2 \right) \\
 &   \quad \oplus \left( 1 \,,\, 0 \,,\, 1 \,,\, 4 \,,\, -2 \right)
     \Big|_{SU(2)_{B_1} \times U(1)_{N_{B_1} + 2 N_{F_1}} \times SU(2)_{B_2} \times U(1)_{N_{B_2} + 2 N_{F_2}} \times
            U(1)}
\end{split}
\label{SU21SU21U1decomp}
\end{equation}

The kinematical supersymmetries in $\mathfrak{g}^{(+1)}$ in this new basis have decomposed as $\mathbf{2} + \mathbf{2}
+ \mathbf{2} + \mathbf{2}$: $Q^{m1} = \left( \yng(1) \,,\, 3 \,,\, 1 \,,\, 0 \,,\, 2 \right)$,
$Q^{m2} = \left( \yng(1) \,,\, 1 \,,\, 1 \,,\, 2 \,,\, 0 \right)$,
$Q^{r1} = \left( 1 \,,\, 2 \,,\, \yng(1) \,,\, 1 \,,\, 0 \right)$ and
$Q^{r2} = \left( 1 \,,\, 0 \,,\, \yng(1) \,,\, 3 \,,\, -2 \right)$.

From the original $\mathfrak{g}^{(0)}$ subspace, we eliminate generators of the form $\mathcal{M}^M_{~~R}$ and
$\mathcal{M}^S_{~~N}$ (10 bosonic and 8 fermionic). Therefore, this pp-wave superalgebra has 24 supersymmetries
(8 in $\hat{\mathfrak{g}}^{(+1)}$, 8 in $\hat{\mathfrak{g}}^{(-1)}$ and 8 in $\hat{\mathfrak{g}}^{(0)} =
\mathfrak{su}(2|1) \oplus \mathfrak{su}(2|1) \oplus \mathfrak{u}(1)$).

The number of bosonic generators that remain in the new dynamical compact subsuperalgebra
$\hat{\mathfrak{g}}^{(0)}$ is 9. Eight of them are rotation generators, that belong to $SU(2) \times SU(2) \times
U(1) \times U1) \approx SO(3) \times SO(3) \times SO(2) \times SO(2)$ and the other one is the Hamiltonian. A
priori it seems that the Hamiltonian of this superalgebra is not unique. There are two $\mathfrak{u}(1)$ charges
inside the two subsuperalgebras (see equations \eqref{firstSU21} and \eqref{secondSU21}) and a third
$\mathfrak{u}(1)$ charge that commutes with both $\mathfrak{su}(2|1)$ (equation \eqref{outsideU1}), so one might
choose any linear combination of them as the Hamiltonian. In fact, what we have in this case is a two-parameter
family of solutions with the same symmetry superalgebra, which differ only in the energies of the states.

Interestingly, this pp-wave superalgebra can be identified with the type IIA pp-wave solution given in
\cite{Mich02a,AGGP02}. If we choose the following linear combination of the three $\mathfrak{u}(1)$ charges as
the Hamiltonian:
\begin{equation}
\begin{split}
\mathcal{H} & = 2 \left( 2 D + F \right) - \left( 2 G + D + F \right) + 3 \left( 2 G - D - F \right) \\
            & = 4 G - 2 F \\
            & = N_{B_1} + N_{B_2} + 4 N_{F_2} \,,
\label{SU21SU21H}
\end{split}
\end{equation}
and assume that the bosonic singlet $\left( 1 \,,\, 0 \,,\, 1 \,,\, 4 \,,\, -2 \right)$ in
$\hat{\mathfrak{g}}^{(+1)}$ - see equation \eqref{SU21SU21U1decomp} - (and its hermitian conjugate in
$\hat{\mathfrak{g}}^{(-1)}$) corresponds to the transverse direction along which the eleven dimensional geometry
is compactified to ten dimensional type IIA, this superalgebra becomes the symmetry algebra of the pp-wave
solution in \cite{Mich02a,AGGP02}.

The six bosonic generators $\left( \yng(1,1) \,,\, 2 \,,\, 1 \,,\, 0 \,,\, 2 \right) \oplus
\left( 1 \,,\, 0 \,,\, \yng(1,1) \,,\, 2 \,,\, -2 \right) \oplus
\left( \yng(1) \,,\, 1 \,,\, \yng(1) \,,\, 1 \,,\, 0 \right)$ in $\hat{\mathfrak{g}}^{(+1)}$ (equation
\eqref{SU21SU21U1decomp}) correspond to the transverse directions $x^3,\dots,x^8$;
$\left( 1 \,,\, 2 \,,\, 1 \,,\, 2 \,,\, 0 \right)$ to $x^2$; and $\left( 1 \,,\, 4 \,,\, 1 \,,\, 0 \,,\, 2 \right)$ to
$x^1$.

The symmetry superalgebra of this pp-wave solution in type IIA geometry is
\begin{equation*}
\left[ \mathfrak{su}(2|1) \oplus \mathfrak{su}(2|1) \oplus \mathfrak{u}(1) \right] ~ \circledS ~ \mathfrak{h}^{8,8} \,.
\end{equation*}

The zero-mode spectrum of this pp-wave superalgebra, in the basis $SU(2)_{B_1} \times$ \linebreak $U(1)_{N_{B_1} + 2
N_{F_1}} \times SU(2)_{B_2} \times U(1)_{N_{B_2} + 2 N_{F_2}} \times U(1)_{N_{B_1} + N_{F_1} - N_{B_2} - N_{F_2}}$,
obtained by starting from the ground state $| \hat{\Omega} \rangle = \ket{0}$ is given in Table
\ref{Table:SU21SU21U1}. It should be noted that in this spectrum, some states occur with multiplicity greater than 1.

\Yboxdim5.0pt
\begin{longtable}[c]{|l|c|c|c||l|c|c|c|}
\kill
\caption[The zero-mode spectrum of type IIA pp-wave superalgebra with 24 supersymmetries, $\left\{ \mathfrak{su}(2|1)
\oplus \mathfrak{su}(2|1) \oplus \mathfrak{u}(1) \right\} ~ \circledS ~ \mathfrak{h}^{8,8}$]{\normalsize The zero-mode
spectrum of type IIA pp-wave superalgebra with 24 supersymmetries, $\left[ \mathfrak{su}(2|1) \oplus
\mathfrak{su}(2|1) \oplus \mathfrak{u}(1) \right] ~ \circledS ~ \mathfrak{h}^{8,8}$.
\label{Table:SU21SU21U1}} \\
\hline
\hline
& & & & & & & \\
{\scriptsize $SU(2)_{B_1} \times U(1)_{N_{B_1} + 2 N_{F_1}} \times$} &
 &
 &
 &
{\scriptsize $SU(2)_{B_1} \times U(1)_{N_{B_1} + 2 N_{F_1}} \times$} &
 &
 &
 \\
{\scriptsize $SU(2)_{B_2} \times U(1)_{N_{B_2} + 2 N_{F_2}} \times$} &
 &
 &
 &
{\scriptsize $SU(2)_{B_2} \times U(1)_{N_{B_2} + 2 N_{F_2}} \times$} &
 &
 &
 \\
{\scriptsize $U(1)_{N_{B_1} + N_{F_1} - N_{B_2} - N_{F_2}}$} &
 &
 &
 &
{\scriptsize $U(1)_{N_{B_1} + N_{F_1} - N_{B_2} - N_{F_2}}$} &
 &
 &
 \\
{\scriptsize Young tableau} &
 &
{\scriptsize $N_{\mathrm{dof}}$} &
{\scriptsize $SU(2)$} &
{\scriptsize Young tableau} &
 &
{\scriptsize $N_{\mathrm{dof}}$} &
{\scriptsize $SU(2)$} \\
{\scriptsize (bosonic states)} &
{\scriptsize $\mathcal{H}$} &
{\scriptsize (B)} &
{\scriptsize spin} &
{\scriptsize (fermionic states)} &
{\scriptsize $\mathcal{H}$} &
{\scriptsize (F)} &
{\scriptsize spin} \\
& & & & & & & \\
\hline
& & & & & & & \\
\endfirsthead
\caption[]{(continued)} \\
\hline
& & & & & & & \\
{\scriptsize $SU(2)_{B_1} \times U(1)_{N_{B_1} + 2 N_{F_1}} \times$} &
 &
 &
 &
{\scriptsize $SU(2)_{B_1} \times U(1)_{N_{B_1} + 2 N_{F_1}} \times$} &
 &
 &
 \\
{\scriptsize $SU(2)_{B_2} \times U(1)_{N_{B_2} + 2 N_{F_2}} \times$} &
 &
 &
 &
{\scriptsize $SU(2)_{B_2} \times U(1)_{N_{B_2} + 2 N_{F_2}} \times$} &
 &
 &
 \\
{\scriptsize $U(1)_{N_{B_1} + N_{F_1} - N_{B_2} - N_{F_2}}$} &
 &
 &
 &
{\scriptsize $U(1)_{N_{B_1} + N_{F_1} - N_{B_2} - N_{F_2}}$} &
 &
 &
 \\
{\scriptsize Young tableau} &
 &
{\scriptsize $N_{\mathrm{dof}}$} &
{\scriptsize $SU(2)$} &
{\scriptsize Young tableau} &
 &
{\scriptsize $N_{\mathrm{dof}}$} &
{\scriptsize $SU(2)$} \\
{\scriptsize (bosonic states)} &
{\scriptsize $\mathcal{H}$} &
{\scriptsize (B)} &
{\scriptsize spin} &
{\scriptsize (fermionic states)} &
{\scriptsize $\mathcal{H}$} &
{\scriptsize (F)} &
{\scriptsize spin} \\
& & & & & & & \\
\hline
& & & & & & & \\
\endhead
& & & & & & & \\
\hline
\endfoot
& & & & & & & \\
\hline
& & & & & & & \\
& & 128 & & & & 128 & \\
& & & & & & & \\
\hline
\hline
\endlastfoot
$\ket{1,0,1,0,0}$                           &  0 & 1 & $(0,0)$                     &
$\ket{1,2,\yng(1),1,0}$                     &  1 & 2 & $(0,\frac{1}{2})$           \\[8pt]
$\ket{1,4,\yng(1,1),2,0}$                   &  2 & 1 & $(0,0)$                     &
$\ket{\yng(1),3,1,0,2}$                     &  1 & 2 & $(\frac{1}{2},0)$           \\[8pt]
$\ket{\yng(1),5,\yng(1),1,2}$               &  2 & 4 & $(\frac{1}{2},\frac{1}{2})$ &
$\ket{\yng(1),7,\yng(1,1),2,2}$             &  3 & 2 & $(\frac{1}{2},0)$           \\[8pt]
$\ket{\yng(1,1),6,1,0,4}$                   &  2 & 1 & $(0,0)$                     &
$\ket{\yng(1,1),8,\yng(1),1,4}$             &  3 & 2 & $(0,\frac{1}{2})$           \\[8pt]
$\ket{\yng(1,1),10,\yng(1,1),2,4}$          &  4 & 1 & $(0,0)$                     &
$\ket{1,0,\yng(1),3,-2}$                    &  5 & 2 & $(0,\frac{1}{2})$           \\[8pt]
$\ket{1,2,\yng(2),4,-2}$                    &  6 & 3 & $(0,1)$                     &
$\ket{\yng(1),1,1,2,0}$                     &  5 & 2 & $(\frac{1}{2},0)$           \\[8pt]
$\ket{1,2,\yng(1,1),4,-2}$                  &  6 & 1 & $(0,0)$                     &
$\ket{1,4,\yng(2,1),5,-2}$                  &  7 & 2 & $(0,\frac{1}{2})$           \\[8pt]
$\ket{\yng(1),3,\yng(1),3,0} \times 2$      &  6 & 8 & $(\frac{1}{2},\frac{1}{2})$ &
$\ket{\yng(1),5,\yng(2),4,0}$               &  7 & 6 & $(\frac{1}{2},1)$           \\[8pt]
$\ket{\yng(2),4,1,2,2}$                     &  6 & 3 & $(1,0)$                     &
$\ket{\yng(1),5,\yng(1,1),4,0} \times 2$    &  7 & 4 & $(\frac{1}{2},0)$           \\[8pt]
$\ket{\yng(1,1),4,1,2,2}$                   &  6 & 1 & $(0,0)$                     &
$\ket{\yng(2),6,\yng(1),3,2}$               &  7 & 6 & $(1,\frac{1}{2})$           \\[8pt]
$\ket{\yng(1),7,\yng(2,1),5,0}$             &  8 & 4 & $(\frac{1}{2},\frac{1}{2})$ &
$\ket{\yng(1,1),6,\yng(1),3,2} \times 2$    &  7 & 4 & $(0,\frac{1}{2})$           \\[8pt]
$\ket{\yng(2),8,\yng(1,1),4,2}$             &  8 & 3 & $(1,0)$                     &
$\ket{\yng(2,1),7,1,2,4}$                   &  7 & 2 & $(\frac{1}{2},0)$           \\[8pt]
$\ket{\yng(1,1),8,\yng(2),4,2}$             &  8 & 3 & $(0,1)$                     &
$\ket{\yng(1,1),10,\yng(2,1),5,2}$          &  9 & 2 & $(0,\frac{1}{2})$           \\[8pt]
$\ket{\yng(1,1),8,\yng(1,1),4,2} \times 2$  &  8 & 2 & $(0,0)$                     &
$\ket{\yng(2,1),11,\yng(1,1),4,4}$          &  9 & 2 & $(\frac{1}{2},0)$           \\[8pt]
$\ket{\yng(2,1),9,\yng(1),3,4}$             &  8 & 4 & $(\frac{1}{2},\frac{1}{2})$ &
$\ket{1,2,\yng(2,1),7,-4}$                  & 11 & 2 & $(0,\frac{1}{2})$           \\[8pt]
$\ket{1,0,\yng(1,1),6,-4}$                  & 10 & 1 & $(0,0)$                     &
$\ket{\yng(1),3,\yng(2),6,-2}$              & 11 & 6 & $(\frac{1}{2},1)$           \\[8pt]
$\ket{\yng(1),1,\yng(1),5,-2}$              & 10 & 4 & $(\frac{1}{2},\frac{1}{2})$ &
$\ket{\yng(1),3,\yng(1,1),6,-2} \times 2$   & 11 & 4 & $(\frac{1}{2},0)$           \\[8pt]
$\ket{\yng(1,1),2,1,4,0}$                   & 10 & 1 & $(0,0)$                     &
$\ket{\yng(2),4,\yng(1),5,0}$               & 11 & 6 & $(1,\frac{1}{2})$           \\[8pt]
$\ket{1,4,\yng(2,2),8,-4}$                  & 12 & 1 & $(0,0)$                     &
$\ket{\yng(1,1),4,\yng(1),5,0} \times 2$    & 11 & 4 & $(0,\frac{1}{2})$           \\[8pt]
$\ket{\yng(1),5,\yng(2,1),7,-2} \times 2$   & 12 & 8 & $(\frac{1}{2},\frac{1}{2})$ &
$\ket{\yng(2,1),5,1,4,2}$                   & 11 & 2 & $(\frac{1}{2},0)$           \\[8pt]
$\ket{\yng(2),6,\yng(2),6,0}$               & 12 & 9 & $(1,1)$                     &
$\ket{\yng(1),7,\yng(2,2),8,-2}$            & 13 & 2 & $(\frac{1}{2},0)$           \\[8pt]
$\ket{\yng(2),6,\yng(1,1),6,0}$             & 12 & 3 & $(1,0)$                     &
$\ket{\yng(2),8,\yng(2,1),7,0}$             & 13 & 6 & $(1,\frac{1}{2})$           \\[8pt]
$\ket{\yng(1,1),6,\yng(2),6,0}$             & 12 & 3 & $(0,1)$                     &
$\ket{\yng(1,1),8,\yng(2,1),7,0} \times 2$  & 13 & 4 & $(0,\frac{1}{2})$           \\[8pt]
$\ket{\yng(1,1),6,\yng(1,1),6,0} \times 3$  & 12 & 3 & $(0,0)$                     &
$\ket{\yng(2,1),9,\yng(2),6,2}$             & 13 & 6 & $(\frac{1}{2},1)$           \\[8pt]
$\ket{\yng(2,1),7,\yng(1),5,2} \times 2$    & 12 & 8 & $(\frac{1}{2},\frac{1}{2})$ &
$\ket{\yng(2,1),9,\yng(1,1),6,2} \times 2$  & 13 & 4 & $(\frac{1}{2},0)$           \\[8pt]
$\ket{\yng(2,2),8,1,4,4}$                   & 12 & 1 & $(0,0)$                     &
$\ket{\yng(2,2),10,\yng(1),5,4}$            & 13 & 2 & $(0,\frac{1}{2})$           \\[8pt]
$\ket{\yng(1,1),10,\yng(2,2),8,0}$          & 14 & 1 & $(0,0)$                     &
$\ket{\yng(1),1,\yng(1,1),8,-4}$            & 15 & 2 & $(\frac{1}{2},0)$           \\[8pt]
$\ket{\yng(2,1),11,\yng(2,1),7,2}$          & 14 & 4 & $(\frac{1}{2},\frac{1}{2})$ &
$\ket{\yng(1,1),2,\yng(1),7,-2}$            & 15 & 2 & $(0,\frac{1}{2})$           \\[8pt]
$\ket{\yng(2,2),12,\yng(1,1),6,4}$          & 14 & 1 & $(0,0)$                     &
$\ket{\yng(1),5,\yng(2,2),10,-4}$           & 17 & 2 & $(\frac{1}{2},0)$           \\[8pt]
$\ket{\yng(1),3,\yng(2,1),9,-4}$            & 16 & 4 & $(\frac{1}{2},\frac{1}{2})$ &
$\ket{\yng(2),6,\yng(2,1),9,-2}$            & 17 & 6 & $(1,\frac{1}{2})$           \\[8pt]
$\ket{\yng(2),4,\yng(1,1),8,-2}$            & 16 & 3 & $(1,0)$                     &
$\ket{\yng(1,1),6,\yng(2,1),9,-2} \times 2$ & 17 & 4 & $(0,\frac{1}{2})$           \\[8pt]
$\ket{\yng(1,1),4,\yng(2),8,-2}$            & 16 & 3 & $(0,1)$                     &
$\ket{\yng(2,1),7,\yng(2),8,0}$             & 17 & 6 & $(\frac{1}{2},1)$           \\[8pt]
$\ket{\yng(1,1),4,\yng(1,1),8,-2} \times 2$ & 16 & 2 & $(0,0)$                     &
$\ket{\yng(2,1),7,\yng(1,1),8,0} \times 2$  & 17 & 4 & $(\frac{1}{2},0)$           \\[8pt]
$\ket{\yng(2,1),5,\yng(1),7,0}$             & 16 & 4 & $(\frac{1}{2},\frac{1}{2})$ &
$\ket{\yng(2,2),8,\yng(1),7,2}$             & 17 & 2 & $(0,\frac{1}{2})$           \\[8pt]
$\ket{\yng(2),8,\yng(2,2),10,-2}$           & 18 & 3 & $(1,0)$                     &
$\ket{\yng(2,1),11,\yng(2,2),10,0}$         & 19 & 2 & $(\frac{1}{2},0)$           \\[8pt]
$\ket{\yng(1,1),8,\yng(2,2),10,-2}$         & 18 & 1 & $(0,0)$                     &
$\ket{\yng(2,2),12,\yng(2,1),9,2}$          & 19 & 2 & $(0,\frac{1}{2})$           \\[8pt]
$\ket{\yng(2,1),9,\yng(2,1),9,0} \times 2$  & 18 & 8 & $(\frac{1}{2},\frac{1}{2})$ &
$\ket{\yng(1,1),4,\yng(2,1),11,-4}$         & 21 & 2 & $(0,\frac{1}{2})$           \\[8pt]
$\ket{\yng(2,2),10,\yng(2),8,2}$            & 18 & 3 & $(0,1)$                     &
$\ket{\yng(2,1),5,\yng(1,1),10,-2}$         & 21 & 2 & $(\frac{1}{2},0)$           \\[8pt]
$\ket{\yng(2,2),10,\yng(1,1),8,2}$          & 18 & 1 & $(0,0)$                     &
$\ket{\yng(2,1),9,\yng(2,2),12,-2}$         & 23 & 2 & $(\frac{1}{2},0)$           \\[8pt]
$\ket{\yng(1,1),2,\yng(1,1),10,-4}$         & 20 & 1 & $(0,0)$                     &
$\ket{\yng(2,2),10,\yng(2,1),11,0}$         & 23 & 2 & $(0,\frac{1}{2})$           \\[8pt]
$\ket{\yng(1,1),6,\yng(2,2),12,-4}$         & 22 & 1 & $(0,0)$                     &
                                            &    &   &                             \\[8pt]
$\ket{\yng(2,1),7,\yng(2,1),11,-2}$         & 22 & 4 & $(\frac{1}{2},\frac{1}{2})$ &
                                            &    &   &                             \\[8pt]
$\ket{\yng(2,2),8,\yng(1,1),10,0}$          & 22 & 1 & $(0,0)$                     &
                                            &    &   &                             \\[8pt]
$\ket{\yng(2,2),12,\yng(2,2),12,0}$         & 24 & 1 & $(0,0)$                     &
                                            &    &   &                             \\[8pt]
\end{longtable}
\Yboxdim7.0pt


\subsection{$\left\{ \mathfrak{su}(2|1) \oplus \mathfrak{su}(1|1) \oplus \mathfrak{u}(1) \right\} ~ \circledS ~
\mathfrak{h}^{9,8}$}
\label{SubSec:SU21SU11U1}

To obtain this particular pp-wave superalgebra, we can start from the superalgebra we just discussed in the previous
section, $\left[ \mathfrak{su}(2|1) \oplus \mathfrak{su}(2|1) \oplus \mathfrak{u}(1) \right] ~ \circledS ~
\mathfrak{h}^{9,8}$, and decompose only one $\mathfrak{su}(2|1)$ into $\mathfrak{su}(1|1) \oplus
\mathfrak{u}(1)$.\footnote{The subsuperalgebra $\mathfrak{su}(1|1)$ consists of two fermionic generators and one
bosonic generator.}

If we are breaking the second $\mathfrak{su}(2|1)$ in equation \eqref{SU42decompSU21SU21U1}, we first obtain two
$\mathfrak{u}(1)$ charges as follows:
\begin{equation}
\begin{split}
\mathfrak{su}(2|1) & \supset \mathfrak{su}(2) \oplus \mathfrak{u}(1)_{\frac{1}{2} N_{B_2} + N_{F_2}} \\
\mathfrak{su}(2)   & \supset \mathfrak{u}(1)_{N_{B_3} - N_{B_4}}
\end{split}
\label{SU21decompSU11U1}
\end{equation}
We have used the notation, $N_{B_3} = \vec{\tilde{a}}^3 \cdot \vec{\tilde{a}}_3 + \vec{\tilde{b}}^3 \cdot
\vec{\tilde{b}}_3$ and $N_{B_4} = \vec{\tilde{a}}^4 \cdot \vec{\tilde{a}}_4 + \vec{\tilde{b}}^4 \cdot
\vec{\tilde{b}}_4$. Therefore, $N_{B_2} = N_{B_3} + N_{B_4}$.

Then we have an $\mathfrak{su}(1|1)$, realized by the bosonic oscillators $\tilde{a}_3(K)$, $\tilde{a}^3(K)$,
$\tilde{b}_3(K)$, $\tilde{b}^3(K)$ and the fermionic oscillators $\tilde{\alpha}_2(K)$, $\tilde{\alpha}^2(K)$,
$\tilde{\beta}_2(K)$, $\tilde{\beta}^2(K)$. The following linear combination of the two $\mathfrak{u}(1)$ charges
in equation \eqref{SU21decompSU11U1} must be part of  this $\mathfrak{su}(1|1)$:
\begin{equation}
\left( \frac{1}{2} N_{B_2} + N_{F_2} \right) + \frac{1}{2} \left( N_{B_3} - N_{B_4} \right)
 = N_{B_3} + N_{F_2} \,.
\end{equation}

The $\mathfrak{g}^{(0)}$ space, therefore, has now decomposed into
\begin{equation}
\mathfrak{su}(4|2) \supset \mathfrak{su}(2|1) \oplus \mathfrak{su}(1|1) \oplus
                           \mathfrak{u}(1)_{N_{B_1} + N_{F_1} - N_{B_4}} \,.
\end{equation}

In the decomposition of $SU(4|2)$ covariant super-oscillators into the $SU(2|1) \times SU(1|1) \times U(1)$ basis we
can retain the same super-oscillators in equation \eqref{SU21Oscillators} that realized the first $\mathfrak{su}(2|1)$
and just decompose the other super-oscillators as
\begin{equation}
\begin{split}
\tilde{\xi}^R(K)  = \left( \begin{matrix} \tilde{a}^r(K) \cr \tilde{\alpha}^2(K) \cr \end{matrix} \right)
 \qquad & \longrightarrow \qquad
  \left( \begin{matrix} \tilde{a}^3(K) \cr \tilde{\alpha}^2(K) \cr \end{matrix} \right) \oplus \tilde{a}^4(K) \\
\tilde{\eta}^R(K) = \left( \begin{matrix} \tilde{b}^r(K) \cr \tilde{\beta}^2(K) \cr \end{matrix} \right)
 \qquad & \longrightarrow \qquad
  \left( \begin{matrix} \tilde{b}^3(K) \cr \tilde{\beta}^2(K) \cr \end{matrix} \right) \oplus \tilde{b}^4(K) \,.
\end{split}
\end{equation}

The decomposition of $\hat{\mathfrak{g}}^{(+1)}$ space, in the $SU(2|1) \times SU(1|1) \times U(1)$ basis takes the
form:
\begin{equation}
\begin{split}
\hat{\mathfrak{g}}^{(+1)}
   = \soneonebox \Big|_{SU(4|2)}
 & = \left( \: \soneonebox \,,\, 1 \,,\, 2 \right) \oplus
     \left( \: \sonebox \,,\, \sonebox \,,\, 1 \right) \oplus
     \left( \: \sonebox \,,\, 1 \,,\, 0 \right) \oplus
     \left( 1 \,,\, \soneonebox \,,\, 0 \right) \\
 &   \quad \oplus \left( 1 \,,\, \sonebox \,,\, -1 \right)
     \Big|_{SU(2|1) \times SU(1|1) \times U(1)} \\
\end{split}
\end{equation}
and therefore, the 8 kinematical supersymmetries in $\hat{\mathfrak{g}}^{(+1)}$ can be identified easily.

Again, from the $\mathfrak{g}^{(0)}$ subspace, we keep only the generators that belong to $\mathfrak{su}(2|1)
\oplus \mathfrak{su}(1|1) \oplus \mathfrak{u}(1)$ (6 bosonic and 6 fermionic), and eliminate the rest. Therefore,
this pp-wave superalgebra has 22 supersymmetries (8 in $\hat{\mathfrak{g}}^{(+1)}$, 8 in
$\hat{\mathfrak{g}}^{(-1)}$ and 6 in $\hat{\mathfrak{g}}^{(0)} = \mathfrak{su}(2|1) \oplus \mathfrak{su}(1|1)
\oplus \mathfrak{u}(1)$). Five of the bosonic generators that remain in $\hat{\mathfrak{g}}^{(0)}$ are rotation
generators that belong to $SU(2) \times U(1) \times U(1)$ and the other generator is the Hamiltonian. Since there
are three $\mathfrak{u}(1)$ charges in $\hat{\mathfrak{g}}^{(0)}$, we again obtain a two-parameter family of
Hamiltonians - a linear combination of those $\mathfrak{u}(1)$ charges (as long as it is bounded from below):
\begin{equation}
\mathcal{H} = N_{B_1} + 2 N_{F_1} + \kappa_1 \left( N_{B_3} + N_{F_2} \right)
              + \kappa_2 \left( N_{B_1} + 2 N_{F_1} - N_{B_4} \right)
\label{SU21SU11U1H}
\end{equation}

The symmetry superalgebra of this pp-wave solution in eleven dimensions is
\begin{equation*}
\left[ \mathfrak{su}(2|1) \oplus \mathfrak{su}(1|1) \oplus \mathfrak{u}(1) \right] ~ \circledS ~ \mathfrak{h}^{9,8} \,.
\end{equation*}

The zero-mode spectrum of this pp-wave superalgebra can be obtained by acting on the lowest weight vector
$\ket{\hat{\Omega}} = \ket{0}$ with the kinematical supersymmetries in $\hat{\mathfrak{g}}^{(+1)}$. We skip
listing the spectrum explicitly in this case , since it is a straightforward exercise when the kinematical
supersymmetries are known.


\subsection{$\mathfrak{su}(2|1) ~ \circledS ~ \mathfrak{h}^{9,8}$}
\label{SubSec:SU21}

By starting from the eleven dimensional maximal dynamical compact subsuperalgebra we discussed in the section
\ref{SubSec:SU21SU21U1}, $\left[ \mathfrak{su}(2|1) \oplus \mathfrak{su}(2|1) \right] ~ \circledS ~
\mathfrak{h}^{9,8}$,\footnote{Note that the third $\mathfrak{u}(1)$ charge that commutes with both
$\mathfrak{su}(2|1)$ is not essential for the closure of that algebra, and we retained it in that section only
for the purpose of identifying that algebra with the work in \cite{AGGP02,Mich02a}.} we can form another
dynamical compact subsuperalgebra by taking the diagonal subsuperalgebra of $\mathfrak{su}(2|1) \oplus
\mathfrak{su}(2|1)$.

In terms of oscillators, we identify the indices $M \leftrightarrow R$ in equations \eqref{SU42decompSU21SU21U1} and
\eqref{SU21Oscillators} in order to form the diagonal subalgebra.

The subspace $\mathfrak{g}^{(+1)}$ decomposes in this new diagonal $SU(2|1)$ as:
\begin{equation}
\begin{split}
\hat{\mathfrak{g}}^{(+1)}
   = \soneonebox \Big|_{SU(4|2)}
 & = \left( \: \soneonebox \, , \, 1 \right) \oplus
     \left( \: \sonebox \, , \, \sonebox \right) \oplus
     \left( 1 \, , \, \soneonebox \right)
     \Big|_{SU(2|1) \times SU(2|1)} \\
 & = ~ \soneonebox \oplus
       \stwobox \oplus
       \soneonebox \oplus
       \soneonebox
     \Big|_{SU(2|1)_{\mathrm{diag}}} \\
 & = ~ 3 \times \left( \yng(1,1) \,,\, 2 \right) \oplus
       4 \times \left( \yng(1) \,,\, 3 \right) \oplus
       3 \times \left( 1 \,,\, 4 \right) \oplus
       \left( \yng(2) \,,\, 2 \right)
     \Big|_{SU(2) \times U(1)}
\end{split}
\end{equation}
where the $U(1)$ charge, which plays the role of the Hamiltonian is given by
\begin{equation}
\mathcal{H} = N_B + 2 N_F \,.
\end{equation}

Now one can easily identify the kinematical supersymmetries in $\mathfrak{g}^{(+1)}$ in this new basis as $4
\times \left( \yng(1) \,,\, 3 \right)$, which transform as four doublets under $SU(2) \times U(1)$ .

Since we formed a diagonal subsuperalgebra from the previous dynamical compact subsuperalgebra
$\hat{\mathfrak{g}}^{(0)} = \mathfrak{su}(2|1) \oplus \mathfrak{su}(2|1)$, we now have only 4 bosonic and 4
fermionic generators left in the new $\hat{\mathfrak{g}}^{(0)} = \mathfrak{su}(2|1)_{\mathrm{diag}}$. Therefore,
this pp-wave superalgebra has 20 supersymmetries (8 in $\hat{\mathfrak{g}}^{(+1)}$, 8 in
$\hat{\mathfrak{g}}^{(-1)}$ and 4 in $\hat{\mathfrak{g}}^{(0)}$).

The new $\hat{\mathfrak{g}}^{(0)}$ contains only 3 rotation generators, and the symmetry superalgebra of this pp-wave
solution is
\begin{equation*}
\mathfrak{su}(2|1) ~ \circledS ~ \mathfrak{h}^{9,8} \,.
\end{equation*}

Since it is possible to obtain the zero-mode spectrum of this pp-wave superalgebra, from that of $\left[
\mathfrak{su}(2|1) \oplus \mathfrak{su}(2|1) \right] ~ \circledS ~ \mathfrak{h}^{9,8}$ in section
\ref{SubSec:SU21SU21U1}, by simply taking the tensor product of the corresponding super Young tableau, we will
not list it explicitly.


\subsection{$\left\{ \mathfrak{su}(1|1) \oplus \mathfrak{u}(1) \right\} ~ \circledS ~ \mathfrak{h}^{9,8}$}
\label{SubSec:SU11U1}

Here we can start from the previous case where the dynamical compact subsuperalgebra is $\mathfrak{su}(2|1)$, and
first decompose it into $\mathfrak{su}(2) \oplus \mathfrak{u}(1)$. Then after breaking $\mathfrak{su}(2)$ further
into its $\mathfrak{u}(1)$ charge, we can combine these two $\mathfrak{u}(1)$ generators to form an
$\mathfrak{su}(1|1)$ subsuperalgebra along with two supersymmetry generators from $\mathfrak{su}(2|1)$ \footnote{
We should stress that there are numerous different ways of getting a dynamical superalgebra $SU(1/1)\times U(1)$
which are , in general, not equivalent.}.

We eliminate the other two supersymmetries, as well as the remaining three bosonic generators. Then, our new
dynamical compact subsuperalgebra becomes
\begin{equation}
\hat{\mathfrak{g}}^{(0)} = \mathfrak{su}(1|1) \,,
\end{equation}
making the total number of supersymmetries in the algebra 18. The $\mathfrak{u}(1)$ charge inside $\mathfrak{su}(1|1)$
acts as the Hamiltonian of this algebra.

The decomposition of $\hat{\mathfrak{g}}^{(+1)}$ subspace with respect to the new basis $SU(1|1)$ has the form:
\begin{equation}
\begin{split}
\hat{\mathfrak{g}}^{(+1)}
   = \soneonebox \Big|_{SU(4|2)}
 & = ~ 3 \times \soneonebox \oplus
       \stwobox
     \Big|_{SU(2|1)_{\mathrm{diag}}} \\
 & = ~ 3 \times \left( \: \soneonebox \,,\, 0 \right) \oplus
       4 \times \left( \: \sonebox \,,\, 1 \right) \oplus
       \left( \: \stwobox \,,\, 0 \right) \oplus
       \left( 1 \,,\, 2 \right)
     \Big|_{SU(1|1) \times U(1)}
\end{split}
\end{equation}

Now one can identify the supersymmetry generators in $\hat{\mathfrak{g}}^{(+1)}$, with which we must act on a lowest
weight vector to generate a UIR, in the same way as before. The zero-mode spectrum once again corresponds to the
lowest weight vector $\ket{\hat{\Omega}} = \ket {0}$.


\subsection{$\left\{ \mathfrak{su}(3|1) \oplus \mathfrak{su}(1|1) \right\} ~ \circledS ~ \mathfrak{h}^{9,8}$}
\label{SubSec:SU31SU11}

In this case, we first break $\mathfrak{su}(4)_B$ into $\mathfrak{su}(3) \oplus \mathfrak{u}(1)_D$ as in equations
\eqref{SU4decompSU3U1} and \eqref{SU3D}. Then we break $\mathfrak{su}(2)_F$ as well, into its $\mathfrak{u}(1)$ charge
as first done in section \ref{SubSec:SU41U1}.

Next we combine $\grave{\alpha}_1(K)$, $\grave{\alpha}^1(K)$, $\grave{\beta}_1(K)$, $\grave{\beta}^1(K)$ oscillators
with $\mathfrak{su}(3)$ to form $\mathfrak{su}(3|1)$ as
\begin{equation}
\mathfrak{su}(3) \oplus \mathfrak{u}(1) \subset \mathfrak{su}(3|1)
\end{equation}
where the $\mathfrak{u}(1)$ charge contained in $\mathfrak{su}(3|1)$ can be identified as
\begin{equation}
G - \frac{1}{3} D + \frac{1}{2} F = \frac{1}{3} N_{B_1} + N_{F_1} \,.
\end{equation}
We recall that $G$, $D$ and $F$ are given by $G = \frac{1}{4} \left( N_B + 2 N_F \right)$, $D = \frac{1}{4} \left( 3
N_{B_2} - N_{B_1} \right)$ and $F = N_{F_1} - N_{F_2}$.

Next we combine $\tilde{\alpha}_2(K)$, $\tilde{\alpha}^2(K)$, $\tilde{\beta}_2(K)$, $\tilde{\beta}^2(K)$
oscillators with the remaining bosonic oscillators (of the type $\tilde{a}$, and $\tilde{b}$) to realize
$\mathfrak{su}(1|1)$. Note that the $\mathfrak{u}(1)$ charge inside this $\mathfrak{su}(1|1)$ is
\begin{equation}
G + D - \frac{1}{2} F = N_{B_2} + N_{F_2} \,.
\end{equation}

Therefore we must now consider the decomposition of $SU(4|2)$ covariant super-oscillators in the $SU(3|1) \times
SU(1|1)$ basis as follows:
\begin{equation}
\begin{split}
\xi^A(K)  = \left( \begin{matrix} a^i(K) \cr \alpha^\mu(K) \cr \end{matrix} \right)
 \qquad & \longrightarrow \qquad
  \grave{\xi}^M(K) \oplus \tilde{\xi}^R(K) \\
\eta^A(K) = \left( \begin{matrix} b^i(K) \cr \beta^\mu(K) \cr \end{matrix} \right)
 \qquad & \longrightarrow \qquad
  \grave{\eta}^M(K) \oplus \tilde{\eta}^R(K) \,, \qquad \mbox{etc.}
\end{split}
\end{equation}
where
\begin{equation}
\begin{split}
\grave{\xi}^M(K)  = \left( \begin{matrix} \grave{a}^m(K) \cr \grave{\alpha}^1(K) \cr \end{matrix} \right)
 \qquad \qquad
\grave{\eta}^M(K) = \left( \begin{matrix} \grave{b}^m(K) \cr \grave{\beta}^1(K) \cr \end{matrix} \right) \\
\tilde{\xi}^R(K)  = \left( \begin{matrix} \tilde{a}^4(K) \cr \tilde{\alpha}^2(K) \cr \end{matrix} \right)
 \qquad \qquad
\tilde{\eta}^R(K) = \left( \begin{matrix} \tilde{b}^4(K) \cr \tilde{\beta}^2(K) \cr \end{matrix} \right) \,.
\end{split}
\end{equation}

The $\mathfrak{g}^{(0)}$ subspace has now decomposed into
\begin{equation}
\mathfrak{su}(4|2) \supset \mathfrak{su}(3|1) \oplus \mathfrak{su}(1|1) \oplus \mathfrak{u}(1) \,.
\label{SU42decompSU31SU11U1}
\end{equation}

Furthermore, the $\hat{\mathfrak{g}}^{(+1)}$ subspace decomposes, in this $SU(3|1) \times SU(1|1)$ basis as:
\begin{equation}
\begin{split}
\hat{\mathfrak{g}}^{(+1)}
   = \soneonebox \Big|_{SU(4|2)}
 & = \left( \: \soneonebox \,,\, 1 \right) \oplus
     \left( \: \sonebox \,,\, \sonebox \right) \oplus
     \left( 1 \,,\, \soneonebox \right)
     \Big|_{SU(3|1) \times SU(1|1)} \\
\end{split}
\end{equation}
and therefore, in the $SU(3) \times U(1)_{N_{B_1} + 3 N_{F_1}} \times U(1)_{N_{B_2} + N_{F_2}}$ basis, we
can write:
\begin{equation}
\begin{split}
\hat{\mathfrak{g}}^{(+1)}
   = \soneonebox \Big|_{SU(4|2)}
 & = \left( \: \soneonebox \,,\, 1 \right) \oplus
     \left( \: \sonebox \,,\, \sonebox \right) \oplus
     \left( 1 \,,\, \soneonebox \right)
     \Big|_{SU(3|1) \times SU(1|1)} \\
 & = \left( \yng(1,1) \,,\, 2 \,,\, 0 \right) \oplus
     \left( \yng(1) \,,\, 4 \,,\, 0 \right) \oplus
     \left( 1 \,,\, 6 \,,\, 0 \right) \\
 &   \quad \oplus 2 \times \left( \yng(1) \,,\, 1 \,,\, 1 \right) \oplus
                  2 \times \left( 1 \,,\, 3 \,,\, 1 \right) \\
 &   \quad \oplus 2 \times \left( 1 \,,\, 0 \,,\, 2 \right)
     \Big|_{SU(3) \times U(1)_{N_{B_1} + 3 N_{F_1}} \times U(1)_{N_{B_2} + N_{F_2}}}
\end{split}
\end{equation}

From this decomposition, one can identify the 8 kinematical supersymmetries in $\hat{\mathfrak{g}}^{(+1)}$ as:
$\left( \yng(1) \,,\, 4 \,,\, 0 \right) \oplus \left( \yng(1) \,,\, 1 \,,\, 1 \right) \oplus \left( 1 \,,\, 3
\,,\, 1 \right) \oplus \left( 1 \,,\, 0 \,,\, 2 \right)$. They transform in the $SU(3) \times U(1)_{N_{B_1} + 3
N_{F_1}} \times U(1)_{N_{B_2} + N_{F_2}}$ basis as $\mathbf{3} + \mathbf{3} + \mathbf{1} + \mathbf{1}$.

Again, from the $\mathfrak{g}^{(0)}$ subspace, we must eliminate generators of the form $\mathcal{M}^M_{~~R}$ and
$\mathcal{M}^S_{~~N}$ (8 bosonic and 8 fermionic), and the $\mathfrak{u}(1)$ generator that commutes with both
$\mathfrak{su}(3|1)$ and $\mathfrak{su}(1|1)$ (see equation \eqref{SU42decompSU31SU11U1}), since we want to keep only
the $\mathfrak{su}(3|1) \oplus \mathfrak{su}(1|1)$ part. Therefore, we obtain a pp-wave superalgebra that has 24
supersymmetries (8 in $\hat{\mathfrak{g}}^{(+1)}$, 8 in $\hat{\mathfrak{g}}^{(-1)}$ and 8 in $\hat{\mathfrak{g}}^{(0)}
= \mathfrak{su}(3|1) \oplus \mathfrak{su}(1|1)$).

The number of bosonic generators that is left in the  dynamical compact subsuperalgebra
$\hat{\mathfrak{g}}^{(0)}$ is 10. Nine of them are rotation generators, that belong to $SU(3) \times SU(1)$ and
the other one is the Hamiltonian. This time we obtain a one-parameter family of Hamiltonians, $\left( \frac{1}{3}
N_{B_1} + N_{F_1} \right) + \kappa \left( N_{B_2} + N_{F_2} \right)$, and if one chooses simply $\kappa = 1$
(just for the purpose of constructing the spectrum) and rescales it so that the energy increases of the states
come in integer steps, we have:
\begin{equation}
\mathcal{H} = \frac{1}{2} \left( N_{B_1} + 3 N_{F_1} + 3 N_{B_2} + 3 N_{F_2} \right)
\end{equation}

Thus, we immediately notice that the 6 kinematical supersymmetries, $Q^{m1} = \left( \yng(1) \,,\, 4 \,,\, 0 \right)$
and $Q^{m2} = \left( \yng(1) \,,\, 1 \,,\, 1 \right)$ increase energy by 2 units, while the other 2 kinematical
supersymmetries $Q^{41} = \left( 1 \,,\, 3 \,,\, 1 \right)$ and $Q^{42} = \left( 1 \,,\, 0 \,,\, 2 \right)$ increase
energy by 3 units.

The symmetry superalgebra of this pp-wave solution in eleven dimensions is
\begin{equation*}
\left[ \mathfrak{su}(3|1) \oplus \mathfrak{u}(1|1) \right] ~ \circledS ~ \mathfrak{h}^{9,8} \,.
\end{equation*}

Finally we give the zero-mode spectrum of this pp-wave superalgebra, in the $SU(3) \times U(1)_{N_{B_1} + 3 N_{F_1}}
\times U(1)_{N_{B_2} + N_{F_2}}$ basis, obtained by starting from the ground state $| \hat{\Omega} \rangle = \ket{0}$
(Table \ref{Table:SU31SU11}). Once again, some states of this spectrum appear with multiplicity greater than 1.

\Yboxdim5.0pt
\begin{longtable}[c]{|l|c|c|c||l|c|c|c|}
\kill
\caption[The zero-mode spectrum of the eleven dimensional pp-wave superalgebra with 24 supersymmetries, $\left\{
\mathfrak{su}(3|1) \oplus \mathfrak{su}(1|1) \right\} ~ \circledS ~ \mathfrak{h}^{9,8}$]{\normalsize The zero-mode
spectrum of the eleven dimensional pp-wave superalgebra with 24 supersymmetries, $\left[ \mathfrak{su}(3|1) \oplus
\mathfrak{su}(1|1) \right] ~ \circledS ~ \mathfrak{h}^{9,8}$.
\label{Table:SU31SU11}} \\
\hline
\hline
& & & & & & & \\
{\scriptsize $SU(3) \times U(1)_{N_{B_1} + 3 N_{F_1}}$} &
 &
 &
 &
{\scriptsize $SU(3) \times U(1)_{N_{B_1} + 3 N_{F_1}}$} &
 &
 &
 \\
{\scriptsize $\times ~ U(1)_{N_{B_2} + N_{F_2}}$} &
 &
 &
{\scriptsize $SU(3)$} &
{\scriptsize $\times ~ U(1)_{N_{B_2} + N_{F_2}}$} &
 &
 &
{\scriptsize $SU(3)$} \\
{\scriptsize Young tableau} &
 &
{\scriptsize $N_{\mathrm{dof}}$} &
{\scriptsize Dynkin} &
{\scriptsize Young tableau} &
 &
{\scriptsize $N_{\mathrm{dof}}$} &
{\scriptsize Dynkin} \\
{\scriptsize (bosonic states)} &
{\scriptsize $\mathcal{H}$} &
{\scriptsize (B)} &
{\scriptsize labels} &
{\scriptsize (fermionic states)} &
{\scriptsize $\mathcal{H}$} &
{\scriptsize (F)} &
{\scriptsize labels} \\
& & & & & & & \\
\hline
& & & & & & & \\
\endfirsthead
\caption[]{(continued)} \\
\hline
& & & & & & & \\
{\scriptsize $SU(3) \times U(1)_{N_{B_1} + 3 N_{F_1}}$} &
 &
 &
 &
{\scriptsize $SU(3) \times U(1)_{N_{B_1} + 3 N_{F_1}}$} &
 &
 &
 \\
{\scriptsize $\times ~ U(1)_{N_{B_2} + N_{F_2}}$} &
 &
 &
{\scriptsize $SU(3)$} &
{\scriptsize $\times ~ U(1)_{N_{B_2} + N_{F_2}}$} &
 &
 &
{\scriptsize $SU(3)$} \\
{\scriptsize Young tableau} &
 &
{\scriptsize $N_{\mathrm{dof}}$} &
{\scriptsize Dynkin} &
{\scriptsize Young tableau} &
 &
{\scriptsize $N_{\mathrm{dof}}$} &
{\scriptsize Dynkin} \\
{\scriptsize (bosonic states)} &
{\scriptsize $\mathcal{H}$} &
{\scriptsize (B)} &
{\scriptsize labels} &
{\scriptsize (fermionic states)} &
{\scriptsize $\mathcal{H}$} &
{\scriptsize (F)} &
{\scriptsize labels} \\
& & & & & & & \\
\hline
& & & & & & & \\
\endhead
& & & & & & & \\
\hline
\endfoot
& & & & & & & \\
\hline
& & & & & & & \\
& & 128 & & & & 128 & \\
& & & & & & & \\
\hline
\hline
\endlastfoot
$\ket{1,0,0}$                     &  0 &  1 & $(0,0)$ &
$\ket{\yng(1),1,1}$               &  2 &  3 & $(1,0)$ \\[8pt]
$\ket{\yng(2),5,1}$               &  4 &  6 & $(2,0)$ &
$\ket{\yng(1),4,0}$               &  2 &  3 & $(1,0)$ \\[8pt]
$\ket{\yng(1,1),2,2}$             &  4 &  3 & $(0,1)$ &
$\ket{1,0,2}$                     &  3 &  1 & $(0,0)$ \\[8pt]
$\ket{\yng(1,1),5,1}$             &  4 &  3 & $(0,1)$ &
$\ket{1,3,1}$                     &  3 &  1 & $(0,0)$ \\[8pt]
$\ket{\yng(1,1),8,0}$             &  4 &  3 & $(0,1)$ &
$\ket{\yng(2,1),6,2}$             &  6 &  8 & $(1,1)$ \\[8pt]
$\ket{\yng(1),1,3}$               &  5 &  3 & $(1,0)$ &
$\ket{\yng(2,1),9,1}$             &  6 &  8 & $(1,1)$ \\[8pt]
$\ket{\yng(1),4,2} \times 2$      &  5 &  6 & $(1,0)$ &
$\ket{\yng(1,1,1),3,3}$           &  6 &  1 & $(0,0)$ \\[8pt]
$\ket{\yng(1),7,1}$               &  5 &  3 & $(1,0)$ &
$\ket{\yng(1,1,1),6,2}$           &  6 &  1 & $(0,0)$ \\[8pt]
$\ket{1,3,3}$                     &  6 &  1 & $(0,0)$ &
$\ket{\yng(1,1,1),9,1}$           &  6 &  1 & $(0,0)$ \\[8pt]
$\ket{\yng(2,2),10,2}$            &  8 &  6 & $(0,2)$ &
$\ket{\yng(1,1,1),12,0}$          &  6 &  1 & $(0,0)$ \\[8pt]
$\ket{\yng(2,1,1),7,3}$           &  8 &  3 & $(1,0)$ &
$\ket{\yng(2),5,3}$               &  7 &  6 & $(2,0)$ \\[8pt]
$\ket{\yng(2,1,1),10,2}$          &  8 &  3 & $(1,0)$ &
$\ket{\yng(2),8,2}$               &  7 &  6 & $(2,0)$ \\[8pt]
$\ket{\yng(2,1,1),13,1}$          &  8 &  3 & $(1,0)$ &
$\ket{\yng(1,1),2,4}$             &  7 &  3 & $(0,1)$ \\[8pt]
$\ket{\yng(2,1),6,4}$             &  9 &  8 & $(1,1)$ &
$\ket{\yng(1,1),5,3} \times 2$    &  7 &  6 & $(0,1)$ \\[8pt]
$\ket{\yng(2,1),9,3} \times 2$    &  9 & 16 & $(1,1)$ &
$\ket{\yng(1,1),8,2} \times 2$    &  7 &  6 & $(0,1)$ \\[8pt]
$\ket{\yng(2,1),12,2}$             &  9 &  8 & $(1,1)$ &
$\ket{\yng(1,1),11,1}$            &  7 &  3 & $(0,1)$ \\[8pt]
$\ket{\yng(1,1,1),3,5}$           &  9 &  1 & $(0,0)$ &
$\ket{\yng(1),4,4}$               &  8 &  3 & $(1,0)$ \\[8pt]
$\ket{\yng(1,1,1),6,4} \times 2$  &  9 &  2 & $(0,0)$ &
$\ket{\yng(1),7,3}$               &  8 &  3 & $(1,0)$ \\[8pt]
$\ket{\yng(1,1,1),9,3} \times 2$  &  9 &  2 & $(0,0)$ &
$\ket{\yng(2,2,1),11,3}$          & 10 &  3 & $(0,1)$ \\[8pt]
$\ket{\yng(1,1,1),12,2} \times 2$ &  9 &  2 & $(0,0)$ &
$\ket{\yng(2,2,1),14,2}$          & 10 &  3 & $(0,1)$ \\[8pt]
$\ket{\yng(1,1,1),15,1}$          &  9 &  1 & $(0,0)$ &
$\ket{\yng(2,2),10,4}$            & 11 &  6 & $(0,2)$ \\[8pt]
$\ket{\yng(2),8,4}$               & 10 &  6 & $(2,0)$ &
$\ket{\yng(2,2),13,3}$            & 11 &  6 & $(0,2)$ \\[8pt]
$\ket{\yng(1,1),1,3}$             & 10 &  3 & $(0,1)$ &
$\ket{\yng(2,1,1),7,5}$           & 11 &  3 & $(1,0)$ \\[8pt]
$\ket{\yng(1,1),5,5}$             & 10 &  3 & $(0,1)$ &
$\ket{\yng(2,1,1),10,4} \times 2$ & 11 &  6 & $(1,0)$ \\[8pt]
$\ket{\yng(1,1),8,4}$             & 10 &  3 & $(0,1)$ &
$\ket{\yng(2,1,1),13,3} \times 2$ & 11 &  6 & $(1,0)$ \\[8pt]
$\ket{\yng(2,2,2),15,3}$          & 12 &  1 & $(0,0)$ &
$\ket{\yng(2,1,1),16,2}$          & 11 &  3 & $(1,0)$ \\[8pt]
$\ket{\yng(2,2,1),11,5}$          & 13 &  3 & $(0,1)$ &
$\ket{\yng(2,1),9,5}$             & 12 &  8 & $(1,1)$ \\[8pt]
$\ket{\yng(2,2,1),14,4} \times 2$ & 13 &  6 & $(0,1)$ &
$\ket{\yng(2,1),12,4}$            & 12 &  8 & $(1,1)$ \\[8pt]
$\ket{\yng(2,2,1),17,3}$          & 13 &  3 & $(0,1)$ &
$\ket{\yng(1,1,1),6,6}$           & 12 &  1 & $(0,0)$ \\[8pt]
$\ket{\yng(2,2),13,5}$            & 14 &  6 & $(0,2)$ &
$\ket{\yng(1,1,1),9,5}$           & 12 &  1 & $(0,0)$ \\[8pt]
$\ket{\yng(2,1,1),10,6}$          & 14 &  3 & $(1,0)$ &
$\ket{\yng(1,1,1),12,4}$          & 12 &  1 & $(0,0)$ \\[8pt]
$\ket{\yng(2,1,1),13,5}$          & 14 &  3 & $(1,0)$ &
$\ket{\yng(1,1,1),15,3}$          & 12 &  1 & $(0,0)$ \\[8pt]
$\ket{\yng(2,1,1),16,4}$          & 14 &  3 & $(1,0)$ &
$\ket{\yng(2,2,2),15,5}$          & 15 &  3 & $(0,0)$ \\[8pt]
$\ket{\yng(2,2,2),18,6}$          & 18 &  1 & $(0,0)$ &
$\ket{\yng(2,2,2),18,4}$          & 15 &  1 & $(0,0)$ \\[8pt]
                                  &    &    &         &
$\ket{\yng(2,2,1),14,6}$          & 16 &  3 & $(0,1)$ \\[8pt]
                                  &    &    &         &
$\ket{\yng(2,2,1),17,5}$          & 16 &  1 & $(0,1)$ \\[8pt]
\end{longtable}
\Yboxdim7.0pt


\section{Summary and Discussion}
\label{Sec:Discussion}

In this paper we have presented an extensive list of pp-wave superalgebras of supergravity theories in eleven
dimensions and constructed the respective zero-mode spectra in some interesting cases.\footnote{A similar
analysis of non-maximally supersymmetric pp-wave algebras that can be obtained by starting from the maximally
supersymmetric IIB pp-wave algebra will be given in \cite{SFMG04}.} Some of these pp-wave solutions have already
been constructed in the literature following field theoretical methods, but for some of these solutions
(especially in non-maximally supersymmetric cases) the underlying pp-wave symmetry superalgebras had not been
identified and/or the corresponding zero mode spectra were not given.

It should also be noted that our prescription using the oscillator formalism to obtain the corresponding pp-wave
algebra of a given superalgebra is quite general. In essence, one can apply this method to \emph{any} superalgebra
that admits a 3-grading with respect to its maximal compact subsuperalgebra as follows:
\begin{itemize}
\item Normal order all the generators of the superalgebra, and identify the $\mathfrak{u}(1)$ generators that
explicitly depend on the number of colors $P$;
\item Re-normalize all the generators in $\mathfrak{g}^{(-1)} \oplus \mathfrak{g}^{(+1)}$ subspace by a factor
proportional to $1/\sqrt{P}$ and the above mentioned $\mathfrak{u}(1)$ generator in $\mathfrak{g}^{(0)}$ that
explicitly depends on $P$ by a factor proportional to $1/P$;
\item Take the limit $P \to \infty$ in all the super-commutation relations;
\end{itemize}
and it produces the corresponding pp-wave superalgebra.

One prominent omission from our list is the eleven dimensional (and type IIA) pp-wave solution with 26
supercharges
 found in \cite{Mich02b}. Its dynamical subsuperalgebra is $\mathfrak{psu}(2|2) \oplus \mathfrak{psu}(1|1) \oplus
 \mathfrak{u}(1) $ \cite{michelson}. It can not be embedded in the superalgebra $\mathfrak{su}(4|2)$ and falls
 outside the class of pp-wave algebras we considered in this paper.

By our construction , all pp-wave solutions discussed above preserve at least half of the supersymmetries (i.e.
all sixteen kinematical supersymmetries), and depending on the  dynamical compact subsuperalgebra they retain
they have some extra dynamical supersymmetries left. Various aspects of many of these solutions have already been
discussed in the literature in great detail \cite{HS02a,HS02b,KS03,Mich02a,AGGP02,CLP02,GH02,OS03,BR02}. We
should stress that the fact that we have sixteen kinematical supersymmetries depended crucially on the fact that
the 3-gradings of $OSp(8|4,\mathbb{R})$ and $OSp(8^*|4)$ with respect to $U(4|2)$ lead to sixteen supersymmetries
in the grade $\pm 1$ subspaces $\mathfrak{g}^{\pm 1}$.\footnote{ One can in general have 3-gradings with respect
to  maximal compact subsuperalgebras that lead to different number of supersymmetries in $\pm 1$ subspaces. For
example, the  triality rotated Lie superalgebra $\mathfrak{osp}_V(6,2|4)$ , in which the supersymmetry generators
transform in the vector representation of $SO(6,2)$, can be given a 3-graded decomposition with respect to its
maximal compact subsuperalgebra $\mathfrak{osp}(6|4) \oplus \mathfrak{u}(1) $. In this case the grade $\pm 1$
subspaces are 10 dimensional with 6 even and 4 odd generators each. Thus the corresponding pp-wave algebra is $
\mathfrak{osp}(6|4) \circledS \mathfrak{h}^{6,4}$,
 corresponding
to 8 kinematical supersymmetries total. The compact superalgebra $\mathfrak{osp}(6|4)$ with the even subalgebra
$\mathfrak{so}(6) \oplus \mathfrak{usp}(4)$ has 24 dynamical supersymmetries.}

All the spectra we discussed in this paper were restricted to the zero-mode sectors of their respective theories.
Even though the oscillator formalism we used in our work originally came from the studies of Kaluza-Klein spectra
of supergravity theories,  we expect it to be  quite straightforward to extend it to the study of higher
(non-zero) modes of M or superstring theories in the pp-wave limit due to the following observation.

The higher modes of type IIB superstring theory in pp-wave background contribute only to the compact dynamical
part of the pp-wave superalgebra \cite{MT02}. The dynamical subsuperalgebra is $\hat{\mathfrak{g}}^{(0)} =
\mathfrak{psu}(2|2) \oplus \mathfrak{psu}(2|2) \oplus \mathfrak{u}(1)$ for IIB theory. In another example, it was
shown in \cite{KS03} that $\mathcal{N} = (4,4)$ type IIA pp-wave solution with 24 supercharges (whose symmetry
superalgebra is $\left[ \mathfrak{su}(2|2) \oplus \mathfrak{su}(2) \right] ~ \circledS ~ \mathfrak{h}^{8,8}$) -
see section \ref{SubSec:SU22SU2} - also displays the same behavior.

Assuming that the same holds true in general for any pp-wave superalgebra, we can write down the ``exact''
realization of a pp-wave algebra (including the non-zero mode sector) in a general case by introducing an extra
index to our oscillators (e.g. $a^i_{(n)}(K)$ to represent the $n^{\mathrm{th}}$ mode) and taking the direct sum
of an infinitely many such sets ($n = 1,2,\dots$). Such non-zero mode spectra would consist of not just short
multiplets, as zero-mode spectra do, but also longer multiplets, even though they contribute to the Hamiltonian
in the same way as the zero-modes.

 {\bf Acknowledgements:} We would like to thank Jeremy Michelson for a correspondence regarding the symmetry
 superalgebra of his solution with 26 supersymmetries. S.H. would
 like to thank the Physics Department at Penn State University for warm
 hospitality during his visit.


\section*{Appendix}
\appendix
\section{Eleven dimensional supergravity spectra on $AdS \times S$ spaces}
\label{ApSec:SupergravitySpectra}

In this appendix, we reproduce the spectra of the eleven dimensional supergravity with maximal supersymmetry, obtained
by using the oscillator method in \cite{GvNW85} and \cite{GW86}.


\subsection{Spectrum of eleven dimensional supergravity on $AdS_7 \times S^4$}
\label{ApSubSec:AdS7S4Sugra}

The oscillator realization of $OSp(8^*|4)$, the symmetry superalgebra of the eleven dimensional supergravity on
$AdS_7 \times S^4$ was reviewed in section \ref{SubSec:AdS7S4}. The spectrum, obtained by starting from the lowest
weight vector $\ket{\Omega} = \ket{0}$ was first constructed in \cite{GvNW85}. We present their results for reference
here in Table \ref{Table:AdS7S4}.

Dynkin labels of an $SU(n)$ representation, which has $m_i$ boxes in the $i^{\mathrm{th}}$ row, are defined as
\begin{equation}
\left( m_1 \,,\, m_2 \,,\, \dots \,,\, m_n \right)_{\mathrm{YT}}
 \equiv
  \left( m_1 - m_2 \,,\, m_2 - m_3 \,,\, \dots \,,\, m_{n-1} - m_n \right)_{\mathrm{D}} \,.
\end{equation}

On the other hand, Dynkin labels of a $USp(2n)$ representation can be written in terms of the Young tableau labels and
the number of `colors' as
\begin{equation}
\left( m_1 \,,\, m_2 \,,\, \dots \,,\, m_n \right)_{\mathrm{YT}}
 \equiv
  \left( m_{n-1} - m_n \,,\, m_{n-2} - m_{n-1} \,,\, \dots \,,\, m_1 - m_2 \,,\, P - m_1 \right)_{\mathrm{D}} \,.
\end{equation}

\Yboxdim5.0pt
\begin{longtable}[c]{|l|c|c|c|c|}
\kill
\caption[The spectrum of the eleven dimensional supergravity compactified on $AdS_7 \times S^4$.]{\normalsize The
spectrum of the eleven dimensional supergravity compactified on $AdS_7 \times S^4$.
\label{Table:AdS7S4}} \\
\hline
\hline
& & & & \\
{\scriptsize $SU(4) \times SU(2)$} &
{\scriptsize $SU(4)_{\mathrm{D}}$} &
{\scriptsize $USp(4)_{\mathrm{D}}$} &
{\scriptsize AdS} &
{\scriptsize Field in} \\
{\scriptsize Young tableau} &
{\scriptsize labels} &
{\scriptsize labels} &
{\scriptsize energy} &
{\scriptsize $d=7$} \\
& & & & \\
\hline
& & & & \\
\endfirsthead
\caption[]{(continued)} \\
\hline
& & & & \\
{\scriptsize $SU(4) \times SU(2)$} &
{\scriptsize $SU(4)_{\mathrm{D}}$} &
{\scriptsize $USp(4)_{\mathrm{D}}$} &
{\scriptsize AdS} &
{\scriptsize Field in} \\
{\scriptsize Young tableau} &
{\scriptsize labels} &
{\scriptsize labels} &
{\scriptsize energy} &
{\scriptsize $d=7$} \\
& & & & \\
\hline
& & & & \\
\endhead
& & & & \\
\hline
\endfoot
& & & & \\
\hline
\hline
\endlastfoot
$P \geqslant 1$                 &           &           &                  &                                \\[8pt]
$\ket{0,0}$                     & $(0,0,0)$ & $(0,P)$   & $2P$             & scalar                         \\[8pt]
$\ket{\yng(1),\yng(1)}$         & $(1,0,0)$ & $(1,P-1)$ & $2P+\frac{1}{2}$ & spinor                         \\[8pt]
$\ket{\yng(2),\yng(1,1)}$       & $(2,0,0)$ & $(0,P-1)$ & $2P+1$           & $\sqrt{a_{\alpha\beta\gamma}}$ \\[8pt]
                                &           &           &                  &                                \\
$P \geqslant 2$                 &           &           &                  &                                \\[8pt]
$\ket{\yng(1,1),\yng(2)}$       & $(0,1,0)$ & $(2,P-2)$ & $2P+1$           & vector                         \\[8pt]
$\ket{\yng(2,1),\yng(2,1)}$     & $(1,1,0)$ & $(1,P-2)$ & $2P+\frac{3}{2}$ & gravitino                      \\[8pt]
$\ket{\yng(2,2),\yng(2,2)}$     & $(0,2,0)$ & $(0,P-2)$ & $2P+2$           & graviton                       \\[8pt]
                                &           &           &                  &                                \\
$P \geqslant 3$                 &           &           &                  &                                \\[8pt]
$\ket{\yng(1,1,1),\yng(3)}$     & $(0,0,1)$ & $(3,P-3)$ & $2P+\frac{3}{2}$ & spinor                         \\[8pt]
$\ket{\yng(2,1,1),\yng(3,1)}$   & $(1,0,1)$ & $(2,P-3)$ & $2P+2$           & $a_{\alpha\beta}$              \\[8pt]
$\ket{\yng(2,2,1),\yng(3,2)}$   & $(0,1,1)$ & $(1,P-3)$ & $2P+\frac{5}{2}$ & gravitino                      \\[8pt]
$\ket{\yng(2,2,2),\yng(3,3)}$   & $(0,0,2)$ & $(0,P-3)$ & $2P+3$           & $\sqrt{a_{\alpha\beta\gamma}}$ \\[8pt]
                                &           &           &                  &                                \\
$P \geqslant 4$                 &           &           &                  &                                \\[8pt]
$\ket{\yng(1,1,1,1),\yng(4)}$   & $(0,0,0)$ & $(4,P-4)$ & $2P+2$           & scalar                         \\[8pt]
$\ket{\yng(2,1,1,1),\yng(4,1)}$ & $(1,0,0)$ & $(3,P-4)$ & $2P+\frac{5}{2}$ & spinor                         \\[8pt]
$\ket{\yng(2,2,1,1),\yng(4,2)}$ & $(0,1,0)$ & $(2,P-4)$ & $2P+3$           & vector                         \\[8pt]
$\ket{\yng(2,2,2,1),\yng(4,3)}$ & $(0,0,1)$ & $(1,P-4)$ & $2P+\frac{7}{2}$ & spinor                         \\[8pt]
$\ket{\yng(2,2,2,2),\yng(4,4)}$ & $(0,0,0)$ & $(0,P-4)$ & $2P+4$           & scalar                         \\[8pt]
\end{longtable}
\Yboxdim7.0pt


\subsection{Spectrum of eleven dimensional supergravity on $AdS_4 \times S^7$}
\label{ApSubSec:AdS4S7Sugra}

In Table \ref{Table:AdS4S7} we give the spectrum of the eleven dimensional supergravity on $AdS_4 \times S^7$,
following the oscillator realization of $OSp(8|4,\mathbb{R})$ reviewed in section \ref{SubSec:AdS4S7}. This was first
constructed in \cite{GW86}.

The Gelfand-Zetlin labels of an $SO(8)$ representation, whose $SU(4)$ Young tableaux is
$\left( m_1 \,,\, m_2 \,,\, m_3 \right)$ is given by
\begin{equation}
\left( n - \frac{N_F}{2} \,,\, \frac{1}{2} ( m_1 + m_2 - m_3 ) \,,\, \frac{1}{2} ( m_1 - m_2 + m_3 ) \,,\,
 \frac{1}{2} ( - m_1 + m_2 + m_3 ) \right)_{\mathrm{G-Z}}
\end{equation}
where $n$ is the number of `colors' and $N_F$ is the fermionic number operator.
The same representation in Dynkin notation can be written as
\begin{equation}
\left( n - \frac{N_F}{2} - \frac{1}{2} ( m_1 + m_2 - m_3 ) \,,\, m_2 - m_3 \,,\, m_1 - m_2 \,,\,
 m_3 \right)_{\mathrm{D}} \,.
\end{equation}

\Yboxdim5.0pt
\begin{longtable}[c]{|l|c|c|c|c|}
\kill
\caption[The spectrum of the eleven dimensional supergravity compactified on $AdS_4 \times S^7$.]{\normalsize The
spectrum of the eleven dimensional supergravity compactified on $AdS_4 \times S^7$.
\label{Table:AdS4S7}} \\
\hline
\hline
& & & & \\
{\scriptsize $SU(2) \times SU(4)$} &
{\scriptsize Spin and} &
{\scriptsize AdS} &
{\scriptsize $SO(8)_{\mathrm{G-Z}}$} &
{\scriptsize $SO(8)_{\mathrm{D}}$} \\
{\scriptsize Young tableau} &
{\scriptsize parity} &
{\scriptsize energy} &
{\scriptsize labels} &
{\scriptsize labels} \\
& & & & \\
\hline
& & & & \\
\endfirsthead
\caption[]{(continued)} \\
\hline
& & & & \\
{\scriptsize $SU(2) \times SU(4)$} &
{\scriptsize Spin and} &
{\scriptsize AdS} &
{\scriptsize $SO(8)_{\mathrm{G-Z}}$} &
{\scriptsize $SO(8)_{\mathrm{D}}$} \\
{\scriptsize Young tableau} &
{\scriptsize parity} &
{\scriptsize energy} &
{\scriptsize labels} &
{\scriptsize labels} \\
& & & & \\
\hline
& & & & \\
\endhead
& & & & \\
\hline
\endfoot
& & & & \\
\hline
\hline
\endlastfoot
$n \geqslant 1$                 &               &                           &
                                                        &               \\[8pt]
$\ket{0,0}$                     & $0^+$         & $\frac{n}{2}$             &
 $(n,0,0,0)$                                            & $(n,0,0,0)$   \\[8pt]
$\ket{\yng(1),\yng(1)}$         & $\frac{1}{2}$ & $\frac{n}{2}+\frac{1}{2}$ &
 $(n-\frac{1}{2},\frac{1}{2},\frac{1}{2},-\frac{1}{2})$ & $(n-1,0,1,0)$ \\[8pt]
                                &               &                           &
                                                        &               \\
$n \geqslant 2$                 &               &                           &
                                                        &               \\[8pt]
$\ket{\yng(2),\yng(1,1)}$       & $1^-$         & $\frac{n}{2}+1$           &
 $(n-1,1,0,0)$                                          & $(n-2,1,0,0)$ \\[8pt]
$\ket{\yng(1,1),\yng(2)}$       & $0^-$         & $\frac{n}{2}+1$           &
 $(n-1,1,1,-1)$                                         & $(n-2,0,2,0)$ \\[8pt]
$\ket{\yng(3),\yng(1,1,1)}$     & $\frac{3}{2}$ & $\frac{n}{2}+\frac{3}{2}$ &
 $(n-\frac{3}{2},\frac{1}{2},\frac{1}{2},\frac{1}{2})$  & $(n-2,0,0,1)$ \\[8pt]
$\ket{\yng(4),\yng(1,1,1,1)}$   & 2             & $\frac{n}{2}+2$           &
 $(n-2,0,0,0)$                                          & $(n-2,0,0,0)$ \\[8pt]
                                &               &                           &
                                                        &               \\
$n \geqslant 3$                 &               &                           &
                                                        &               \\[8pt]
$\ket{\yng(2,1),\yng(2,1)}$     & $\frac{1}{2}$ & $\frac{n}{2}+\frac{3}{2}$ &
 $(n-\frac{3}{2},\frac{3}{2},\frac{1}{2},-\frac{1}{2})$ & $(n-3,1,1,0)$ \\[8pt]
$\ket{\yng(3,1),\yng(2,1,1)}$   & $1^+$         & $\frac{n}{2}+2$           &
 $(n-2,1,1,0)$                                          & $(n-3,0,1,1)$ \\[8pt]
$\ket{\yng(4,1),\yng(2,1,1,1)}$ & $\frac{3}{2}$ & $\frac{n}{2}+\frac{5}{2}$ &
 $(n-\frac{5}{2},\frac{1}{2},\frac{1}{2},-\frac{1}{2})$ & $(n-3,0,1,0)$ \\[8pt]
                                &               &                           &
                                                        &               \\
$n \geqslant 4$                 &               &                           &
                                                        &               \\[8pt]
$\ket{\yng(2,2),\yng(2,2)}$     & $0^+$         & $\frac{n}{2}+2$           &
 $(n-2,2,0,0)$                                          & $(n-4,2,0,0)$ \\[8pt]
$\ket{\yng(3,2),\yng(2,2,1)}$   & $\frac{1}{2}$ & $\frac{n}{2}+\frac{5}{2}$ &
 $(n-\frac{5}{2},\frac{3}{2},\frac{1}{2},\frac{1}{2})$  & $(n-4,1,0,1)$ \\[8pt]
$\ket{\yng(4,2),\yng(2,2,1,1)}$ & $1^-$         & $\frac{n}{2}+3$           &
 $(n-3,1,0,0)$                                          & $(n-4,1,0,0)$ \\[8pt]
$\ket{\yng(3,3),\yng(2,2,2)}$   & $0^-$         & $\frac{n}{2}+3$           &
 $(n-3,1,1,1)$                                          & $(n-4,0,0,2)$ \\[8pt]
$\ket{\yng(4,3),\yng(2,2,2,1)}$ & $\frac{1}{2}$ & $\frac{n}{2}+\frac{7}{2}$ &
 $(n-\frac{7}{2},\frac{1}{2},\frac{1}{2},\frac{1}{2})$  & $(n-4,0,0,1)$ \\[8pt]
$\ket{\yng(4,4),\yng(2,2,2,2)}$ & $0^+$         & $\frac{n}{2}+4$           &
 $(n-4,0,0,0)$                                          & $(n-4,0,0,0)$ \\[8pt]
\end{longtable}
\Yboxdim7.0pt


\end{document}